%% file: rcw103.tex
\documentclass[iop]{emulateapj}
\usepackage{natbib}
\usepackage{subfigure}

\usepackage[backref,breaklinks,colorlinks,citecolor=blue]{hyperref}
\usepackage[all]{hypcap}

\newcommand{\myemail}{kafrank@psu.edu}

\newcommand{\rcw}{RCW 103}

\newcommand{\nely}{Ne Ly$\alpha$}
\newcommand{\nehefe}{Ne He$\alpha$+Fe L}
\newcommand{\tauunit}{$10^{11}$ s cm$^{-3}$}
\newcommand{\nhunit}{$10^{22}\mbox{cm}^{-2}$}
\newcommand{\nh}{$N_{\mbox{\footnotesize{H}}}$}
\newcommand{\vpshock}{{\tt vpshock}}

\newcommand{\einstein}{{\it Einstein}}

\newcommand{\chandra}{{\it Chandra}}
\newcommand{\chandrafull}{{\it Chandra X-ray Observatory}}
\newcommand{\fermi}{{\it Fermi}}
\newcommand{\spitzer}{{\it Spitzer}}
\newcommand{\acis}{{\rm ACIS}}
\newcommand{\xspec}{XSPEC}

\shorttitle{\chandra\ observations of SNR RCW 103}
\shortauthors{Frank et al.}

\slugcomment{{\sc Accepted to ApJ:} August 5, 2015}

\begin{document}

\title{\chandra\ observations of SNR RCW 103}
\author{Kari A. Frank\altaffilmark{1}, David N. Burrows\altaffilmark{1}}
\altaffiltext{1}{Department of Astronomy and Astrophysics, Pennsylvania State University, University Park, PA 16802, USA; \email{\myemail}}
\author{Sangwook Park\altaffilmark{2}} 
\altaffiltext{2}{Department of Physics, University of Texas at Arlington, Arlington, TX 76019, USA}

\begin{abstract}
We analyze three \chandra\ observations, with a combined exposure time of 99 ks, of the Galactic supernova remnant \rcw, a young supernova remnant, previously with no clear detection of metal-rich ejecta.  Based on our imaging and spectral analyses of these deep \chandra\ data, we find evidence for metal-rich ejecta emission scattered throughout the remnant.  X-ray emission from the shocked ejecta is generally weak, and the shocked circumstellar medium (CSM) is a largely dominant component across the entire remnant.  The CSM component shows abundances of $\sim$0.5 solar, while Ne, Mg, Si, S, and Fe abundances of the ejecta are up to a few times solar. Comparison of these ejecta abundances with yields from supernova nucleosynthesis models suggests, together with the existence of a central neutron star, a progenitor mass of $\sim$ 18-20 M$_\odot$, though the Fe/Si ratios are larger than predicted.  The shocked CSM emission suggests a progenitor with high mass-loss rate and subsolar metallicity.
\end{abstract}


\section{Introduction}
\label{section:intro}
\rcw\ is a young Galactic supernova remnant (SNR) that is bright at X-ray, radio, optical, and infrared wavelengths.   \citet{Xing2014} have also recently reported a likely $\gamma$-ray detection with \fermi.  Its age is estimated as $\sim$2000 yrs \citep{Carter1997,Nugent1984}, and a number of studies suggest a distance of $\sim$3.3 kpc \citep{Westerlund1969,Caswell1975,Nugent1984,Reynoso2004,Paron2006}. The presence of a central object \citep{Tuohy1980} indicates a core-collapse origin.  

\citet{Oliva1990,Oliva1999} performed infrared (IR) spectroscopy on the southern edge of \rcw\ and found emission from both molecular and ionized gas, mainly in the form of H$_2$ and [\ion{Fe}{2}] lines.  They found the molecular emission to be displaced to the outside of the atomic emission.  Multi-band imaging observations from the Two Micron All Sky Survey \citep{Rho2001} and \spitzer\ \citep{Reach2006,Pinheiro-Goncalves2011} confirm these findings for the southern edge of the remnant and find similar, but fainter, emission in the northwest.  The IR emission is mostly filamentary, though \spitzer\ 24 $\mu$m images also show diffuse emission throughout most of the remnant that closely follows the X-ray morphology.  The optical morphology follows that of the filamentary IR emission, with a bright region of filamentary emission in the south and a smaller region in the northwest \citep{Leibowitz1983,Ruiz1983,Carter1997}.  \citet{Carter1997} find an optical expansion rate of 1100 km s$^{-1}$.

Radio observations of \rcw\ at 2372 and 1362 MHz reveal a diffuse nearly circular shell of emission, with additional bright filamentary emission in the south and northwest that matches the optical and IR emission \citep{Dickel1996}. $^{12}$CO and HCO$^+$ observations provide further evidence for interaction with a molecular cloud in the south \citep{Paron2006}, as does nearby OH(1720 MHz) maser line emission \citep{Frail1996}.

Most X-ray observations have targeted the unusual central compact object (CCO).   Visible only in the X-ray \citep{DeLuca2008} and with a periodicity of 6.68 hours \citep{Esposito2011}, the nature of this CCO is still being debated \citep{Reynoso2004,Li2007,DeLuca2008,Pizzolato2008,Bhadkamkar2009,Esposito2011,Ikhsanov2013}.  X-ray observations also reveal a nearly circular shell $\sim$8$'$ in diameter, similar in morphology to the diffuse radio emission.  \citet{Nugent1984}, using \einstein\ observations, fit non-equilibrium ionization plasma models to the X-ray spectrum and found Mg, Si, S, and Fe abundances near solar, leading them to conclude that the X-ray emission is primarily due to shocked interstellar medium.  The IR analysis of \citet{Oliva1999} also found no enhanced abundances.  On the other hand, a more recent analysis of a \chandra\ observation by \citet{Lopez2011} found moderately supersolar abundances of Mg, Si, and Fe (abundances of all other elements were fixed to solar).

The overall picture of \rcw\ is that of a young remnant just beginning to interact with the surrounding CSM and molecular cloud, particularly in the south.  However, despite this interaction, the X-ray and radio morphology is nearly circular.  Combined with measurements of the magnetic field orientation \citep{Dickel1996} and optical expansion rate \citep{Carter1997}, this suggests \rcw\ has only transitioned from the ejecta-dominated double-shock phase to the swept-up CSM-dominated phase of its evolution within the last few hundred years.  The lack of a clear detection of metal-rich ejecta is thus particularly puzzling, as one would still expect emission from ejecta so close to the transition time.

\rcw\ has been observed several times with the \chandrafull\ \citep{Weisskopf1996}.  We report here results of a spectral and spatial analysis that combines the three deepest \chandra\ observations, with a total exposure of $\sim$100 ks, to provide an updated, comprehensive picture of the diffuse X-ray emission of \rcw.  The observations and data reduction procedure are described in \S\ref{section:data}.  Analyses of X-ray images, equivalent width images, and spectra of 27 regions are presented in \S\ref{section:analysisresults} and discussed in \S\ref{section:discussion}.  Conclusions are presented in \S\ref{section:conclusions}.    

\section{Observations and Data Reduction}
\label{section:data}
here are numerous available \chandra\ observations of \rcw\ taken with the Advanced CCD Imaging Spectrometer (ACIS).  The majority of these have short exposures (a few ks) and are targeted at the CCO, utilizing only a 1/4 subarray of a single ACIS-I CCD; they therefore cover only a fraction of the surface area of the remnant and the combined exposure time over most of the remnant is short compared to the other observations. The earliest observation, ObsID 123, was taken with a warmer focal plane temperature of -110 C, and we therefore do not include it as no CTI correction is available for such spectra.  We use the remaining three \acis\ observations, providing a total of 99.2 ks of exposure time.  Details are given in Table \ref{table:observations}.  ObsIDs  11823 and 12224 were imaged with the ACIS-I array.  \rcw, with a diameter of $\sim$$8'$, fits easily within the full ACIS-I field of view.  ObsID 970 was imaged with the back-illuminated ACIS-S3 chip.  \rcw\ is slightly wider than the ACIS-S3 field of view, resulting in portions of the eastern and western edges of the remnant falling outside the chip boundary.

Reduction and analysis of the observations was carried out with CIAO 4.5 and CALDB 4.4.10.  The CIAO task {\tt chandra\_repro} with standard parameters was used to create new level two event files.  Examination of the background light curves revealed only one period of high background, in ObsID 970.  The 2.6 ks interval was removed prior to further analysis.  Event files were reprojected to the common tangent point of ObsID 11823 for the imaging analysis.  Merged, exposure-corrected images were created from the reprojected event files.
\capstartfalse
\begin{deluxetable}{cccc}[h]
\tabletypesize{\footnotesize}
\tablecaption{Observation Parameters \label{table:observations}}
\tablewidth{0pt}
\tablehead{\colhead{ObsID} & \colhead{Date} & \colhead{Exposure (ks)} & \colhead{Instrument} } \\
\startdata
970 & 2000-02-08 & 18.9 & ACIS-S3\\
12224 & 2010-06-27 & 17.8 & ACIS-I \\
11823 & 2010-06-01 & 62.5 & ACIS-I 
\enddata
\end{deluxetable}
\capstarttrue
\vspace{5mm}
\section{Analysis and Results}
\label{section:analysisresults}
\subsection{X-ray Images}
\label{section:xrayimages}
\begin{figure}[htbp]
\includegraphics[width=\columnwidth]{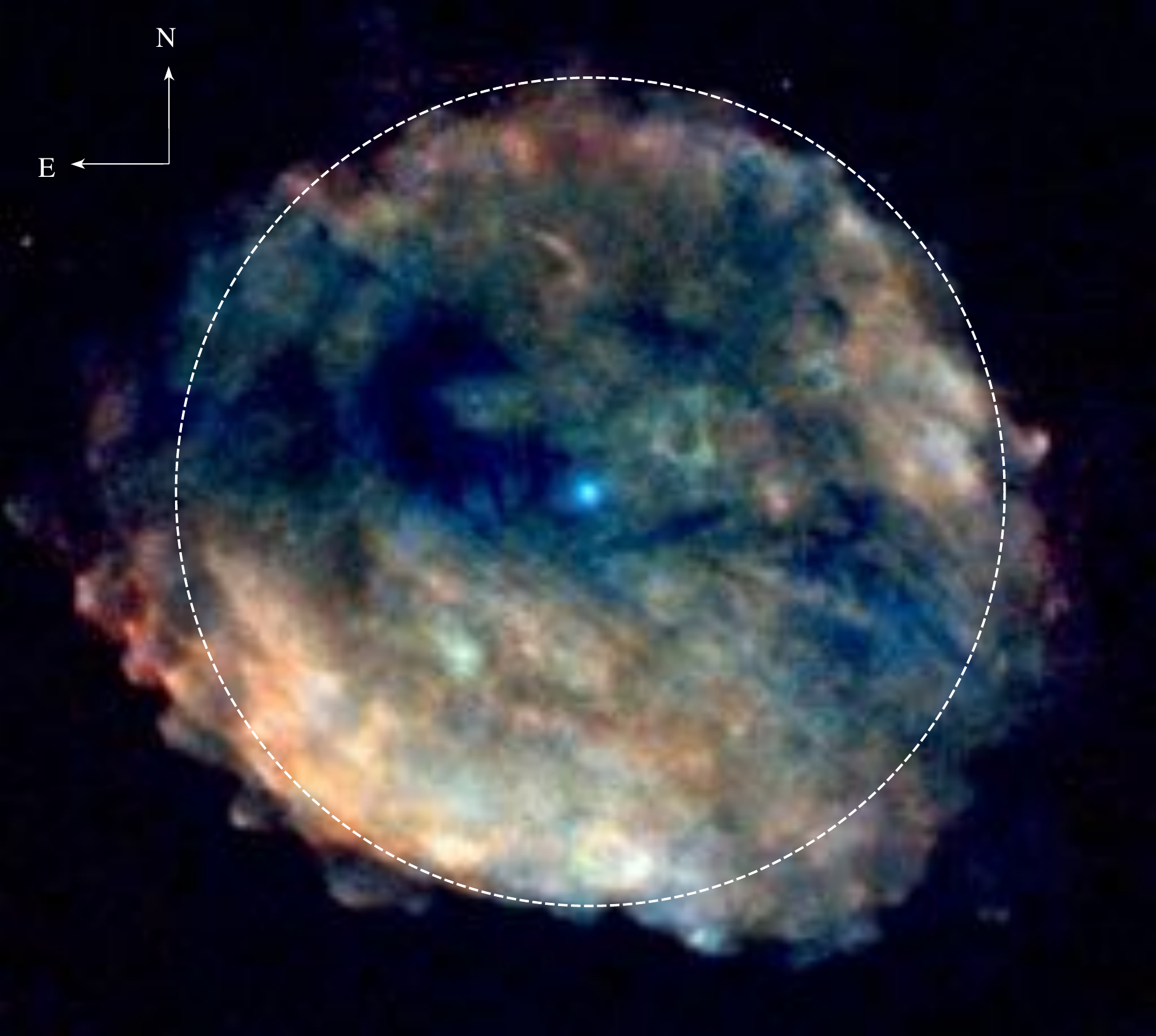}
\caption{\footnotesize Three-color, square-root scale, image of \rcw, smoothed with a two pixel gaussian.  Red $= 0.3 - 0.85$ keV, green $= 0.85 - 1.70$ keV, blue $= 1.7 - 3.0$ keV.  Green and blue are weighted by factors of 1.75 and 7, respectively, relative to red.  An 8$'$ diameter circle centered on the CCO is shown in white.}
\label{fig:rgb}
\end{figure}

\begin{figure}[ht]
\begin{center}
\subfigure {\includegraphics[width=0.45\textwidth]{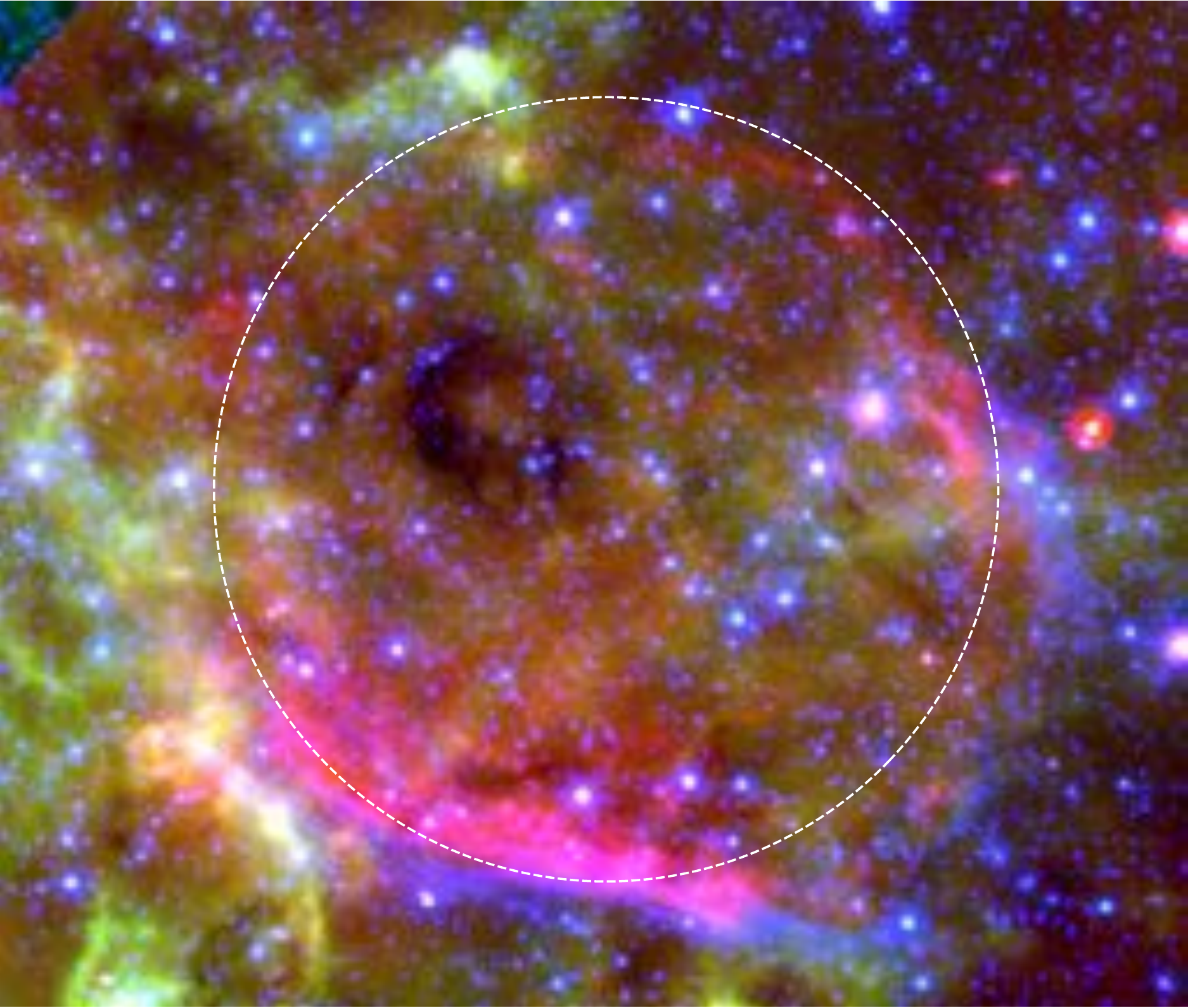}}
\caption{\footnotesize Archival \spitzer\ images of \rcw\ at 24 $\mu$m (red), 8 $\mu$m (green), and 4.5 $\mu$m (blue) with the 8$'$ circle from Figure \ref{fig:rgb} in white.  The 24 $\mu$m emission traces heated dust, the 8 $\mu$m image shows the location of ambient unshocked gas, and the diffuse 4.5 $\mu$m emission approximately traces heated molecular gas \citep[see ][]{Reach2006,Pinheiro-Goncalves2011}.}
\label{fig:spitzer}
\end{center}
\end{figure}

A three-color image of \rcw\  (Figure \ref{fig:rgb}) reveals the CCO, which dominates the $\gtrsim1.7$ keV emission.  There is no evidence for an associated pulsar wind nebula.  Large swaths of the soft emission in the northeast quadrant, and to a lesser extent the southwest, appear to be absorbed or very weak.  In particular, there is a distinctive C-shaped `hole' just northeast of the CCO, in which nearly all the X-ray emission has been absorbed, a feature also visible in the mid-infrared \citep[Figure \ref{fig:spitzer}, see also][]{Pinheiro-Goncalves2011}.  


The portions of the outermost edge that correspond to the brightest emission at 8 $\mu$m (Figure \ref{fig:spitzer}) are also very soft in X-rays (red in Figure \ref{fig:rgb}).  Overall the X-ray emission gets harder with decreasing radius.  The brightest emission comes from the southern region, where the remnant is interacting with a molecular cloud and ambient atomic gas.  The overall morphology is quite circular and $\sim$8$'$ in diameter, though the remnant appears to be elongated toward the southwest and the east.  The 24 $\mu$m emission, tracing shock-heated dust, has an almost identical morphology.

In many young SNRs, strong ejecta emission is found in distinct metal-rich X-ray features with abundances that are several times solar, such as the knots in G292.0+1.8 \citep{Park2004}, the ejecta `bullets' in the Vela SNR \citep{Miyata2001,Katsuda2006} and the `Head' region in N49 \citep{Park2012}.  There are several knots and filaments throughout \rcw, including several distinctive protrusions from the southeast edge of the remnant and one in the west, that are reminiscent of these ejecta features seen in other remnants.  We investigate the spectral characteristics of these and other features to search for evidence of ejecta in \S\ref{section:spectral} and \S\ref{section:ejecta} (individual regions are discussed in the appendix).

\subsection{Atomic Line Equivalent Width Images}
\label{section:ewi}
\begin{figure}[htb]
\begin{center}
\includegraphics[width=\columnwidth]{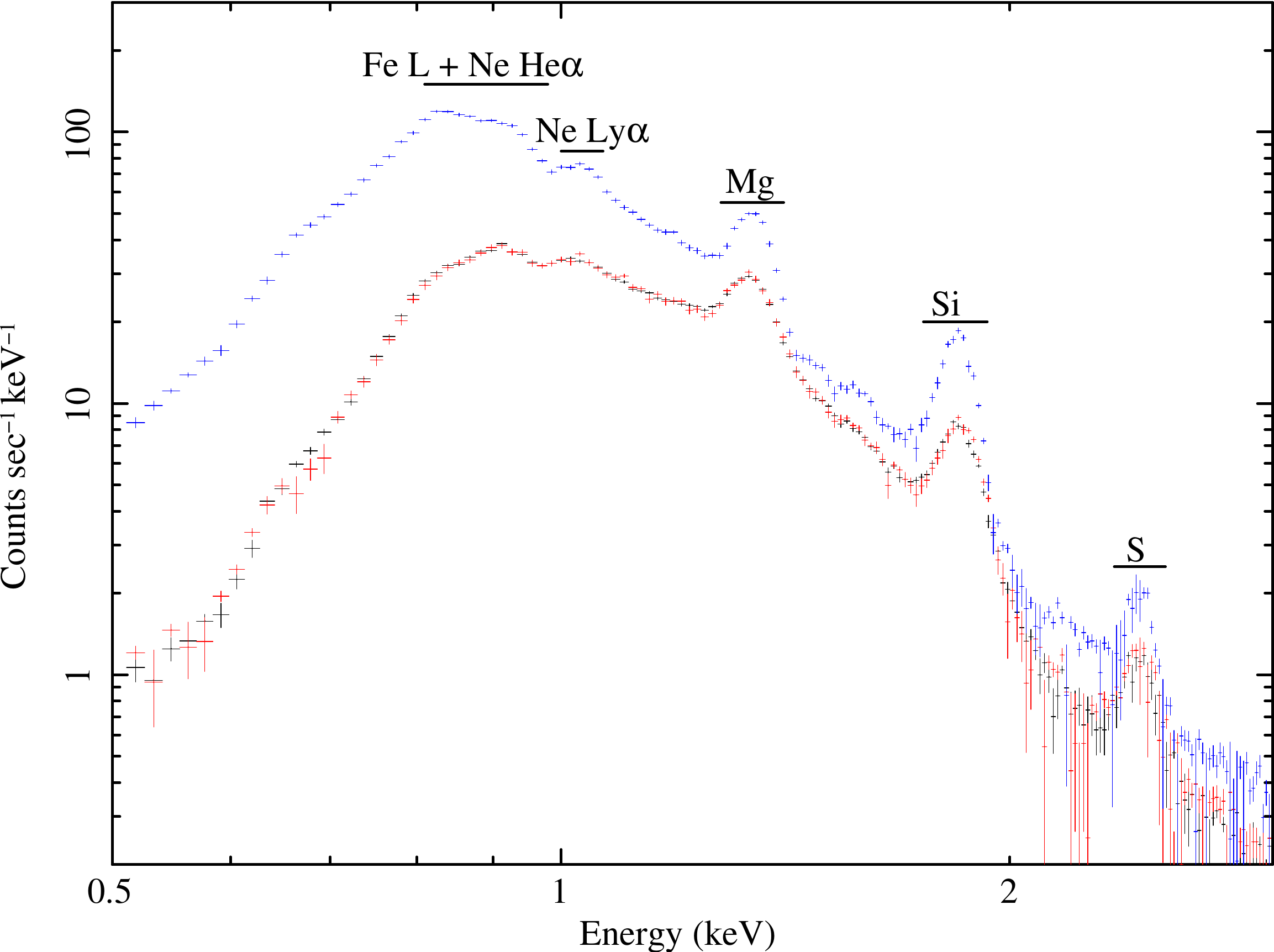}
\caption{\footnotesize Integrated spectra of \rcw\  from $0.5 - 3.0$ keV, for ObsID 11823 (black), 12224 (red), and 970 (blue).  Bands used for constructing the EWIs are shown above each line (or line combination in the case of \nehefe).}
\label{fig:integratedspectrum}
\end{center}
\end{figure}

\begin{figure*}[ht]
\begin{center}
\subfigure{\includegraphics[width=0.323\textwidth]{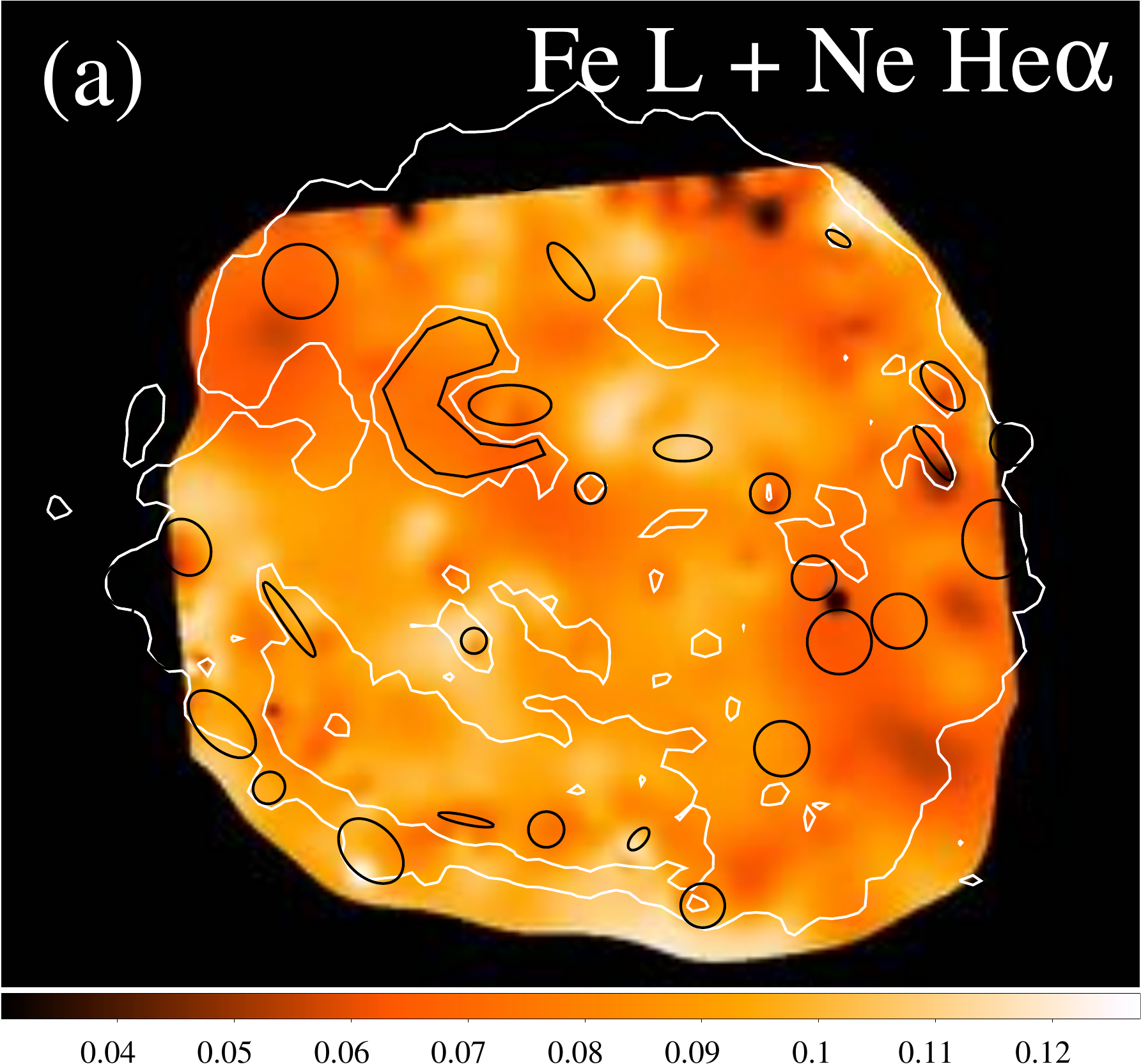}}
\subfigure{\includegraphics[width=0.330\textwidth]{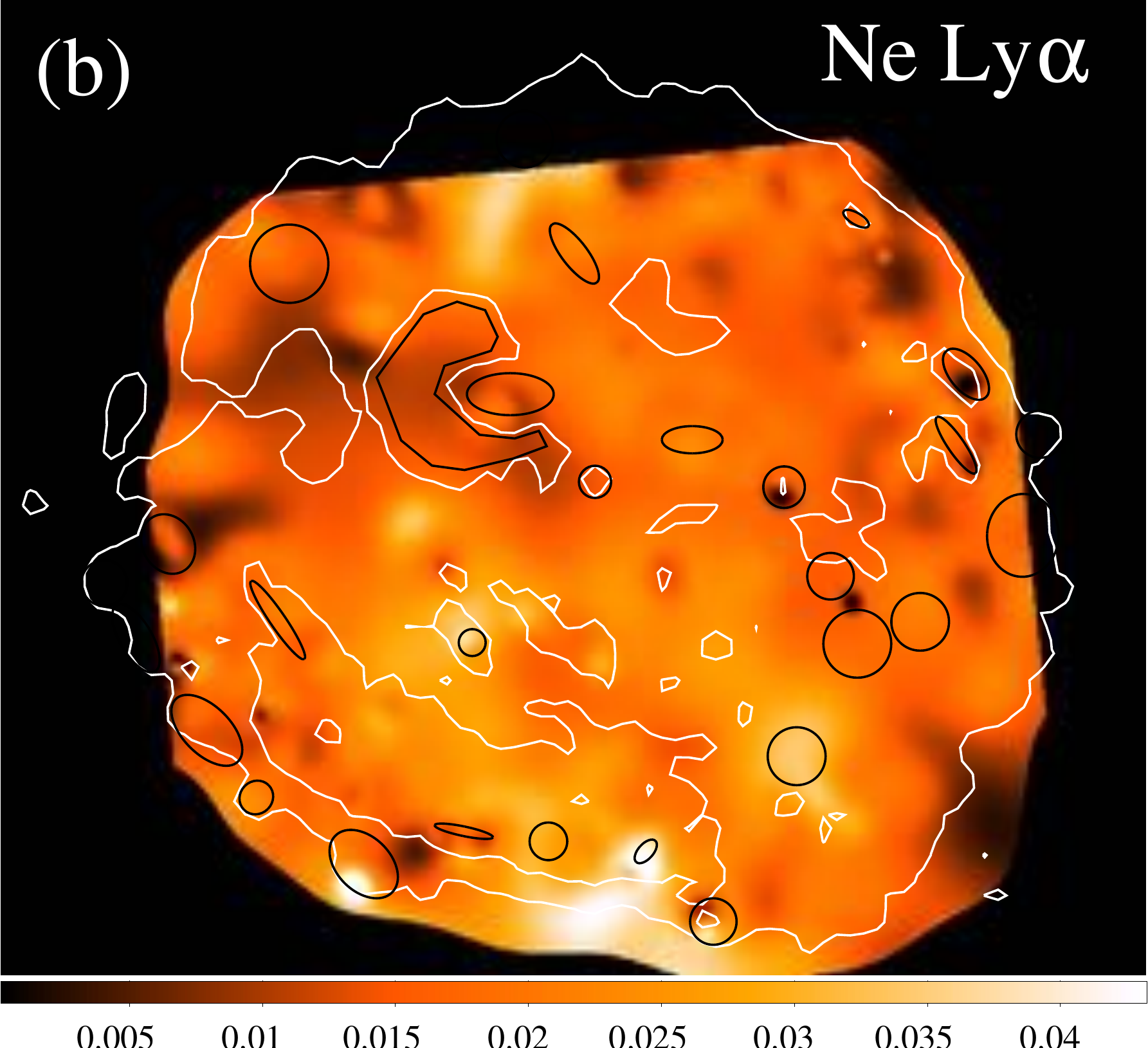}}
\subfigure{\includegraphics[width=0.329\textwidth]{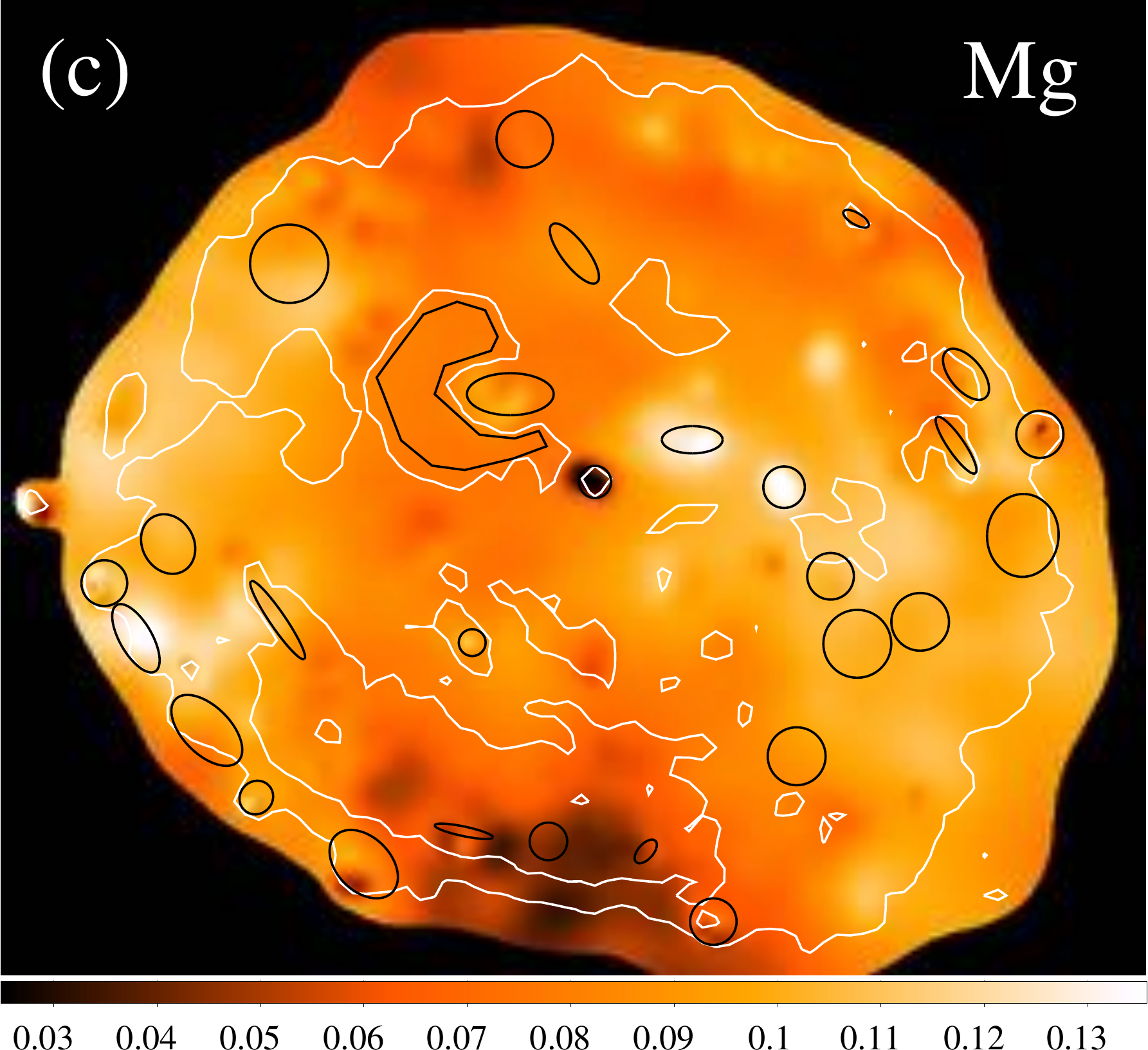}} 

\subfigure{\includegraphics[width=0.332\textwidth]{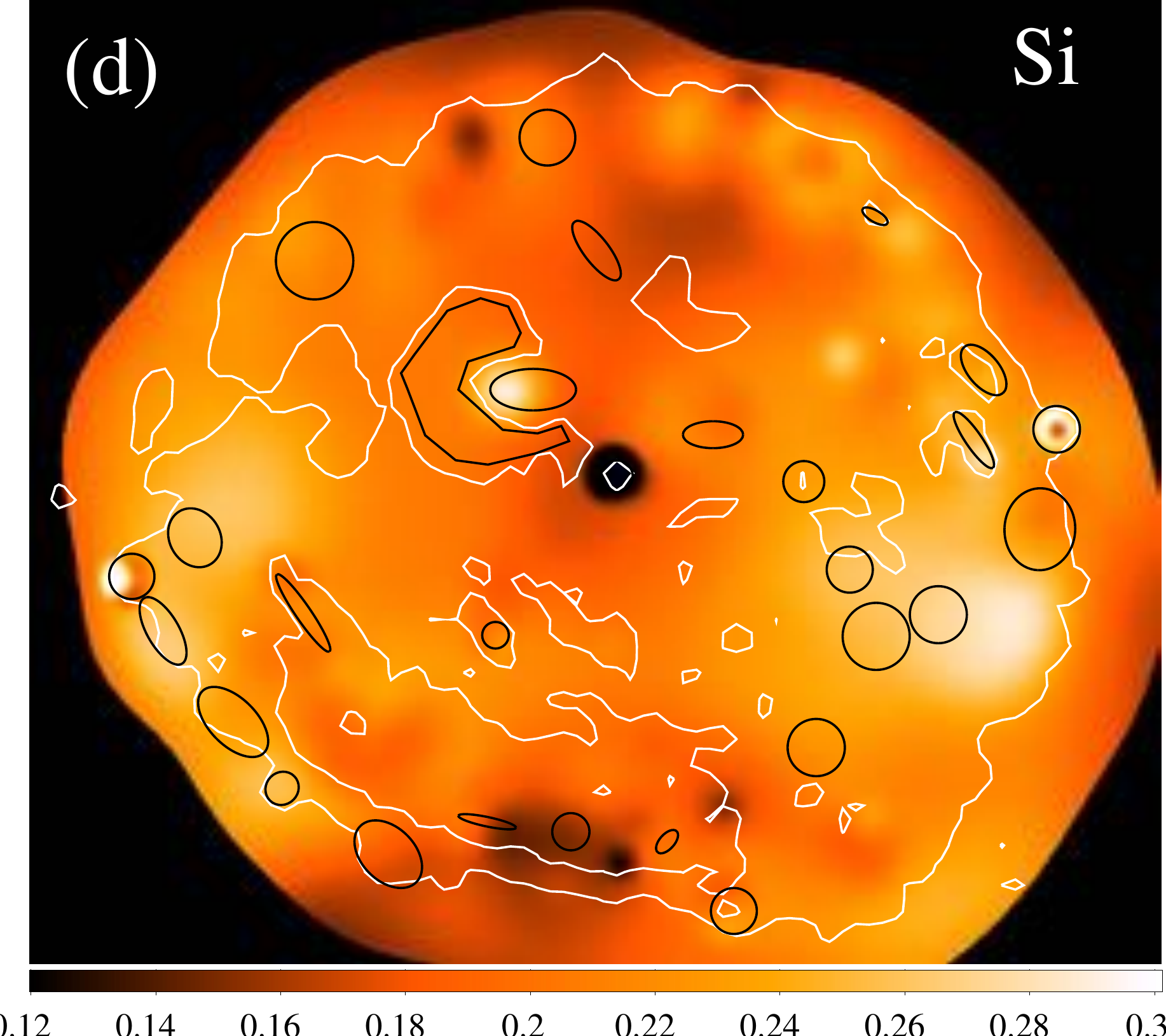}} 
\subfigure{\includegraphics[width=0.324\textwidth]{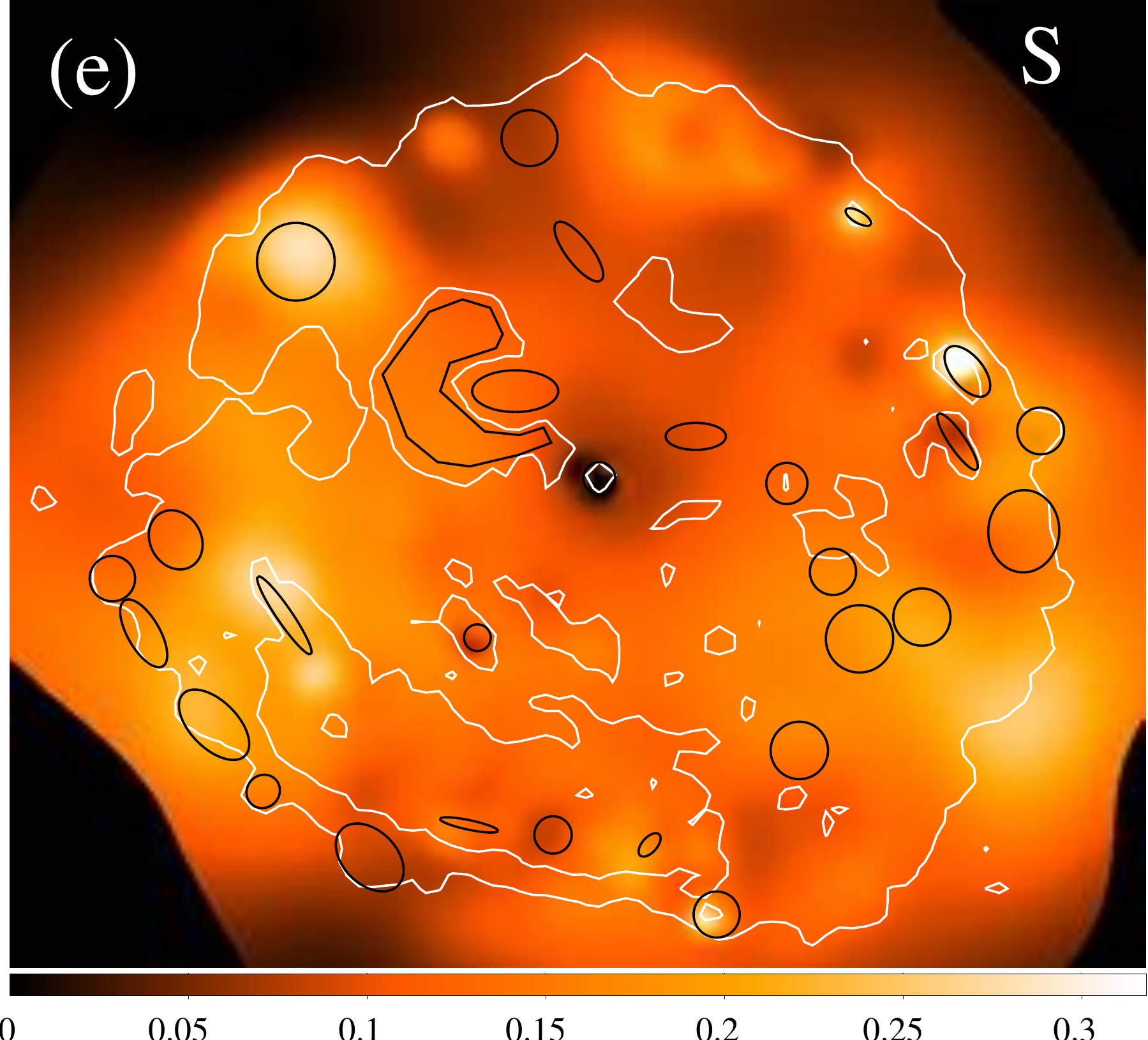}} 
\subfigure{\includegraphics[width=0.325\textwidth]{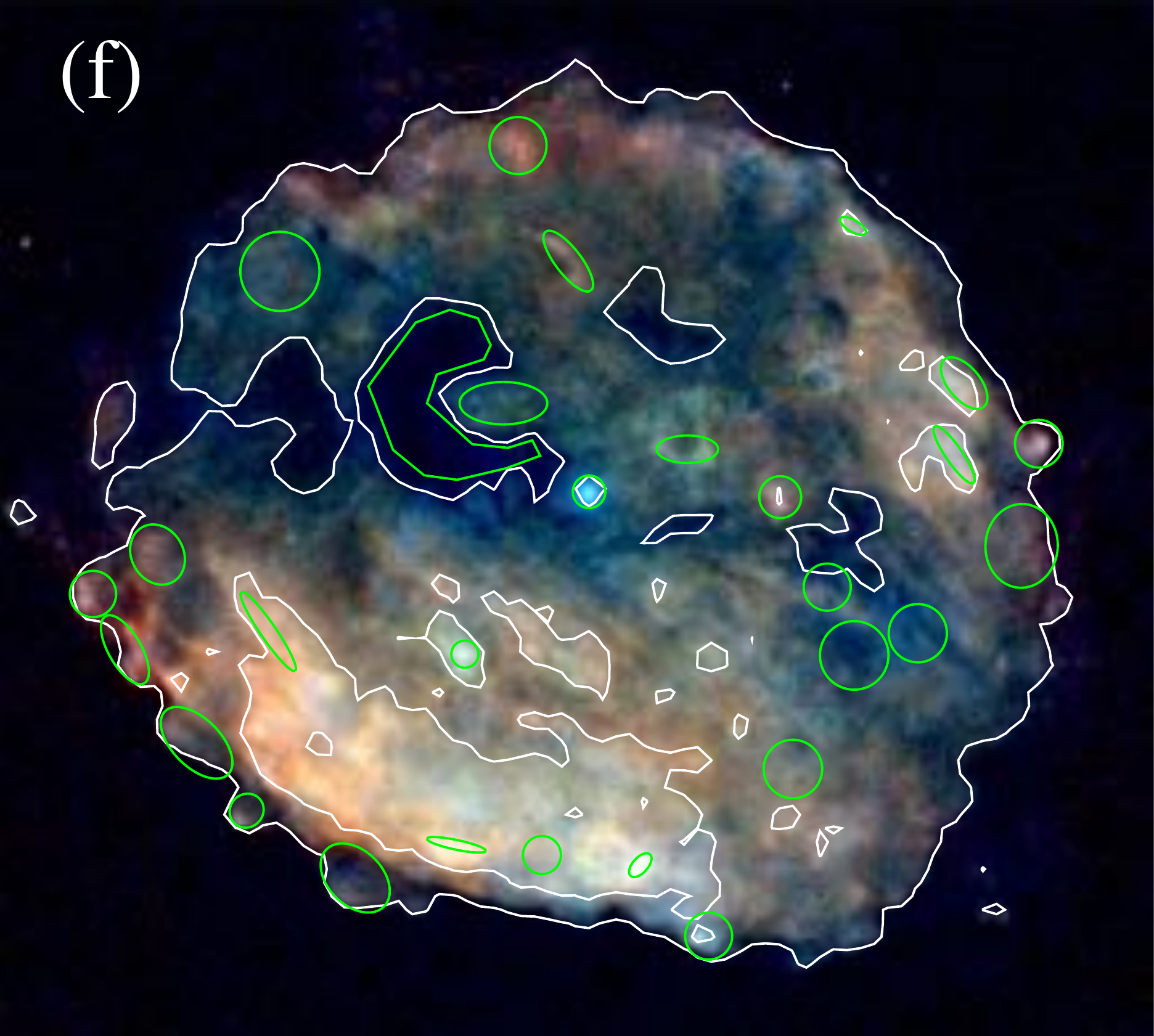}} 

\caption{\footnotesize (a) - (e): Equivalent width images with broadband X-ray contours overlaid in white.  Scales are in keV.  The regions chosen for spectral analysis are shown in black (see Figure \ref{fig:regions} for region labels).  The \nehefe, and \nely\ EWI were constructed using only ObsID 970.  The Mg, Si, and S EWI utilized all three observations.  The central holes in the Mg, Si, and S EWIs are the result of the strong continuum emission of the CCO above $\sim$1 keV.  (f):  The three-color X-ray image from Figure \ref{fig:rgb} with the same regions and contours (green and white, respectively) as shown in (a) - (f).}
\label{fig:ewi}
\end{center}
\end{figure*}

To investigate the distribution of atomic lines throughout the remnant we followed the method of \citet{Hwang2000} \citep[see also][]{Park2002} to construct equivalent width images (EWIs) for lines visible in the integrated spectrum (Figure \ref{fig:integratedspectrum}): S, Si, Mg, \nely, and \nehefe.  EWIs for the \nehefe\ and \nely\ lines were constructed using ObsID 970 only, as the ACIS-S3 chip has substantially larger effective area than ACIS-I at low energies.  Mg, Si, and S EWIs utilized all three observations.  The line bands used to create the EWIs are given in Table \ref{table:ewibands}, along with the associated continuum bands.  The equivalent width was set to zero in pixels where the estimated continuum flux was $<15\%$ of the mean to avoid noise due to poor photon statistics in the faint edge regions of the remnant.  The resulting EWIs are shown in Figure \ref{fig:ewi}.  The EWIs serve primarily as a qualitative guide for a more effective regional spectral analysis (\S\ref{section:spectral}), and no attempt is made to interpret them quantitatively.

\capstartfalse
\begin{deluxetable}{cccc}[htb]
\tabletypesize{\footnotesize}
\tablecaption{Photon Energy Bands Used for EWIs \label{table:ewibands}}
\tablewidth{0pt}
\tablehead{\colhead{Atomic Lines} & \colhead{Line} & \colhead{Low Continuum} & \colhead{High Continuum} \\
& (keV) & (keV) & (keV) }
\startdata
\nehefe & $0.81 - 0.98$ & $0.75 - 0.80$ & $1.10 - 1.14$ \\
\nely & $1.00 - 1.07$ & $0.75 - 0.80$ & $1.10 - 1.14$ \\
Mg & $1.28 - 1.41$ & $1.21 - 1.25$ & $1.43 - 1.48$ \\
Si & $1.75 - 1.93$ & $1.68 - 1.73$ & $1.95 - 2.05$ \\
S & $2.35 - 2.54$ & $1.95 - 2.05$ & $2.58 - 2.68$
\enddata
\end{deluxetable}
\capstarttrue

The S, Mg, and Si EWIs look broadly similar, distributed mainly in two large concentrations in the east and west and weaker in the north and south.  S is limited to smaller regions on the outskirts, while Si and Mg are more widely spread and extend further into the interior. Mg in particular has significant emission almost to the very center.  All three are anti-correlated with the bright southeast region. 

Unlike S, Si, and Mg, both the \nehefe\ and \nely\ emission tends to be more uniform throughout the remnant.  The \nely\ emission is strongest in the southern half of the remnant,  while \nehefe\ is strongest in the southeast and the north.  However, it is important to note that due to their close spacing and position near the peak of the X-ray spectrum, it is more difficult to determine the local continuum for these lines; the resulting EWIs are thus more uncertain than those of Mg, Si, and S.

\subsection{Spectral Analysis}
\label{section:spectral}
\begin{figure}[h]
\includegraphics[width=\columnwidth]{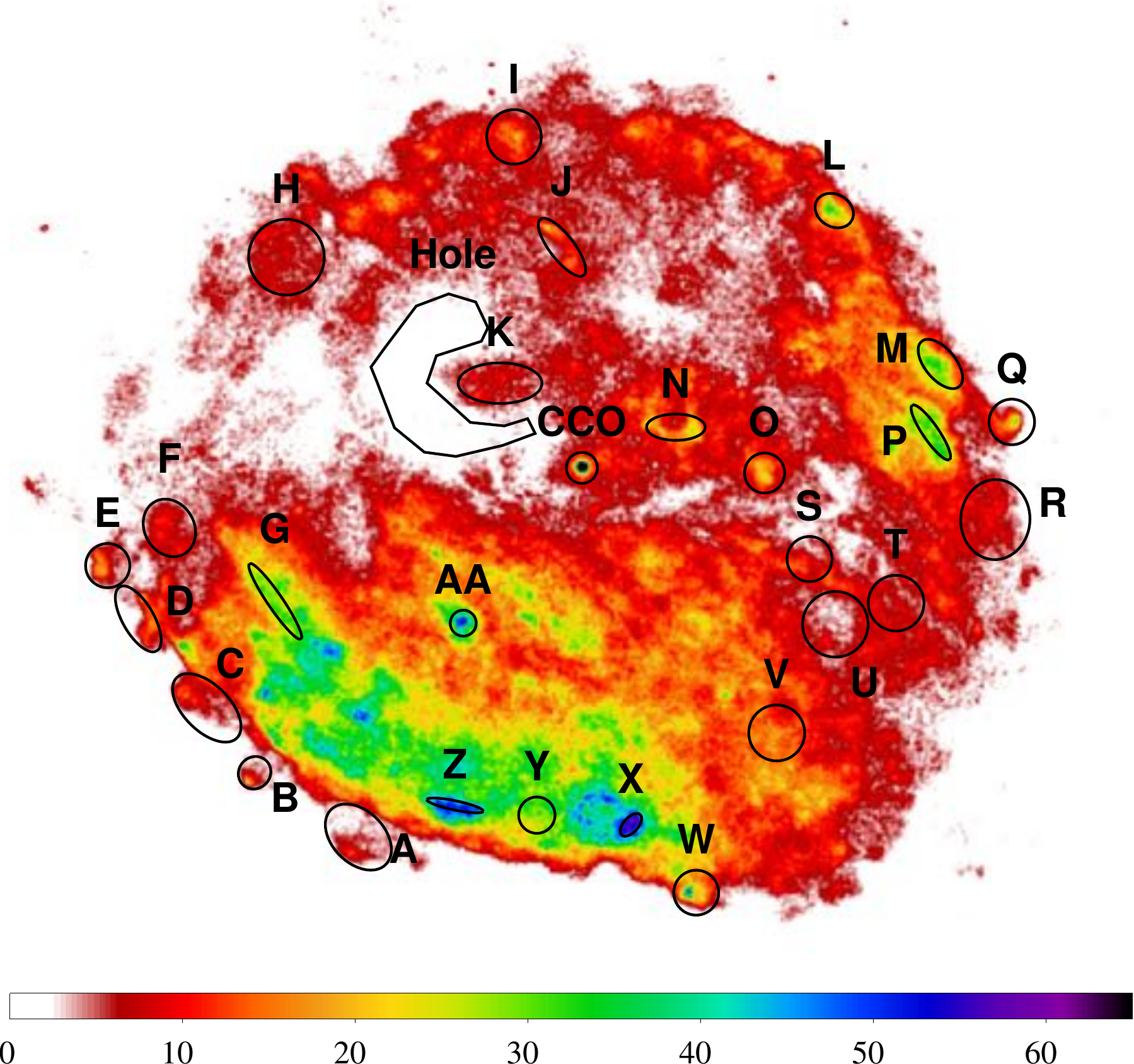} 
\caption{\footnotesize Regions chosen for spectral analysis overlaid on the X-ray image. Scale is in counts per pixel, where the raw image has been binned by a factor of 2.}
\label{fig:regions}
\end{figure}

\begin{figure*}[!htp]
\subfigure {\includegraphics[width=0.31\textwidth]{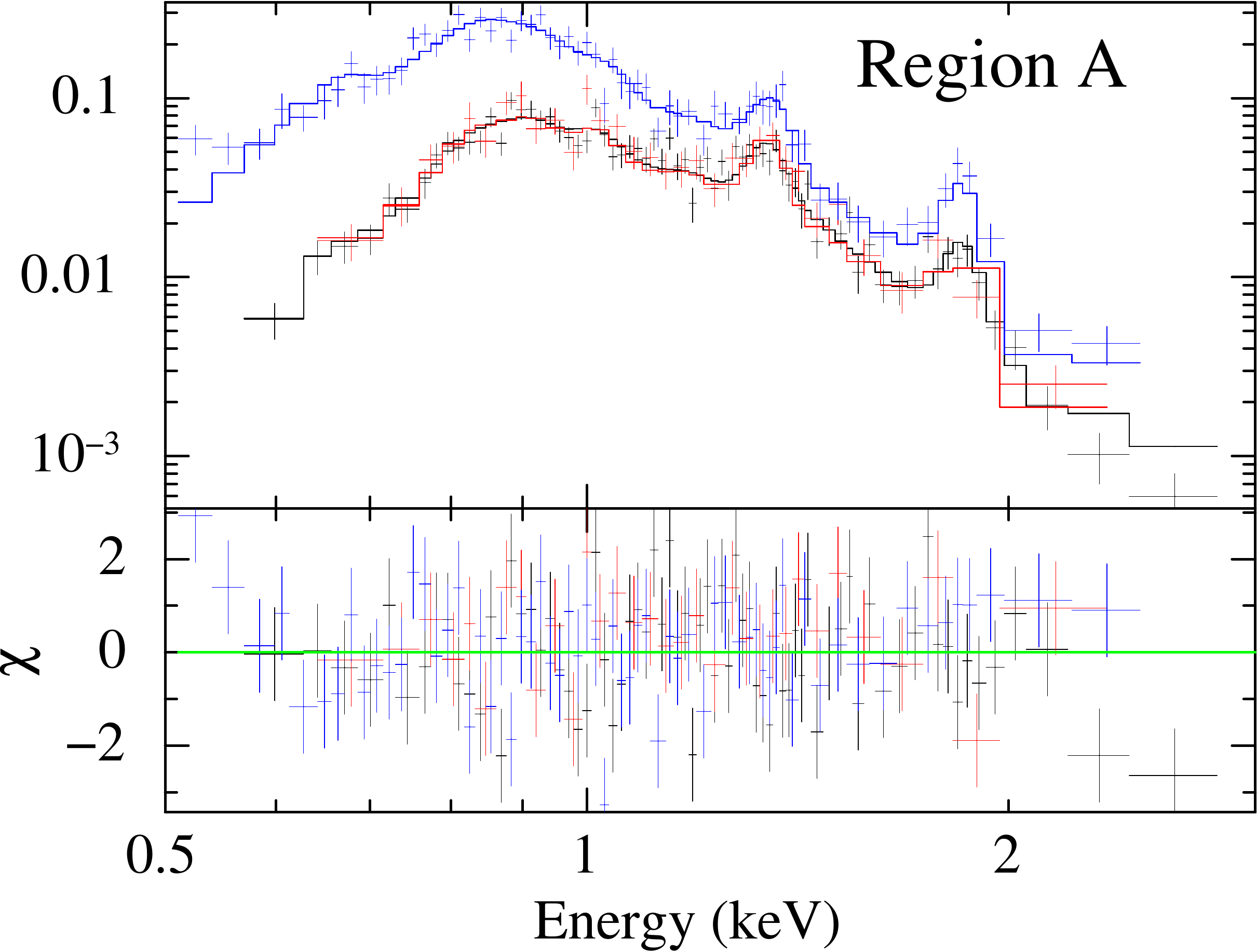}}\hspace{-2em}
\qquad
\subfigure {\includegraphics[width=0.31\textwidth]{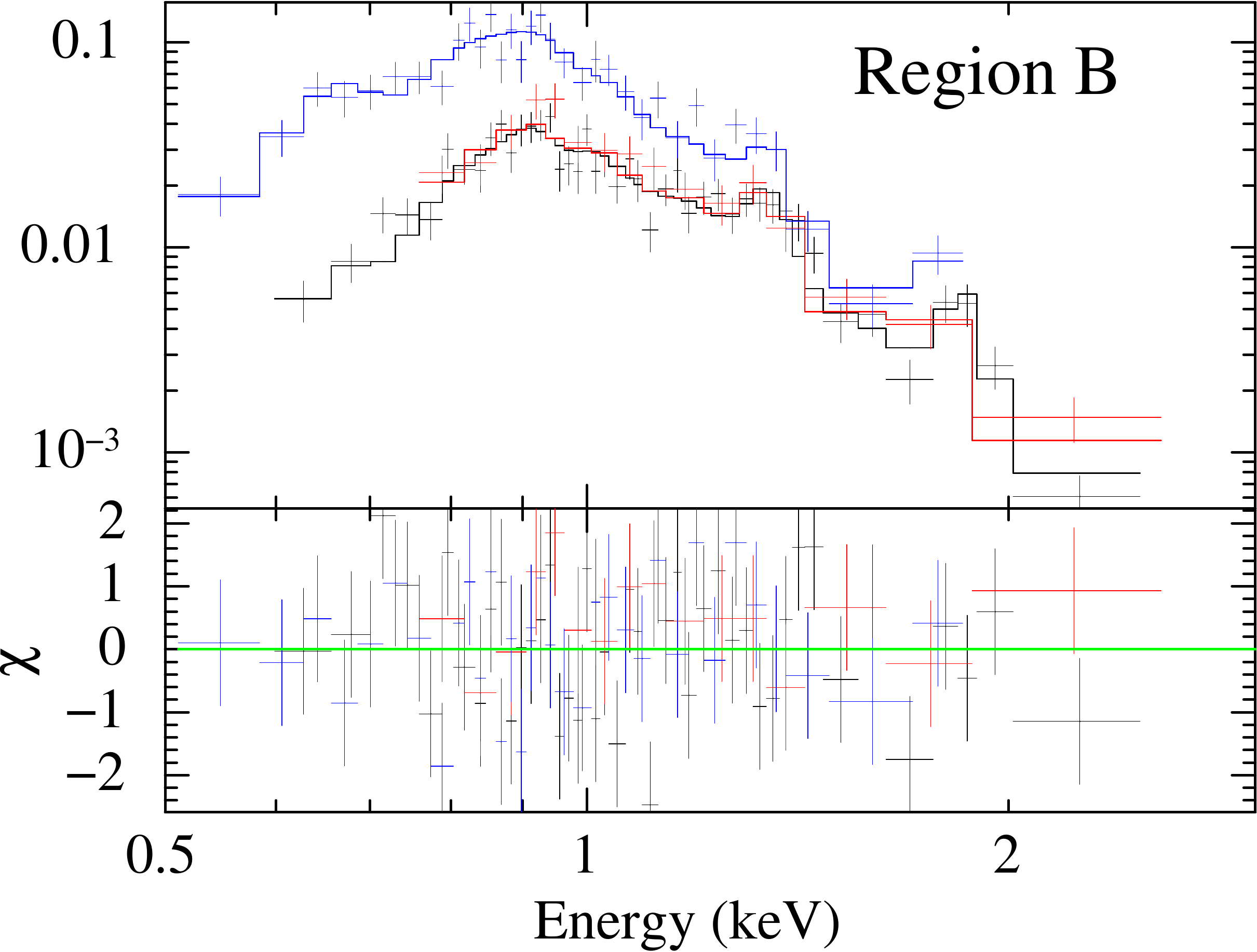}}\hspace{-2em}
\qquad
\subfigure {\includegraphics[width=0.31\textwidth]{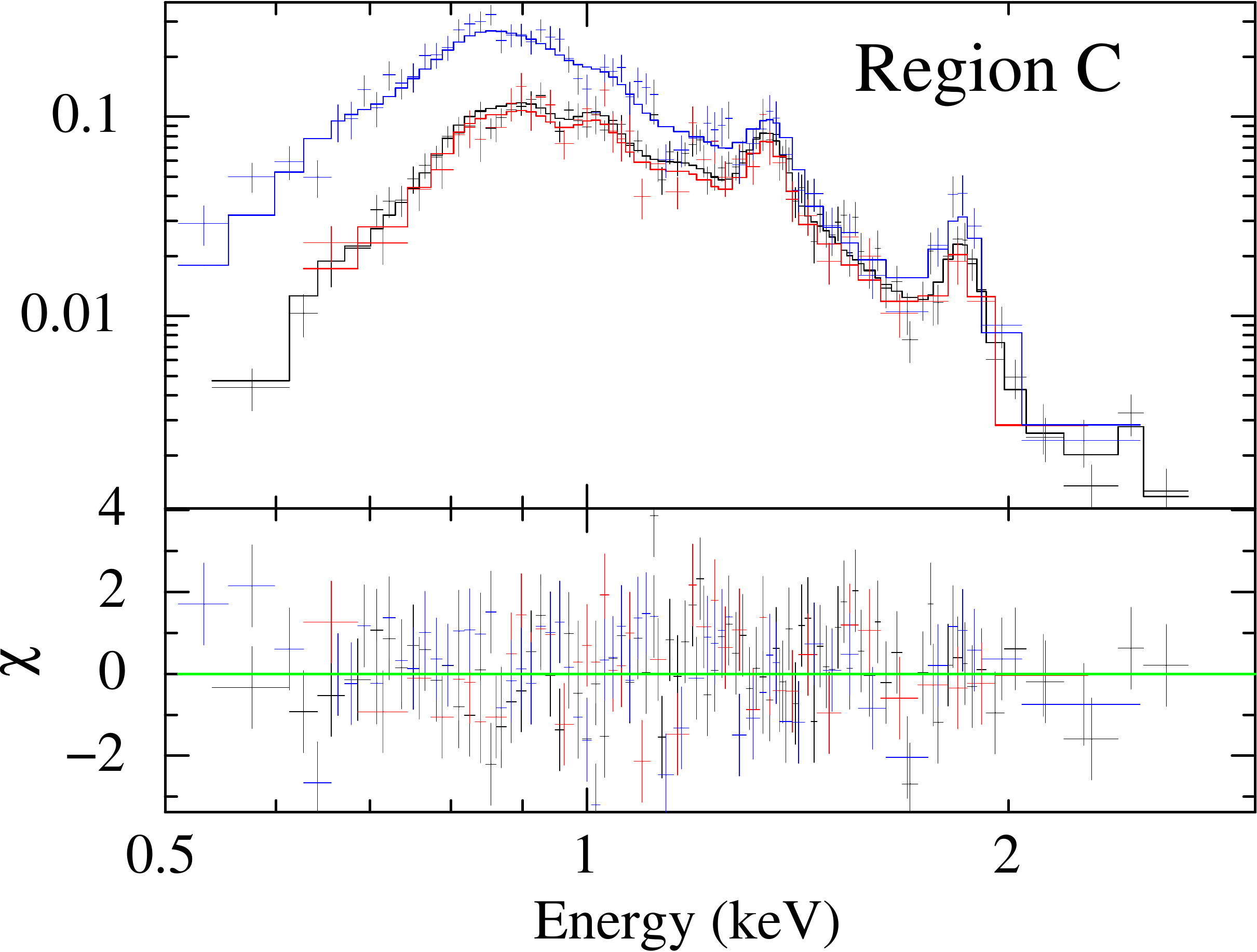}}

\subfigure {\includegraphics[width=0.31\textwidth]{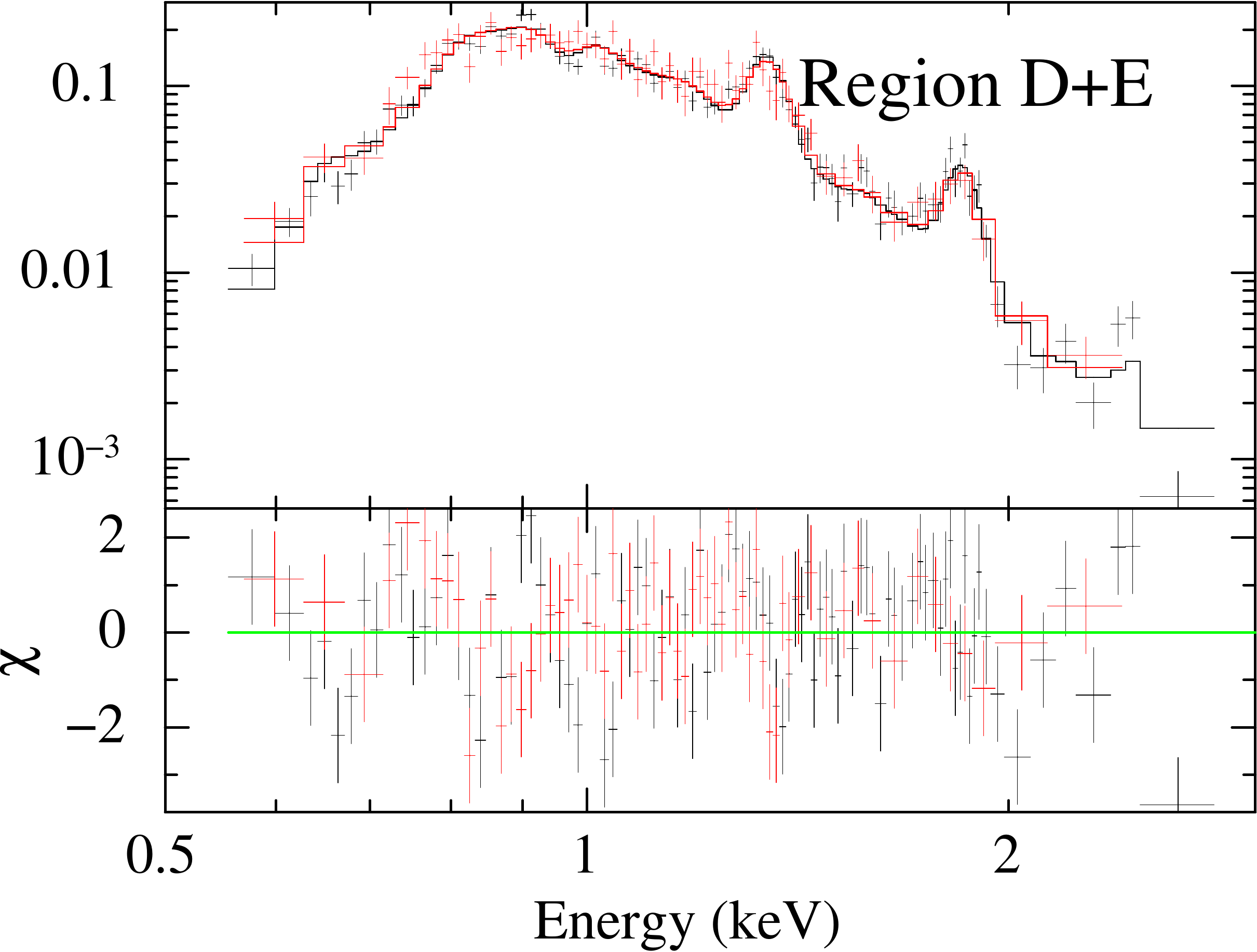}}\hspace{-2em}
\qquad
\subfigure {\includegraphics[width=0.31\textwidth]{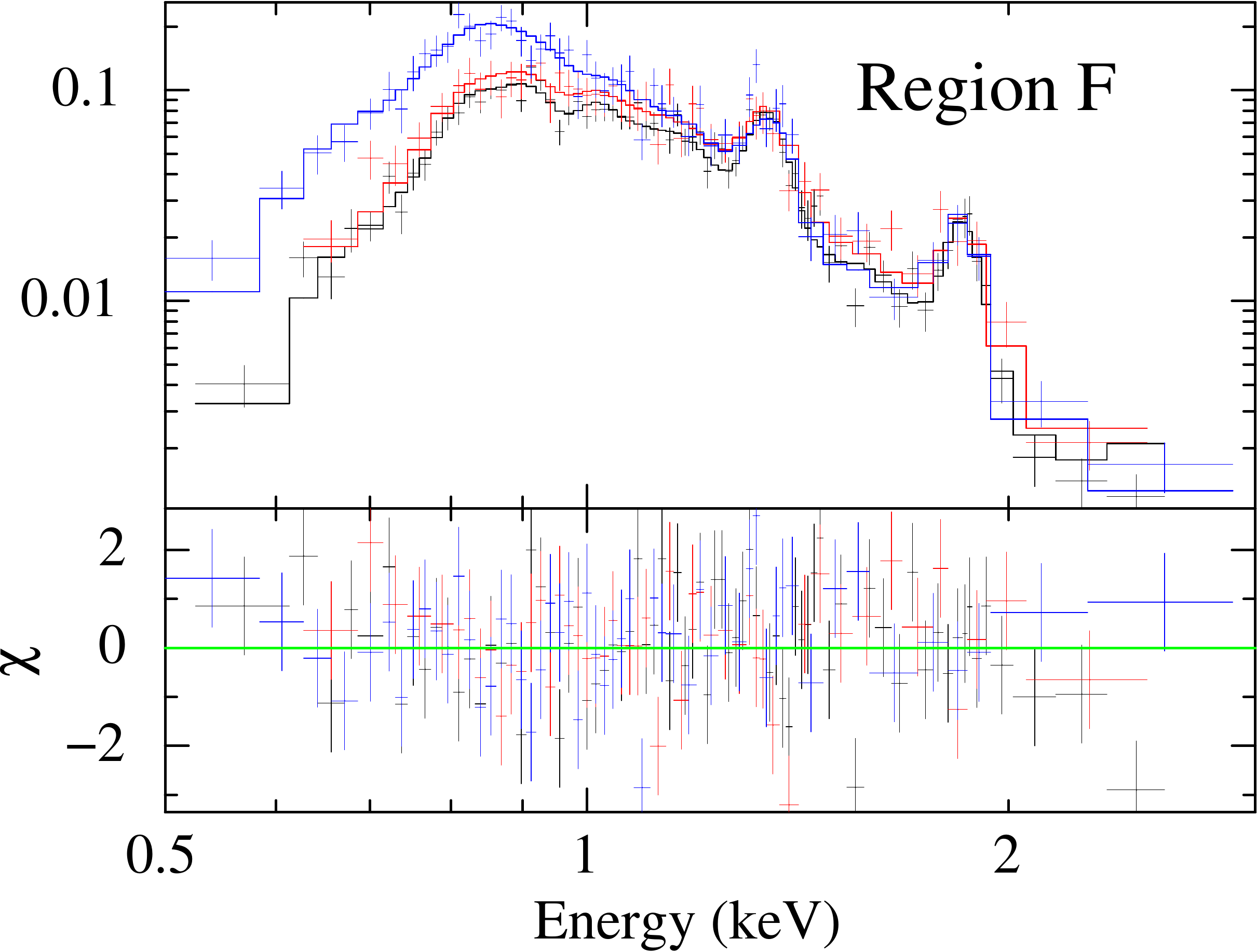}}\hspace{-2em}
\qquad
\subfigure {\includegraphics[width=0.31\textwidth]{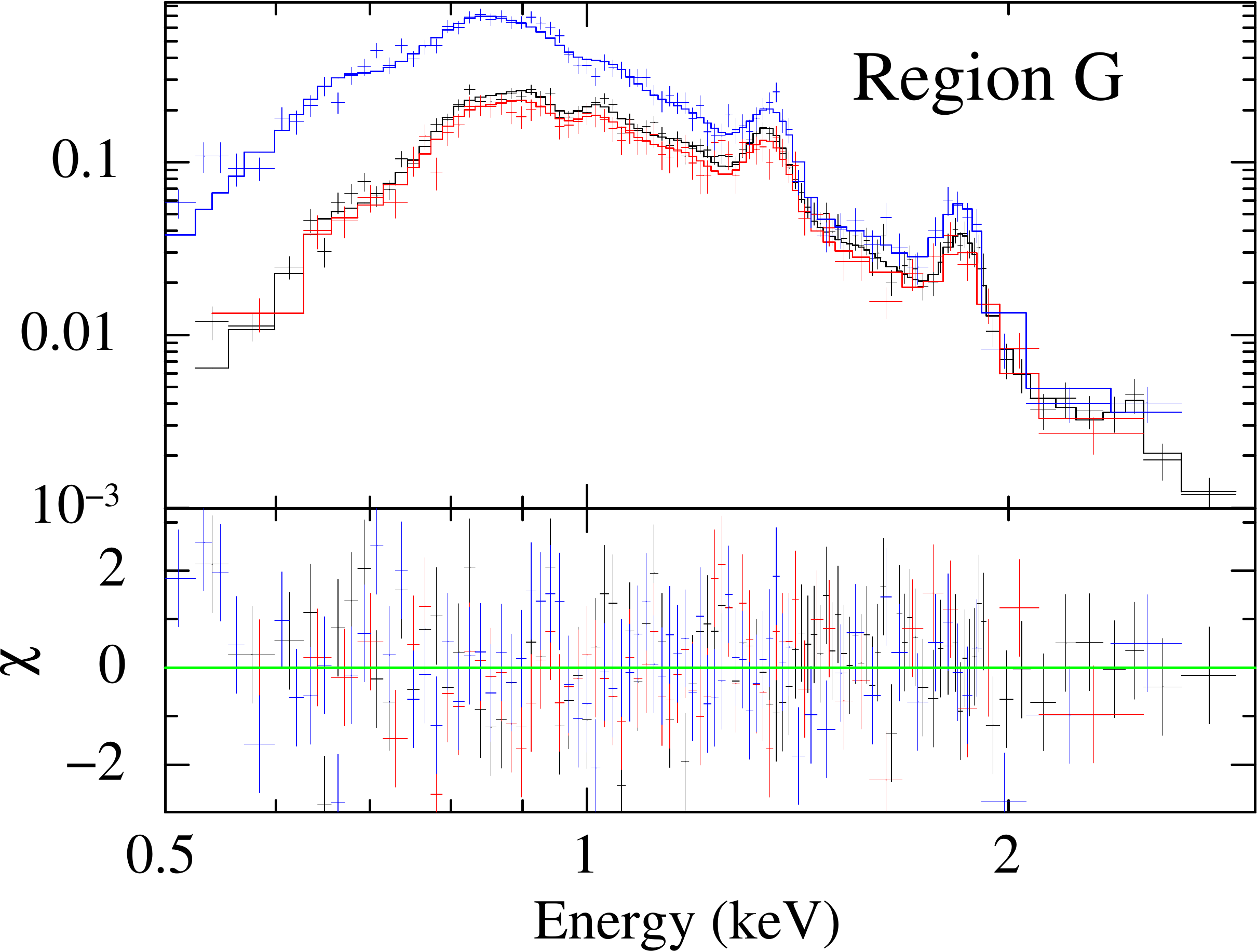}}

\subfigure {\includegraphics[width=0.31\textwidth]{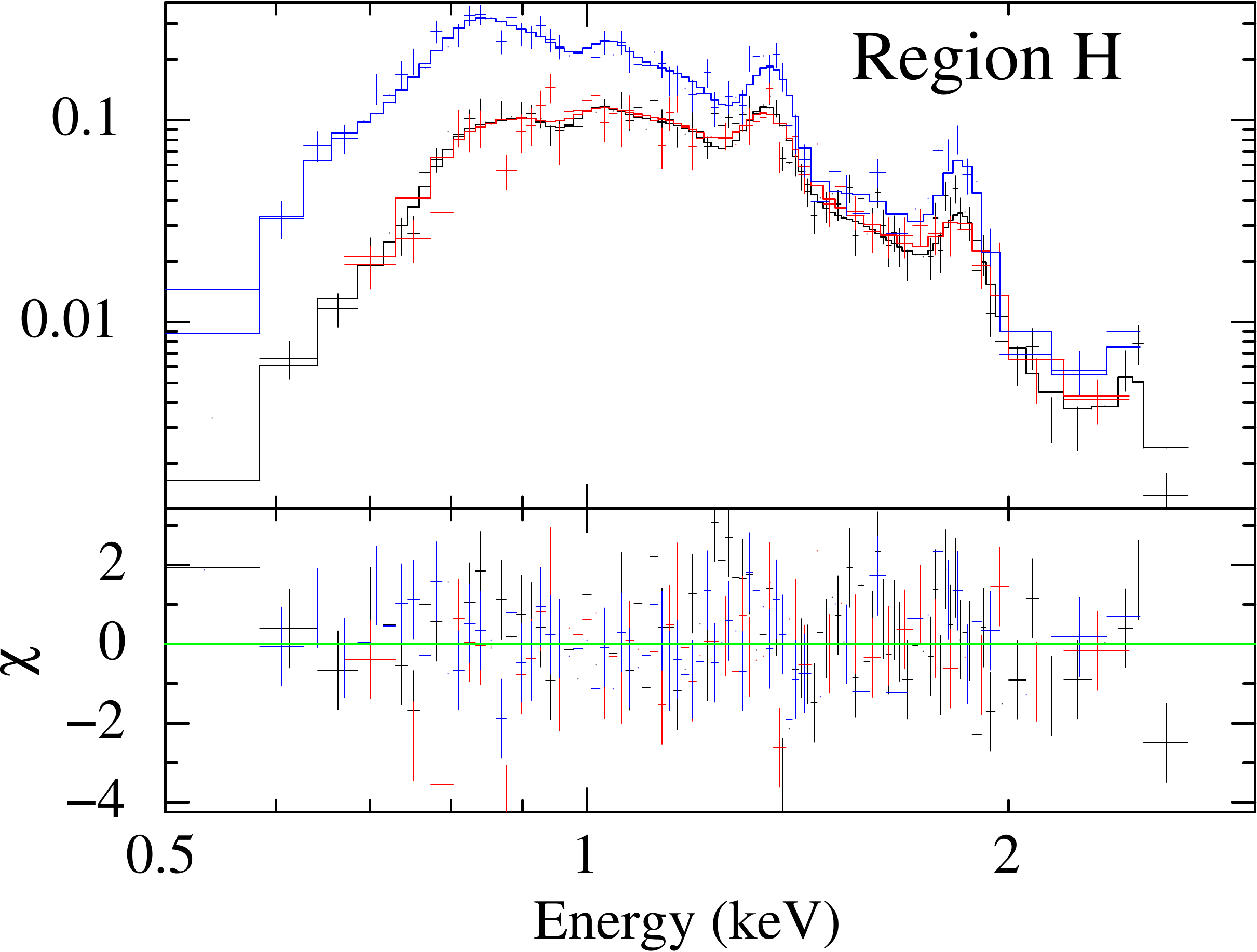}}\hspace{-2em}
\qquad
\subfigure {\includegraphics[width=0.31\textwidth]{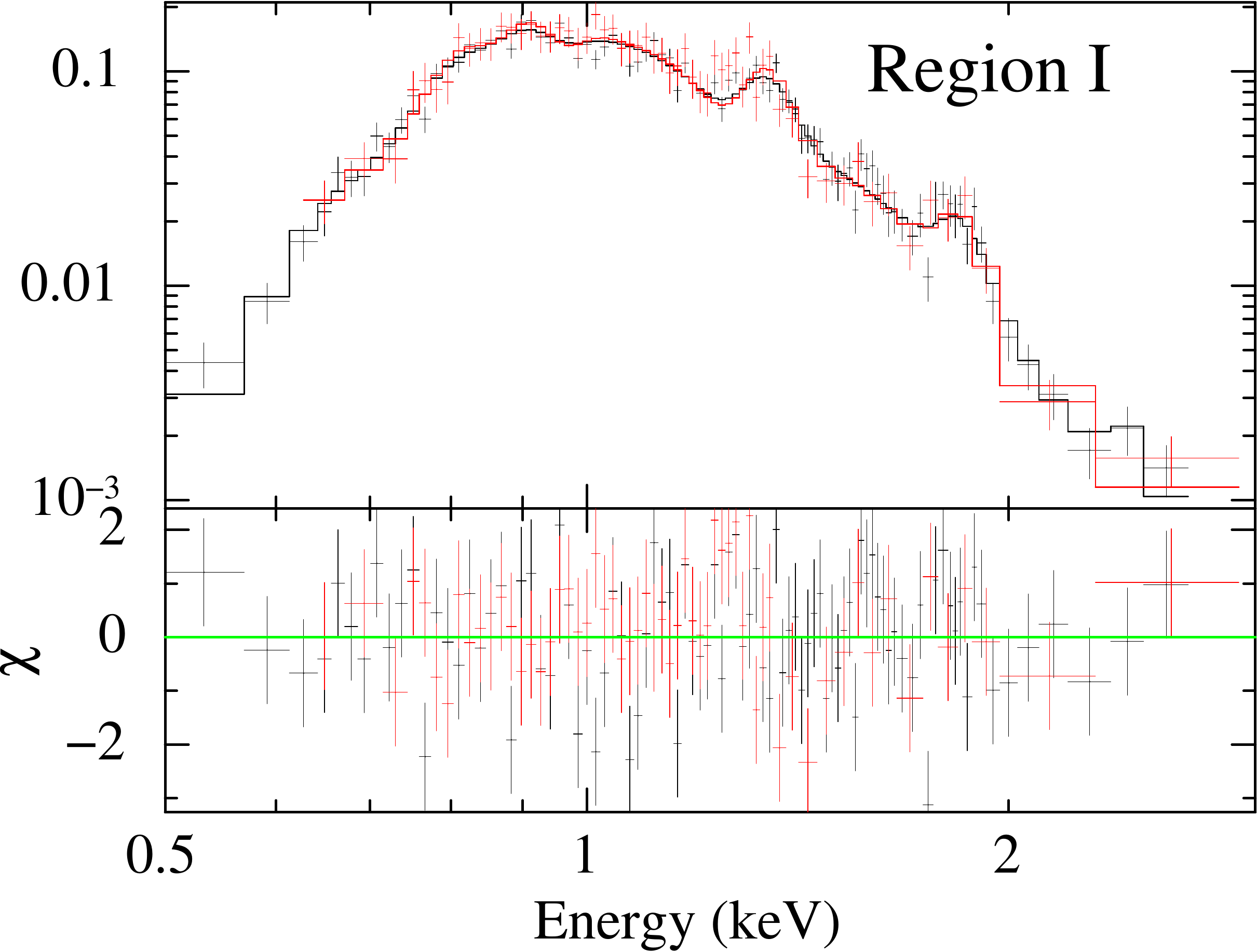}}\hspace{-2em}
\qquad
\subfigure {\includegraphics[width=0.31\textwidth]{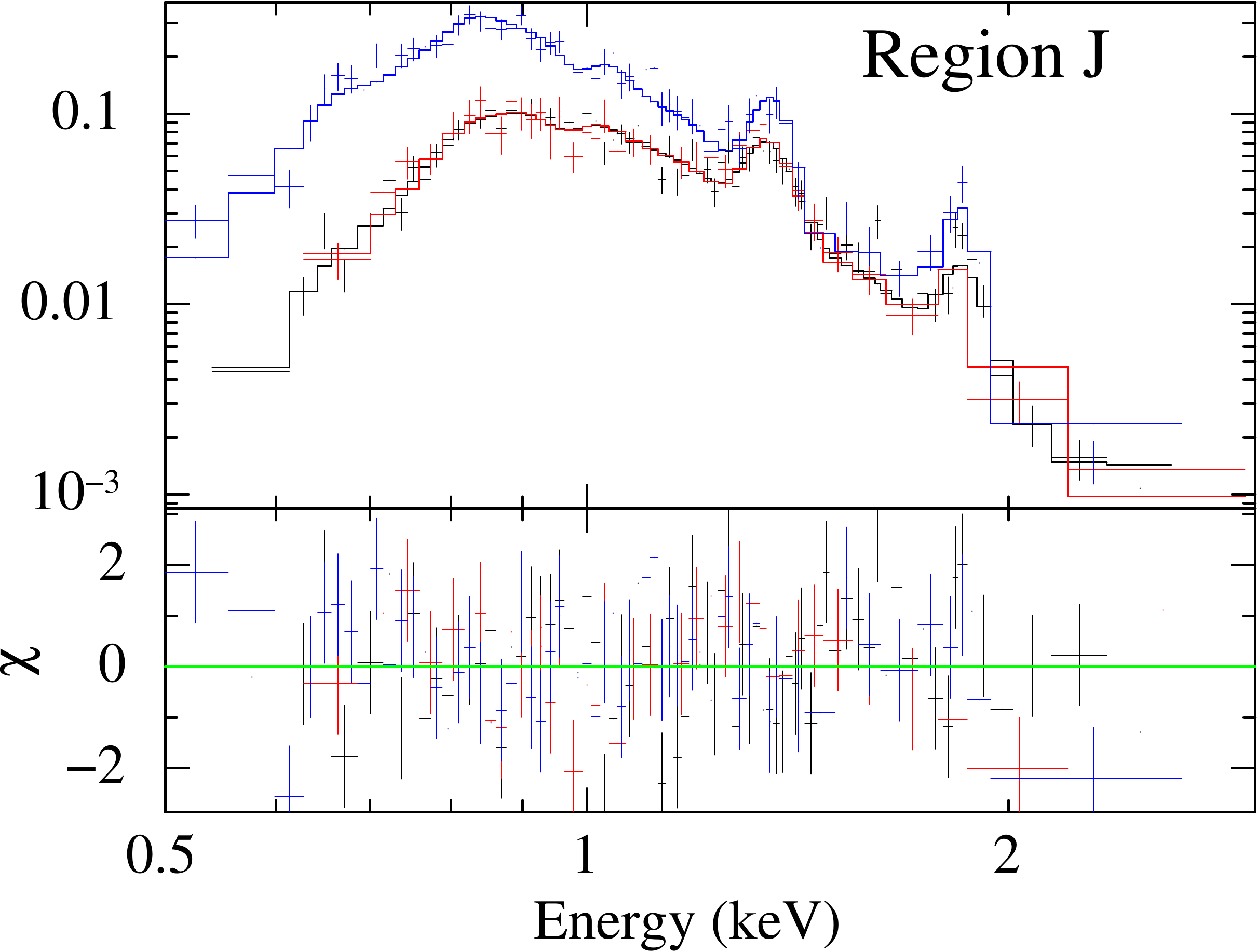}}

\subfigure {\includegraphics[width=0.31\textwidth]{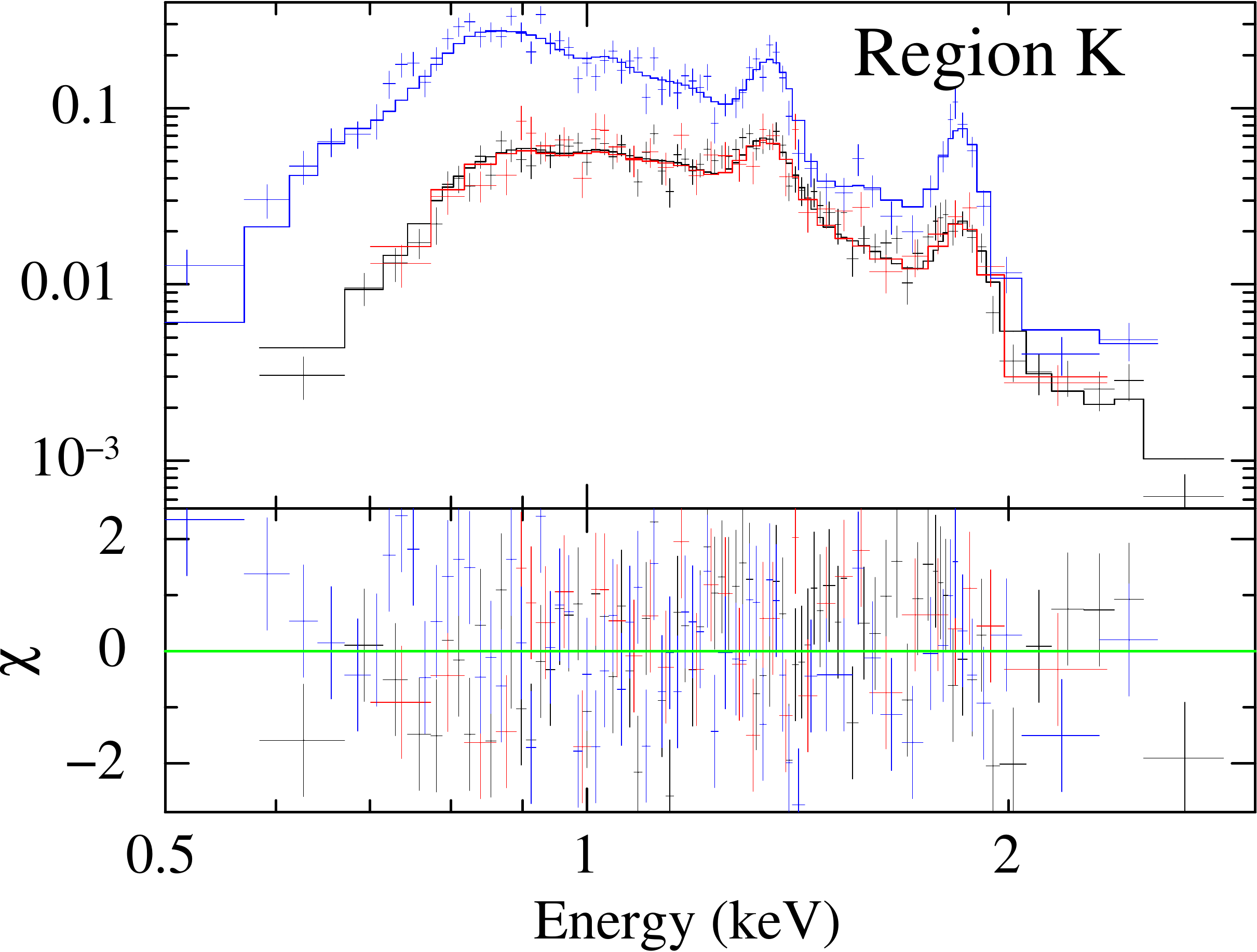}}\hspace{-2em}
\qquad
\subfigure {\includegraphics[width=0.31\textwidth]{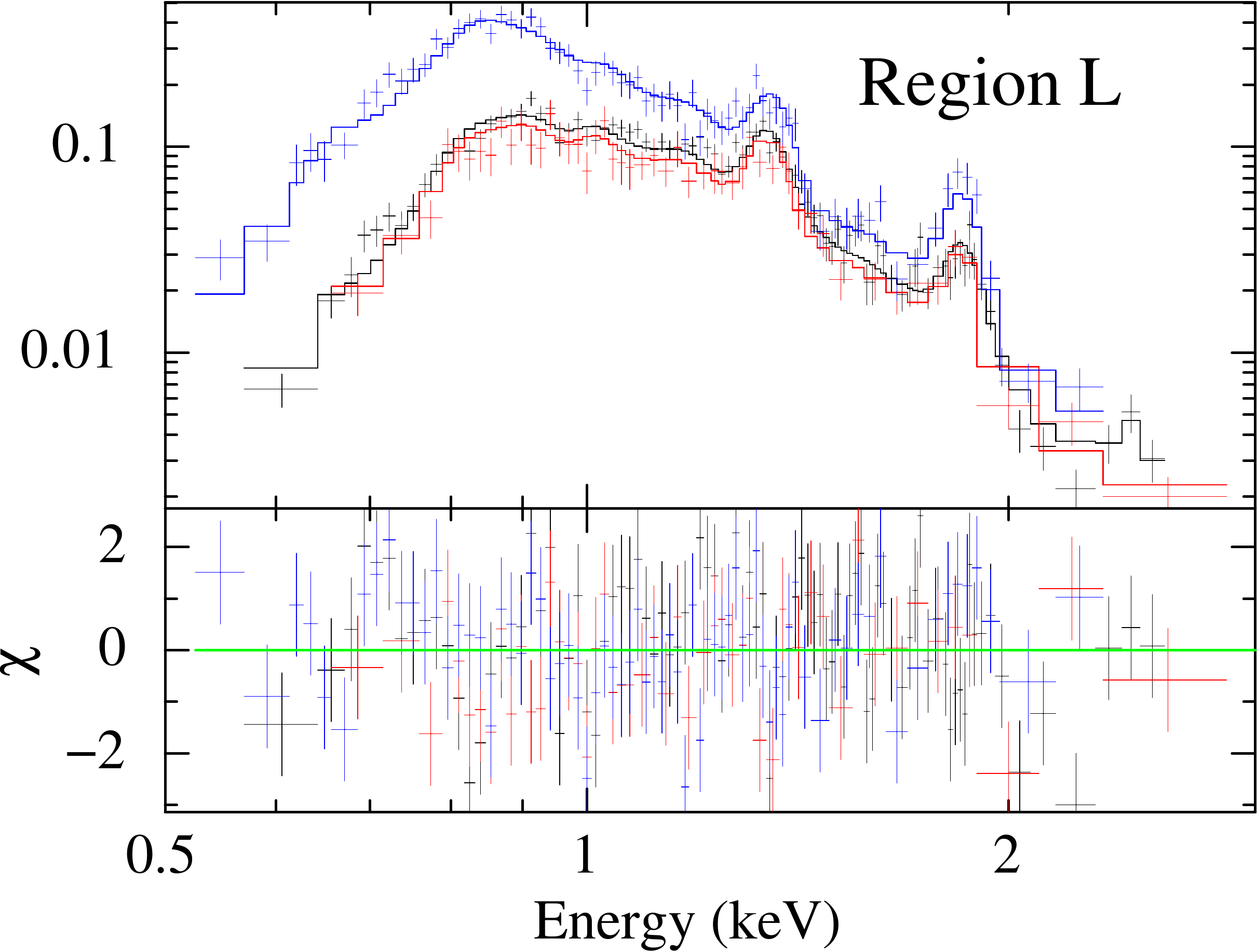}}\hspace{-2em}
\qquad
\subfigure {\includegraphics[width=0.31\textwidth]{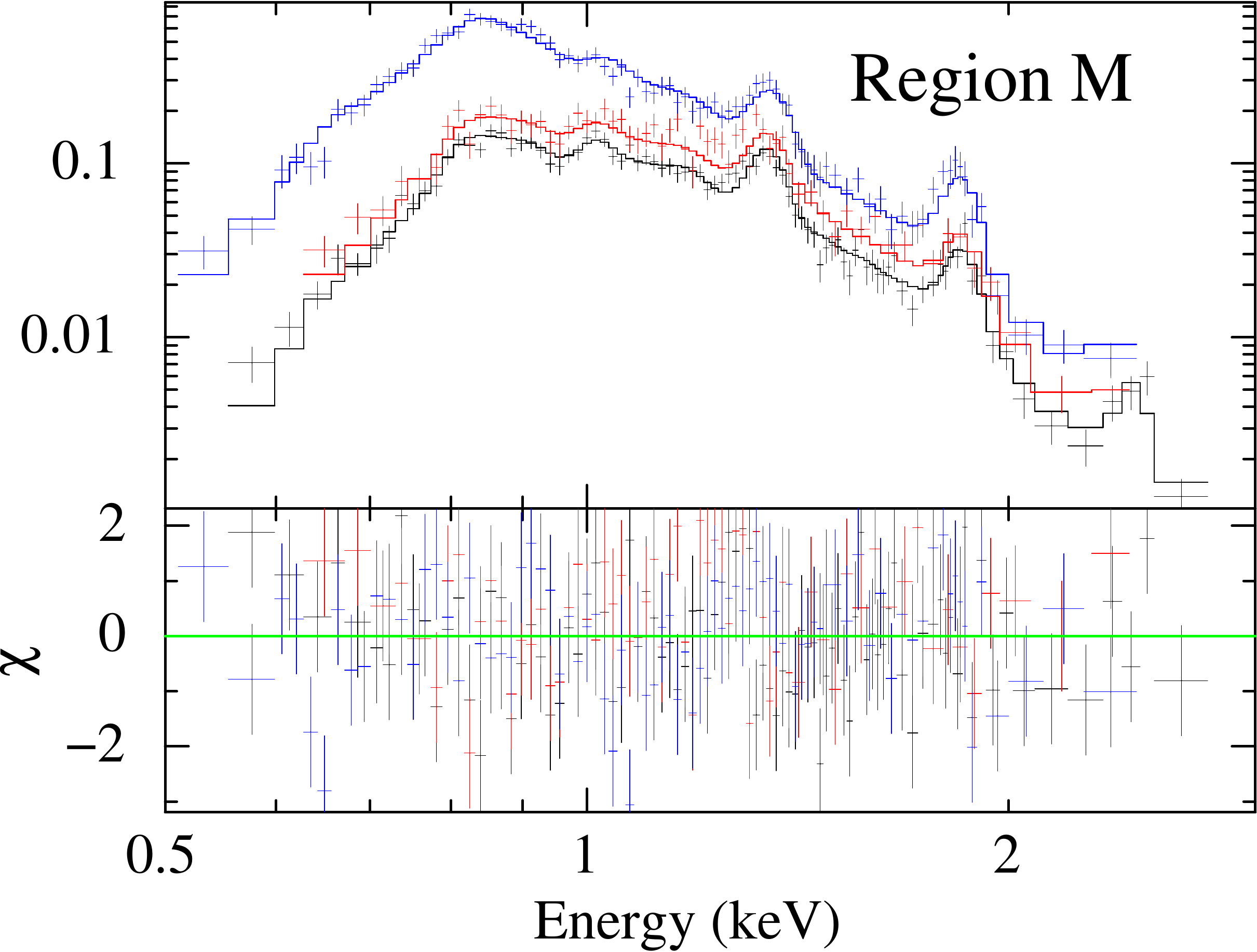}}

\subfigure {\includegraphics[width=0.31\textwidth]{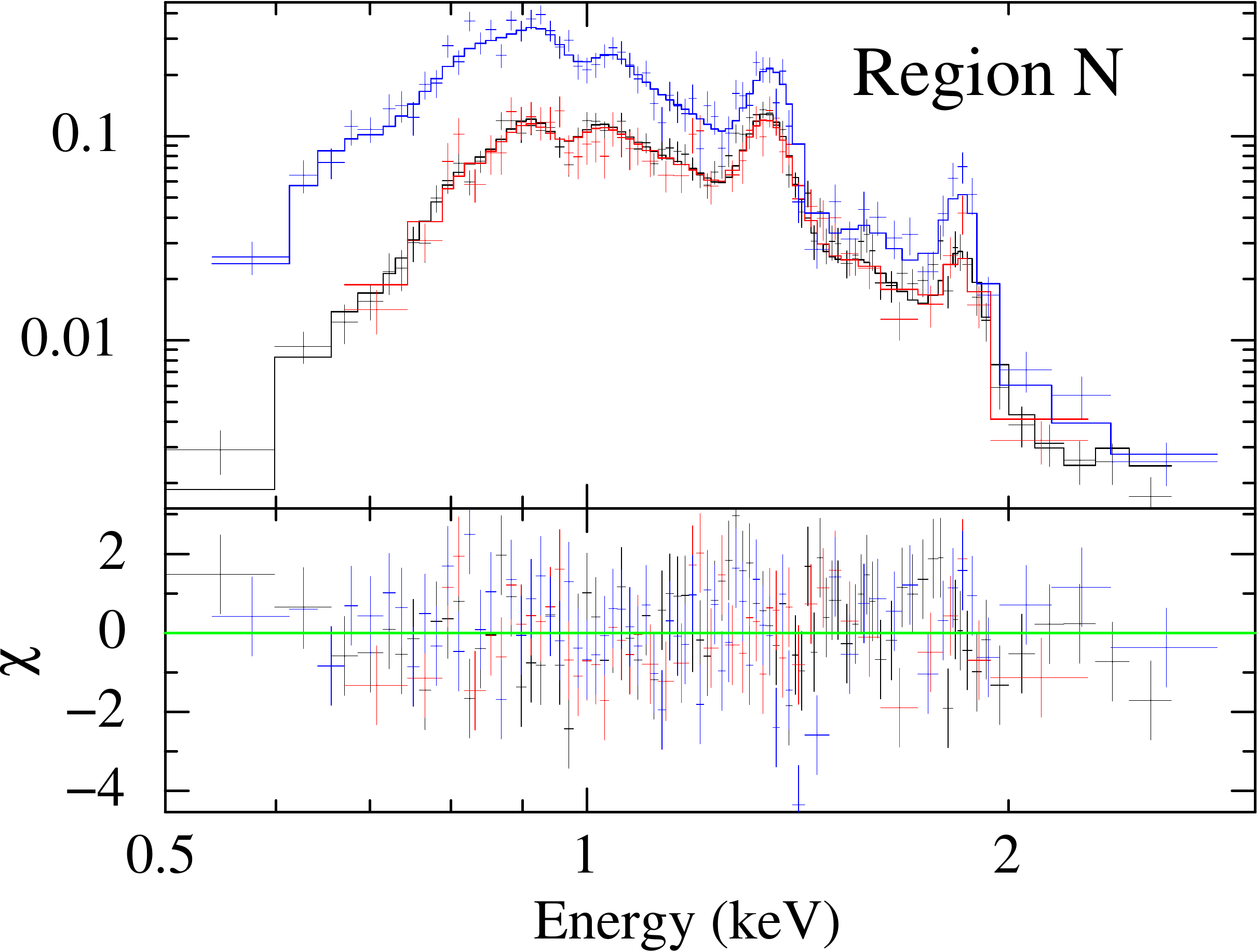}}\hspace{-2em}
\qquad
\subfigure {\includegraphics[width=0.31\textwidth]{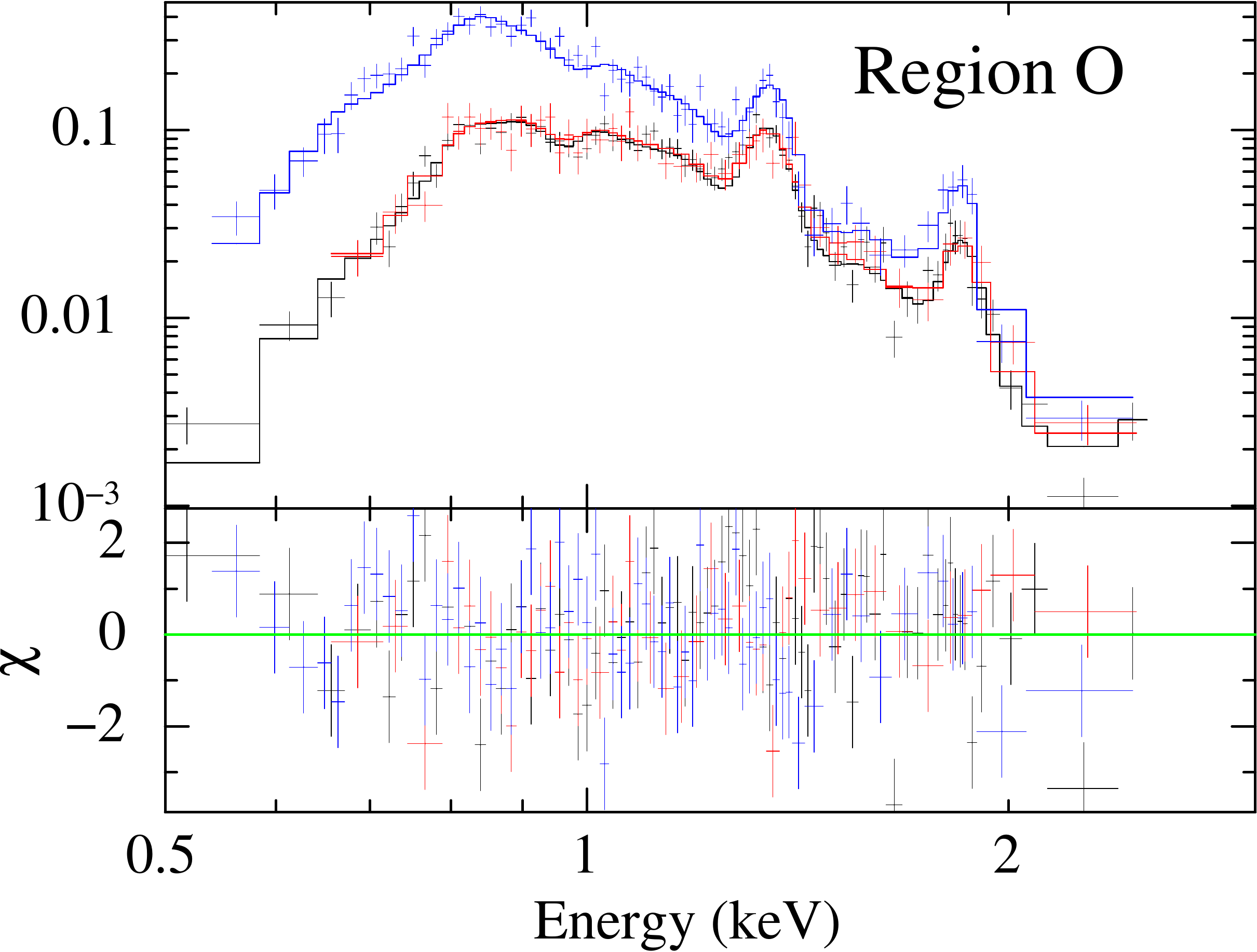}}\hspace{-2em}
\qquad
\subfigure {\includegraphics[width=0.31\textwidth]{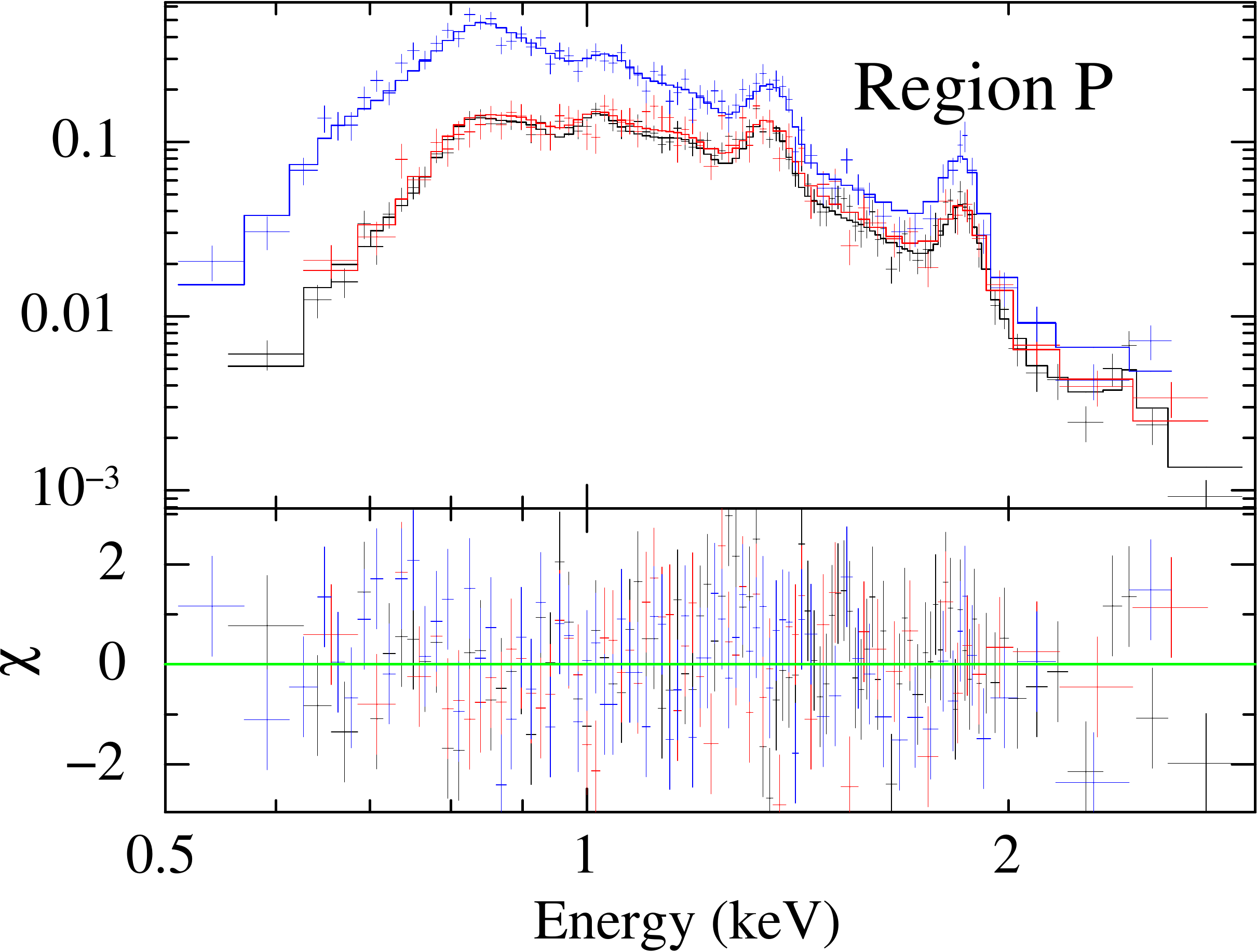}}
\caption{\footnotesize Spectra and best-fit \vpshock\ models for regions A $-$ P.  All observations were fit simultaneously: ObsID 11823 (black), 12224 (red), and 970 (blue). Vertical axis is count rate in units of  counts s$^{-1}$ keV$^{-1}$.}
\label{fig:specfits1}
\end{figure*}

\begin{figure*}[!htp]
\subfigure {\includegraphics[width=0.31\textwidth]{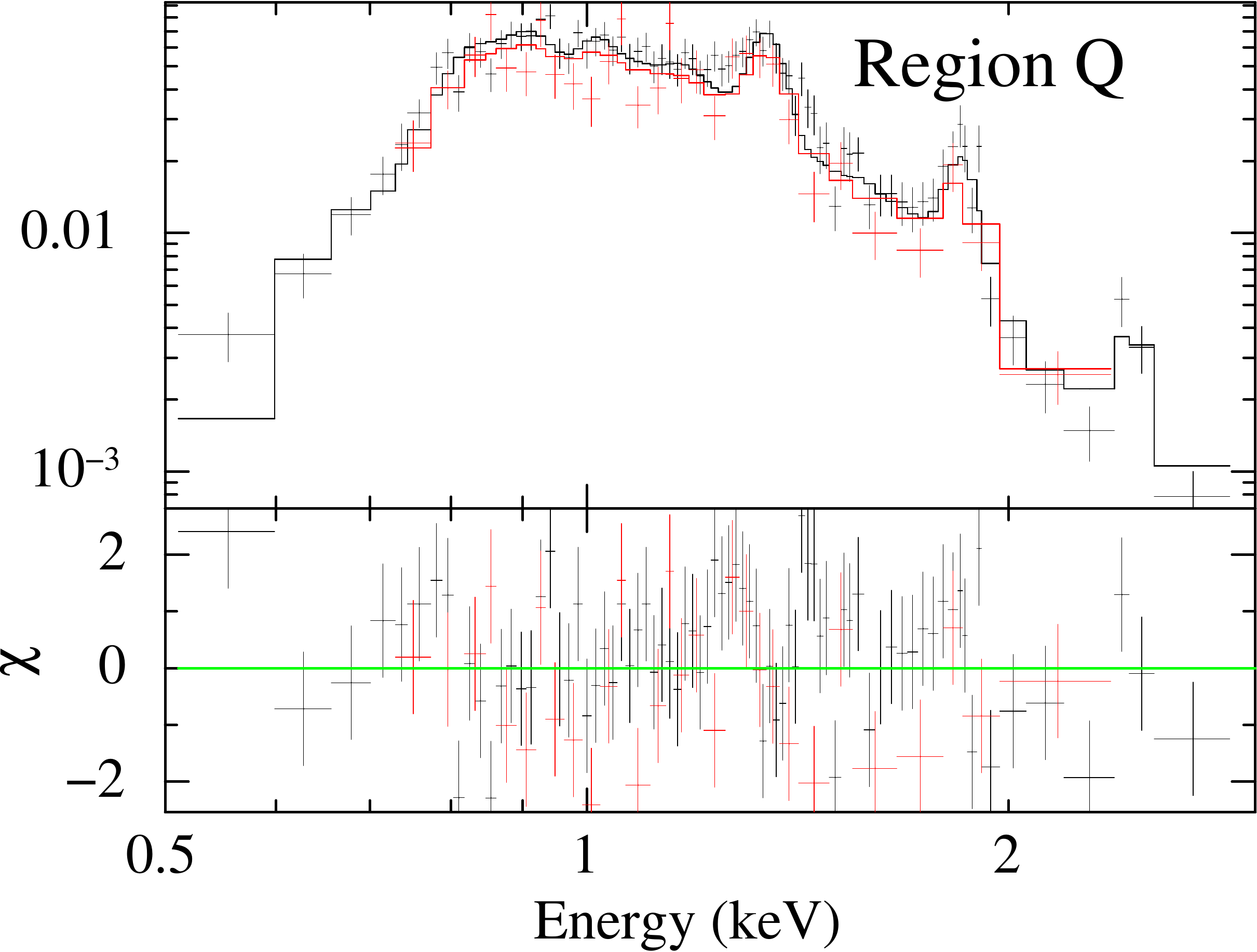}}\hspace{-2em}
\qquad
\subfigure {\includegraphics[width=0.31\textwidth]{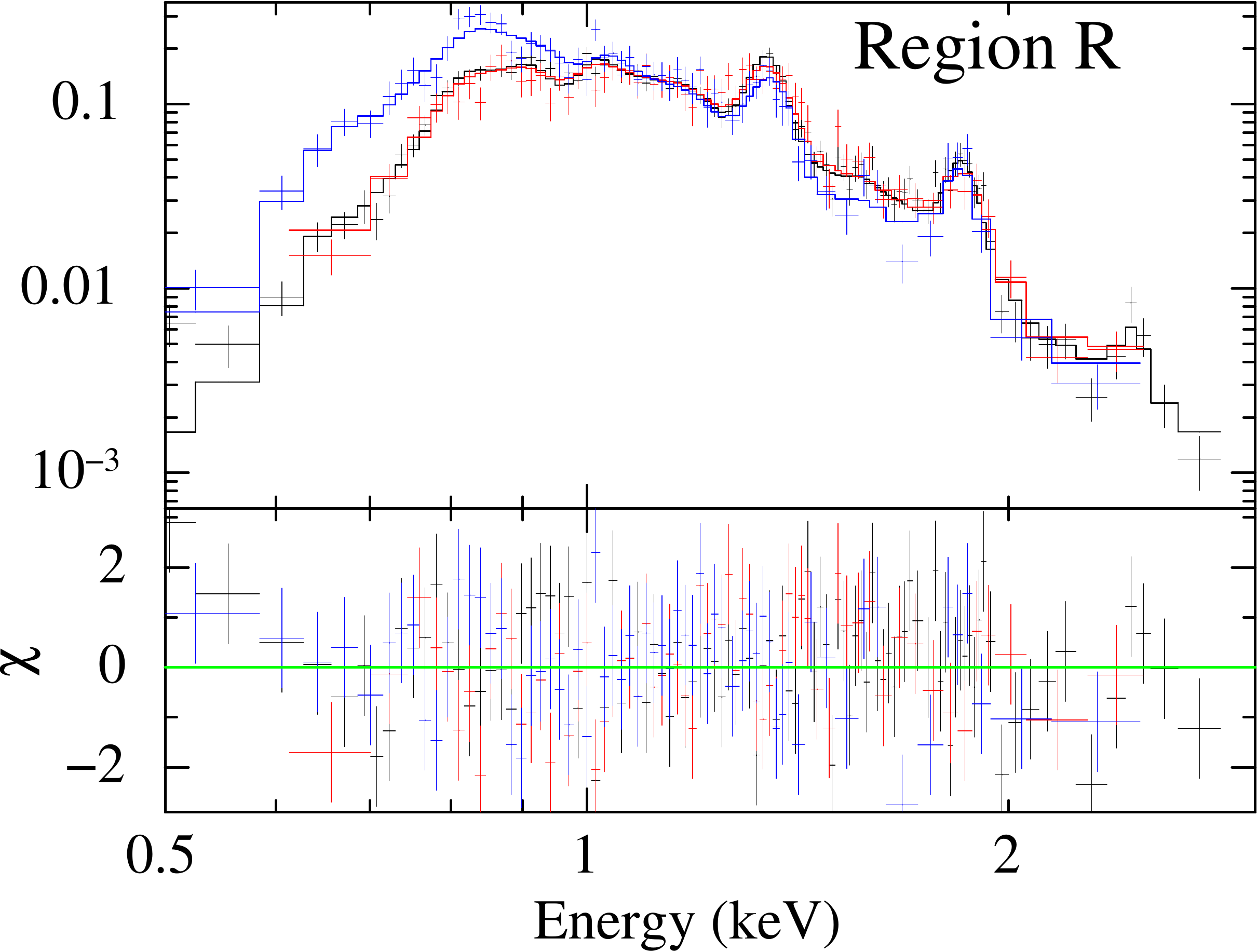}}\hspace{-2em}
\qquad
\subfigure {\includegraphics[width=0.31\textwidth]{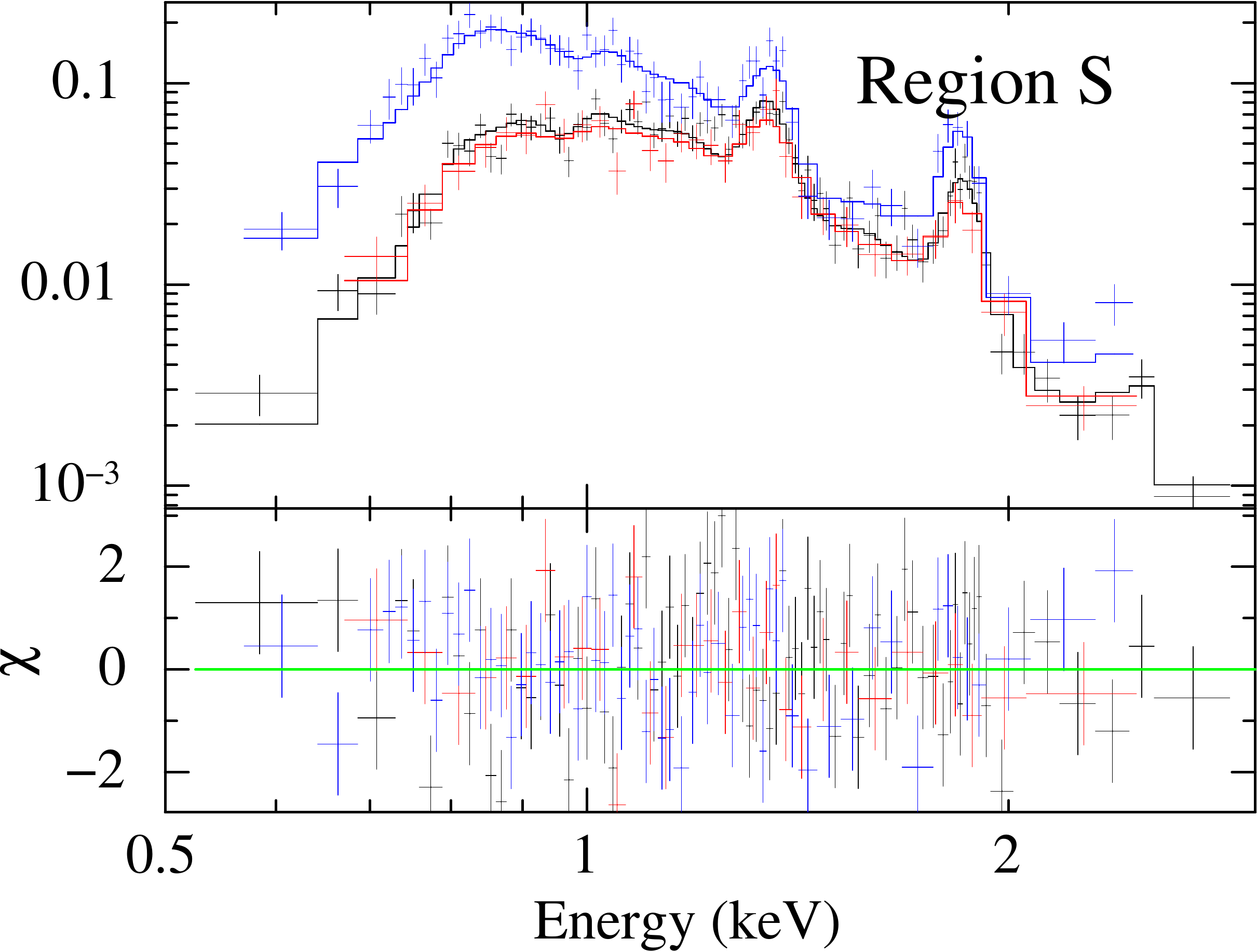}}

\subfigure {\includegraphics[width=0.31\textwidth]{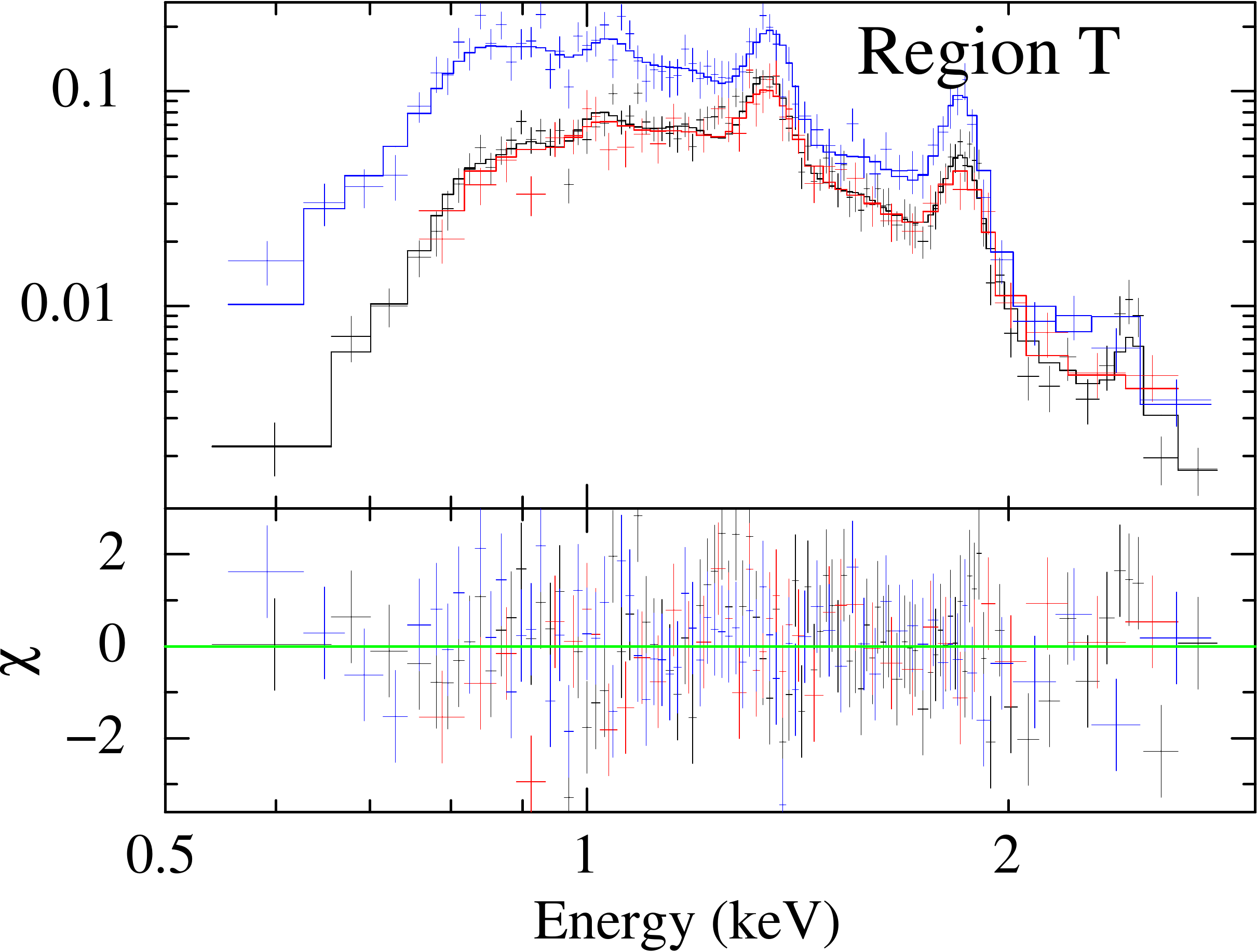}}\hspace{-2em}
\qquad
\subfigure {\includegraphics[width=0.31\textwidth]{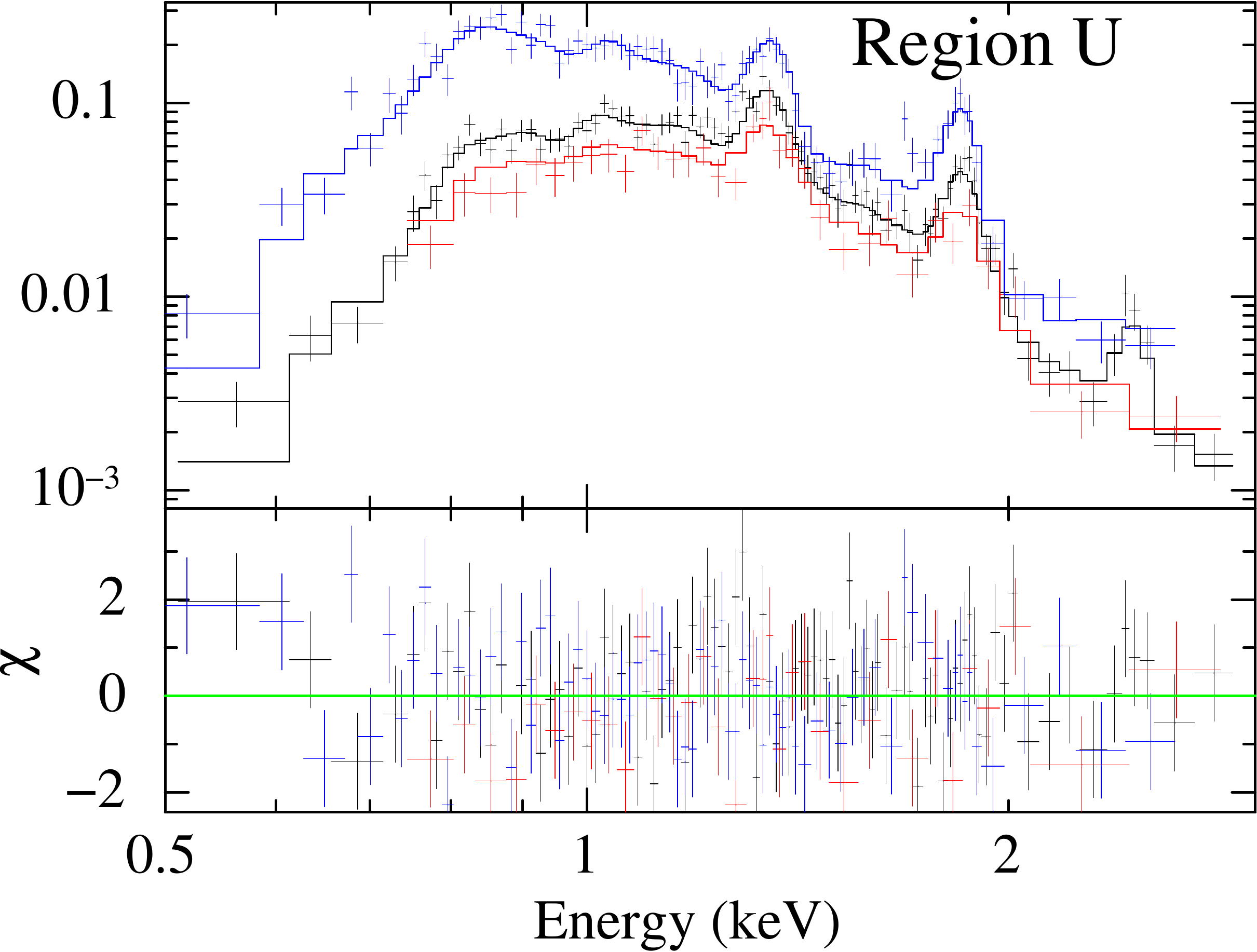}}\hspace{-2em}
\qquad
\subfigure {\includegraphics[width=0.31\textwidth]{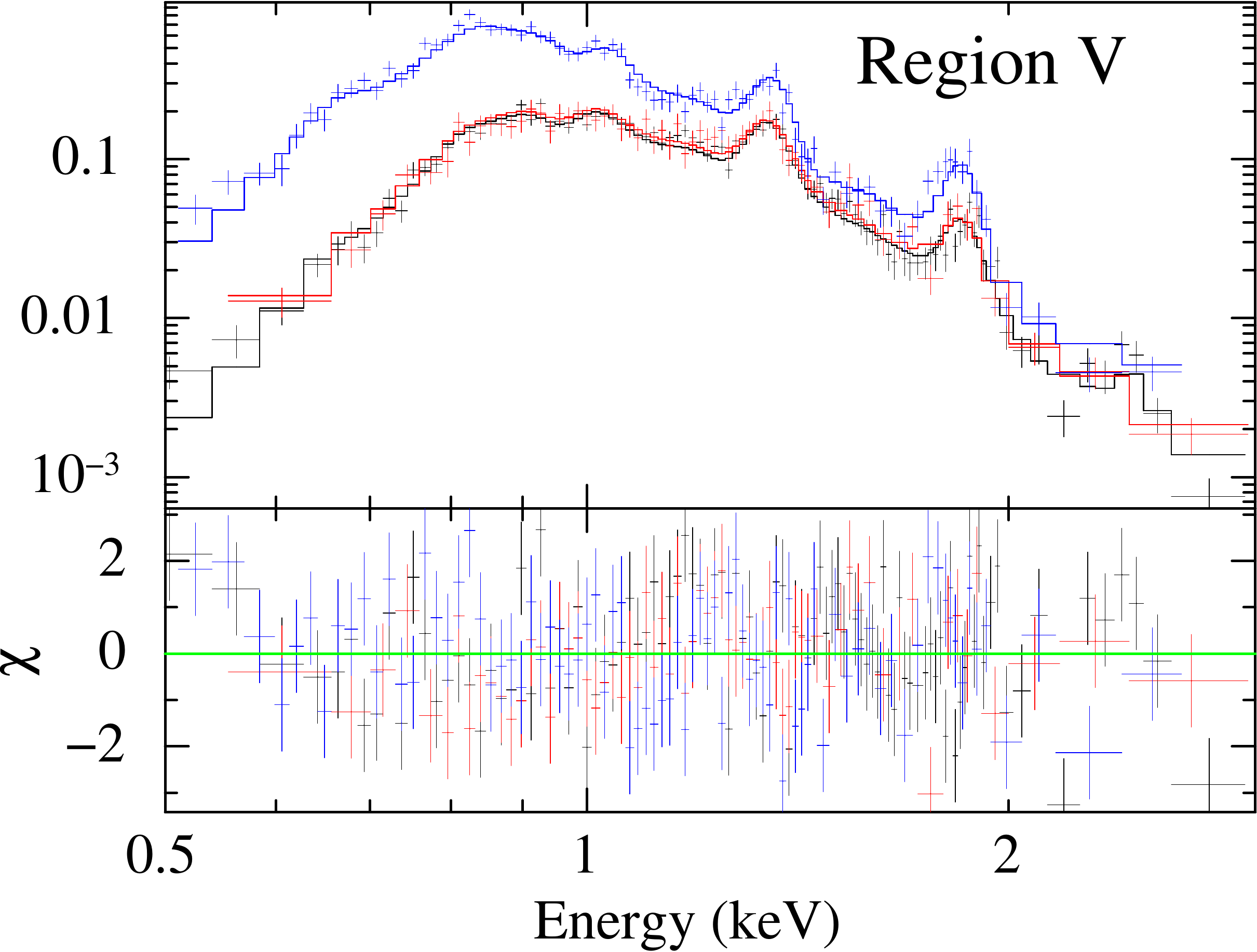}}

\subfigure {\includegraphics[width=0.31\textwidth]{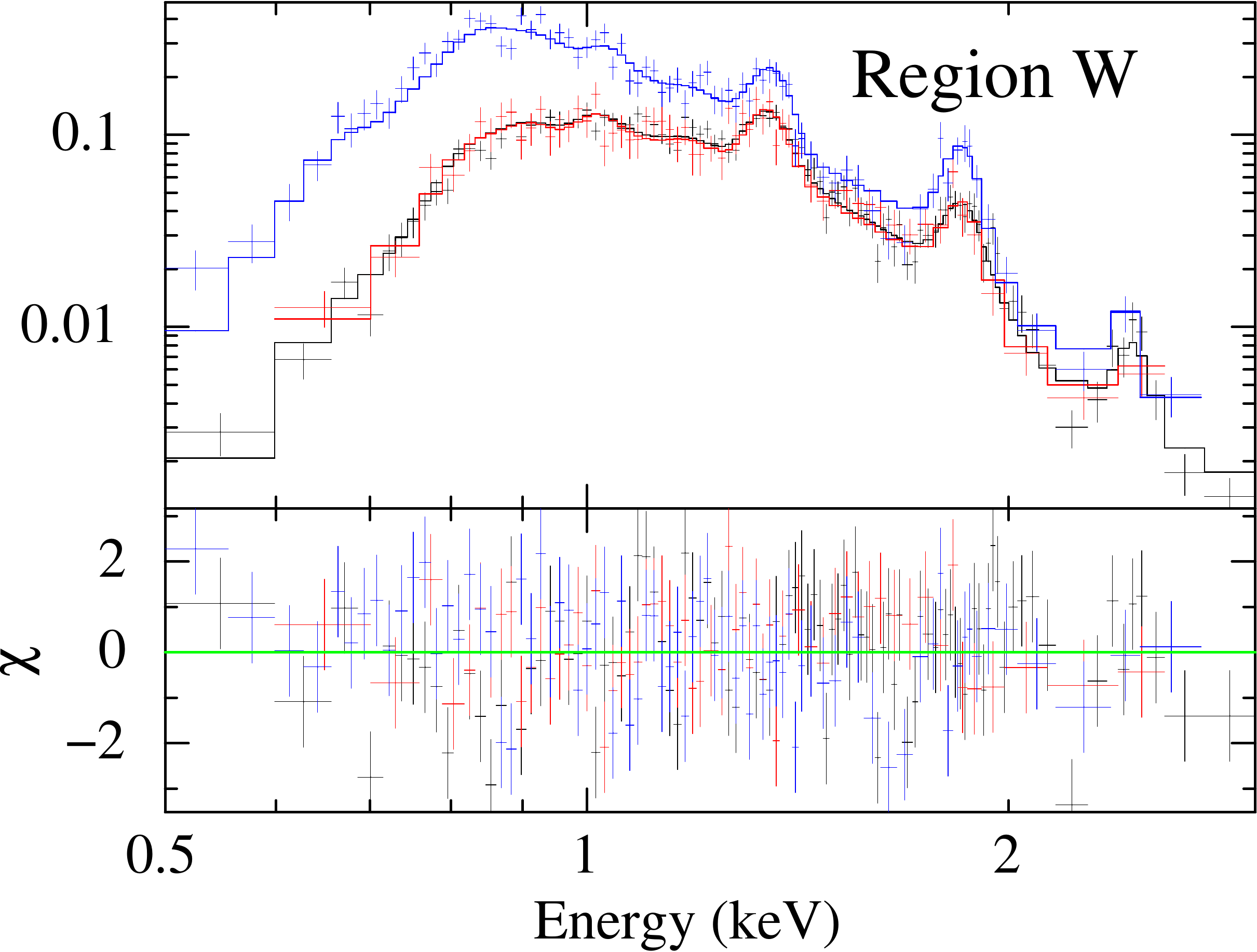}}\hspace{-2em}
\qquad
\subfigure {\includegraphics[width=0.31\textwidth]{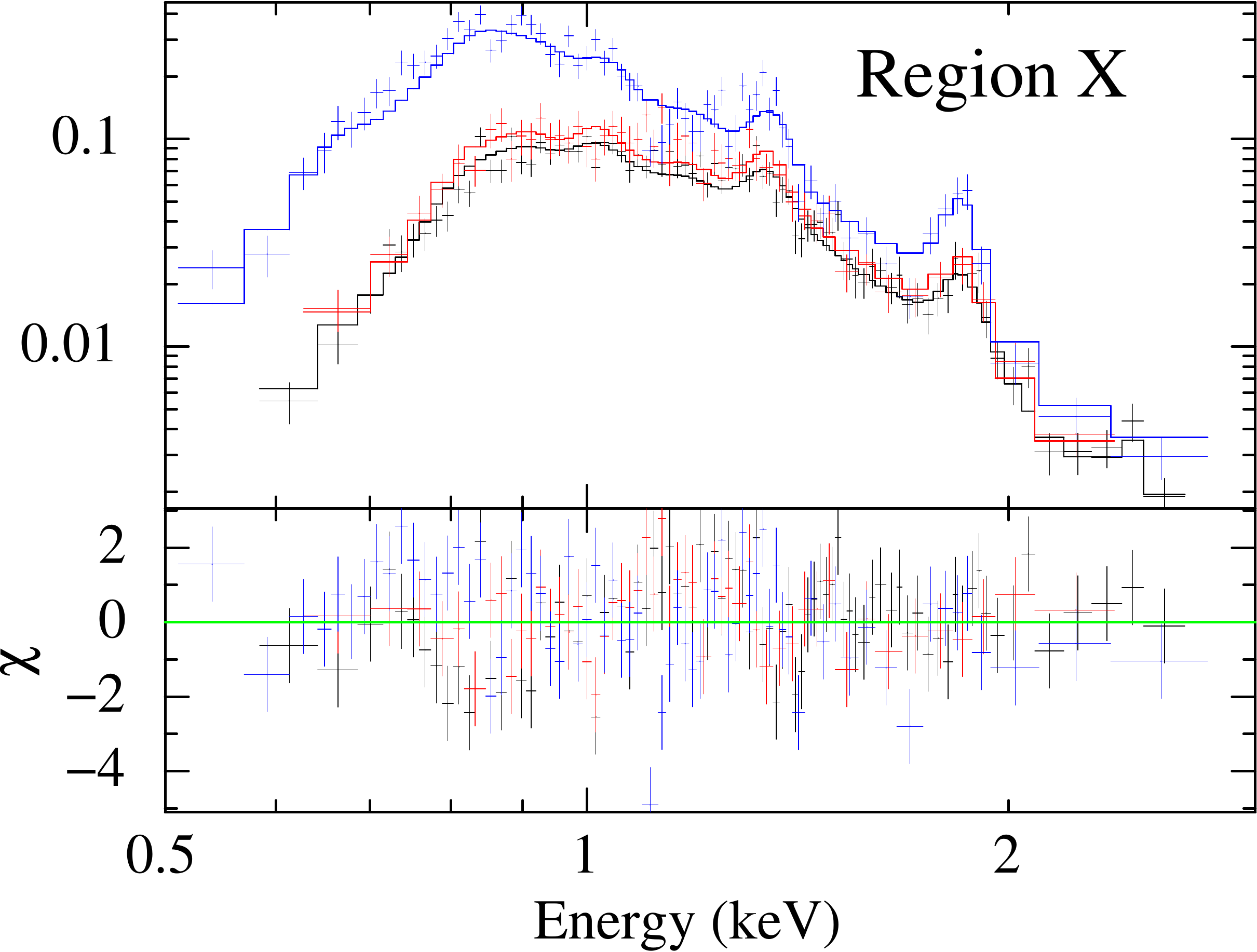}}\hspace{-2em}
\qquad
\subfigure {\includegraphics[width=0.31\textwidth]{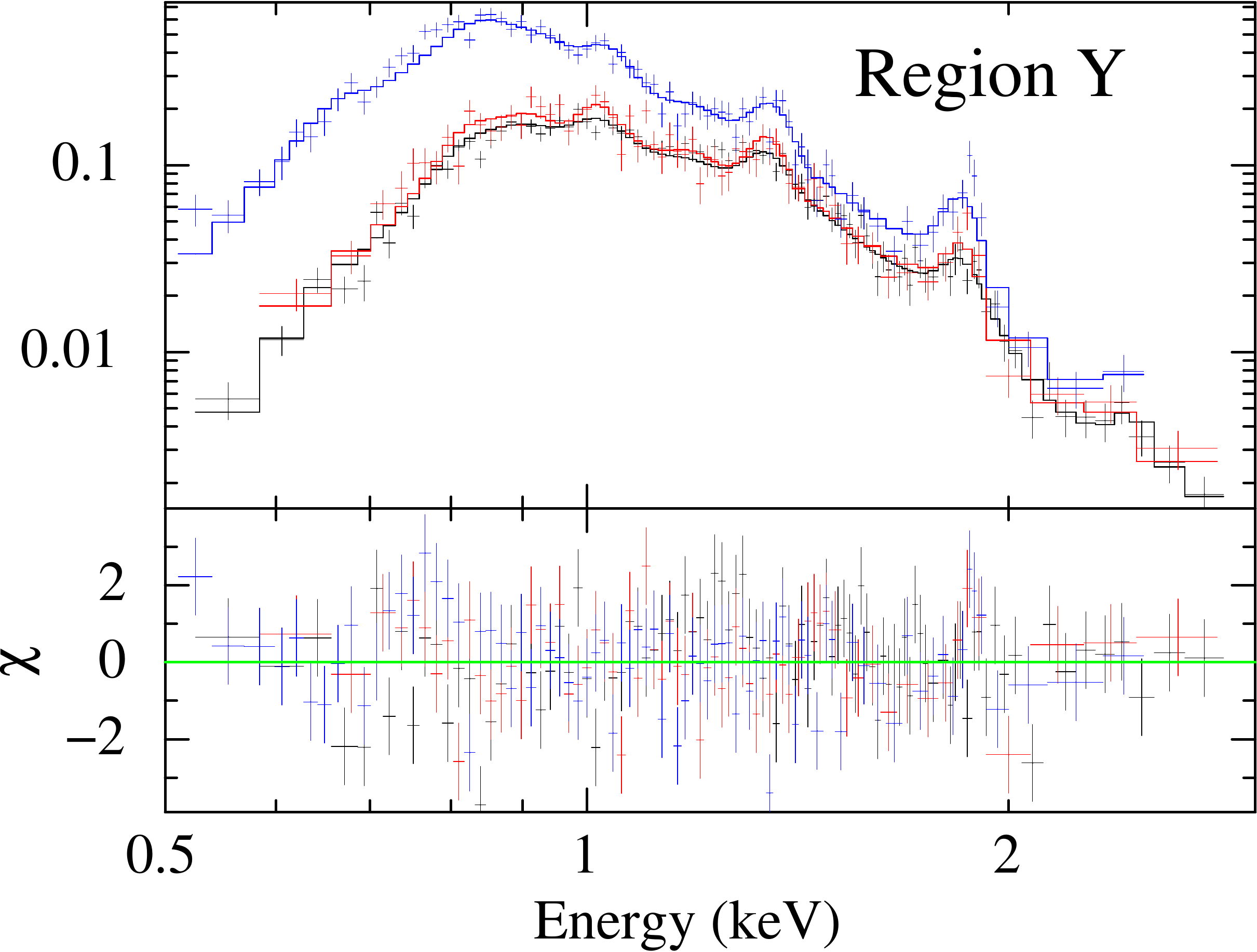}}

\subfigure {\includegraphics[width=0.31\textwidth]{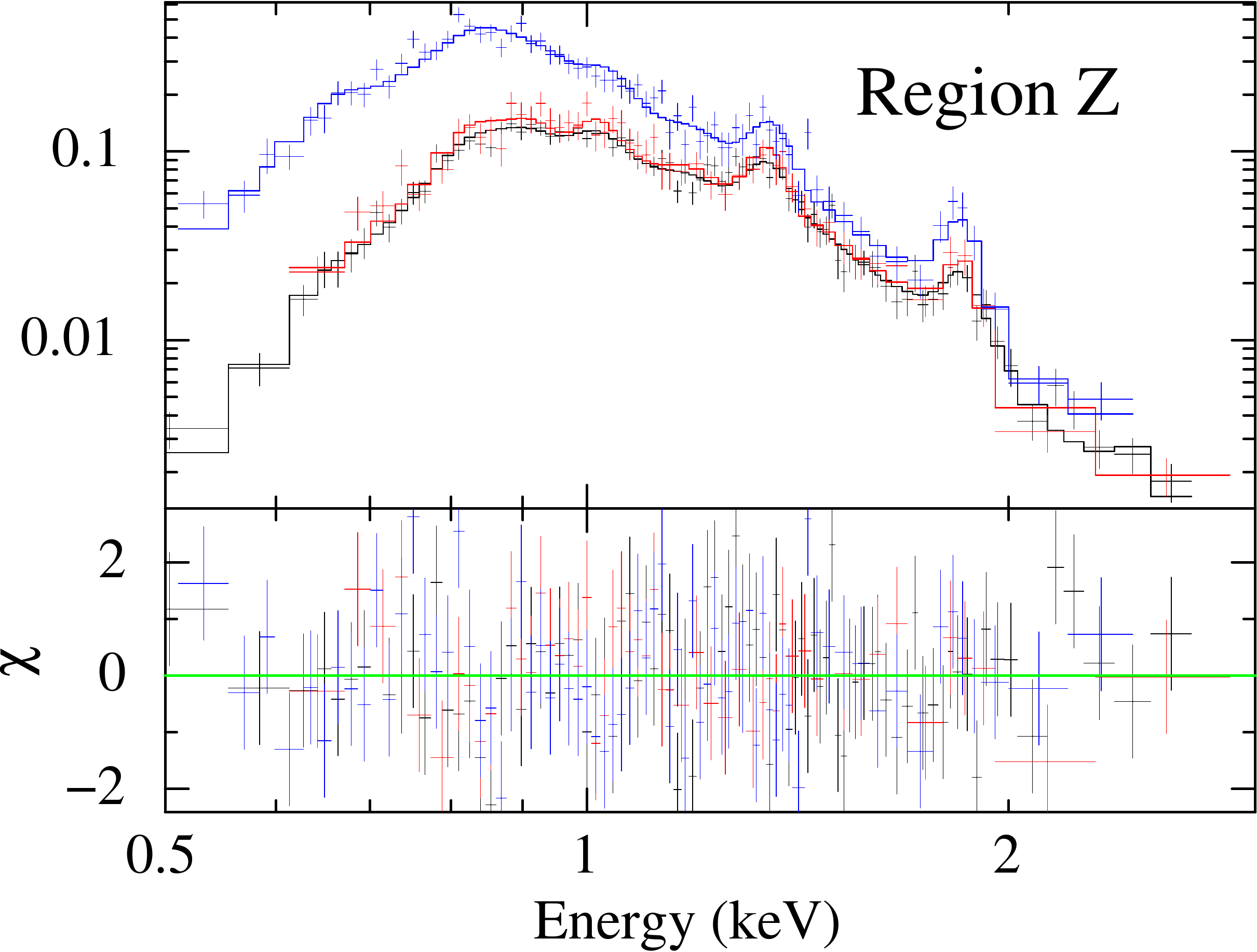}}\hspace{-2em}
\qquad
\subfigure {\includegraphics[width=0.31\textwidth]{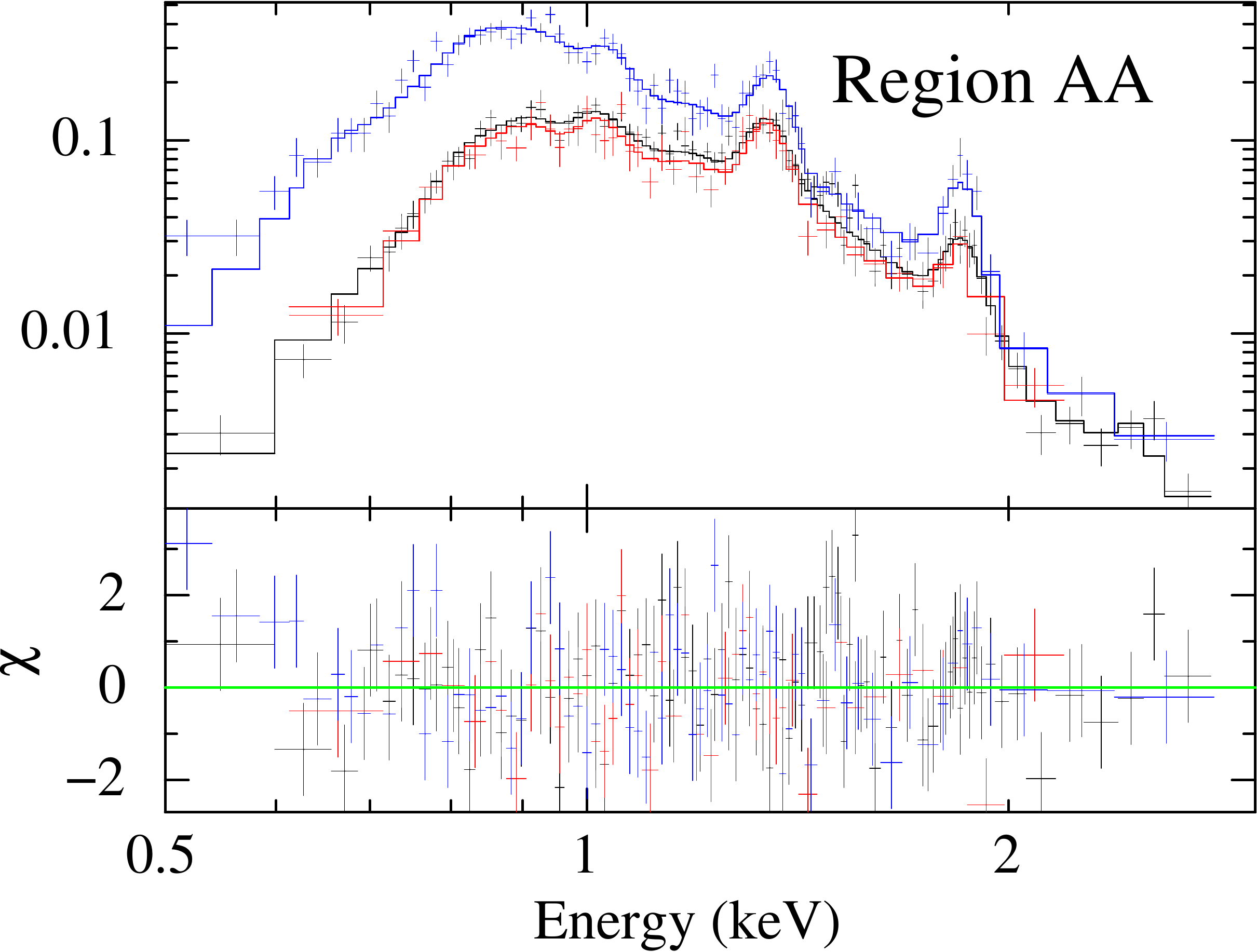}}\hspace{-2em}
\qquad
\subfigure {\includegraphics[width=0.31\textwidth]{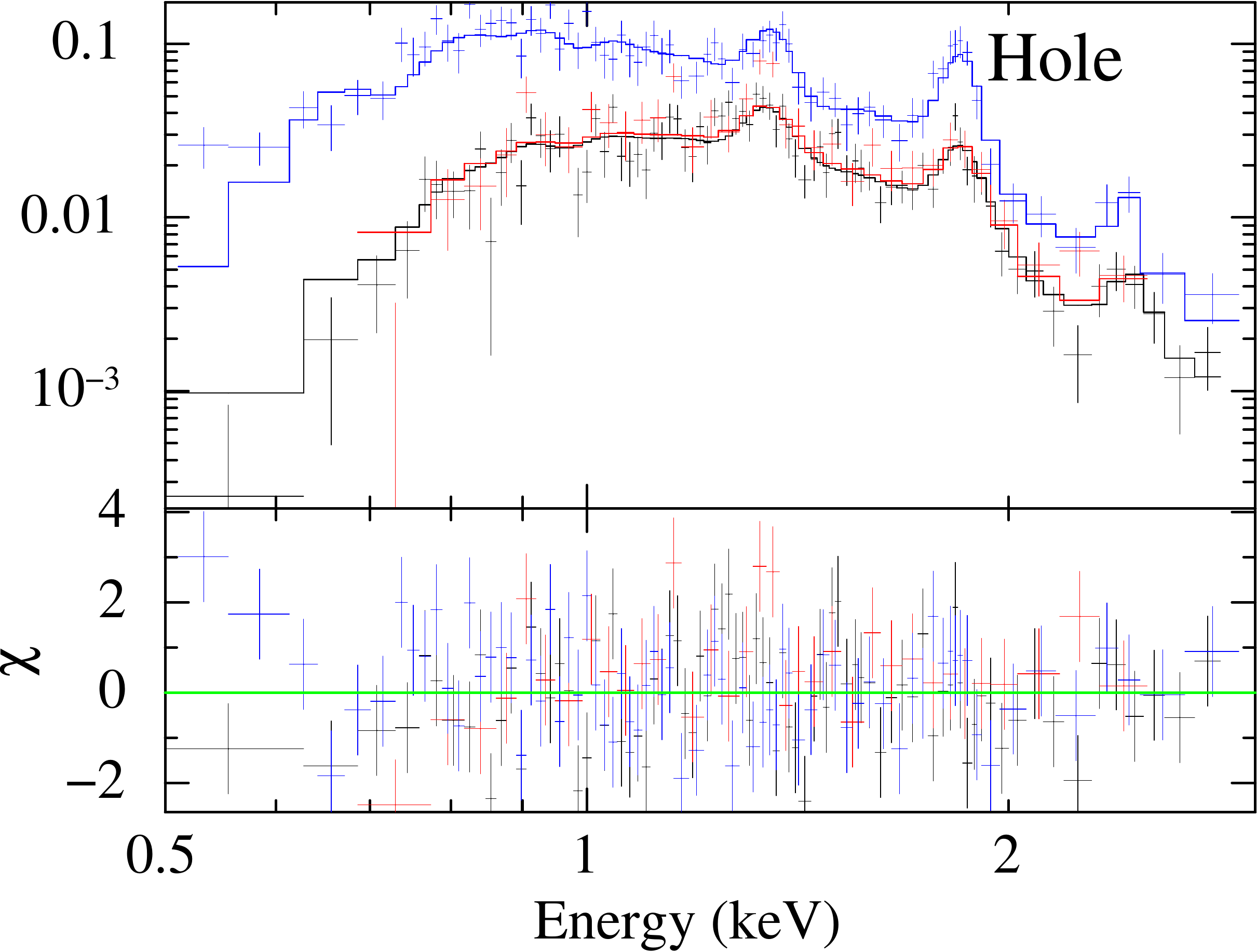}}
\caption{\footnotesize Spectra and best-fit \vpshock\ models for regions Q $-$ Z, AA, and the Hole.  All observations were fit simultaneously: ObsID 11823 (black), 12224 (red), and 970 (blue). Vertical axis is count rate in units of  counts s$^{-1}$ keV$^{-1}$.}
\label{fig:specfits2}
\end{figure*}

Based on the X-ray and equivalent width images, we chose 27 regions for spectral analysis (Figure \ref{fig:regions}), with $2500-16000$ counts per region (Table \ref{table:specfits}).  For each region, individual spectra were extracted from each observation using the CIAO script {\tt specextract} with standard parameters for the full 0.1 to 8.0 keV band.  Spectra were grouped to have a minimum of 20 counts per bin.  Corresponding background spectra were extracted from a source-free region outside of the remnant, $\sim$$6'$ north of the central object, and subtracted for the purposes of spectral fitting.  Spectra for each region are shown in Figures \ref{fig:specfits1} and \ref{fig:specfits2}.

All regional spectra show clear K$\alpha$ line features from highly ionized He-like Mg and Si ions. This confirms the indications from the Mg and Si EWIs that they are widely distributed throughout the remnant.   The S line visible in the integrated \rcw\ spectrum is absent or very weak in many of the regional spectra, suggesting, as does the S EWI, that sulfur has a patchier distribution than Mg or Si and/or is weak throughout the remnant.  

\begin{figure}[!htb]
\subfigure{\includegraphics[width=0.45\textwidth]{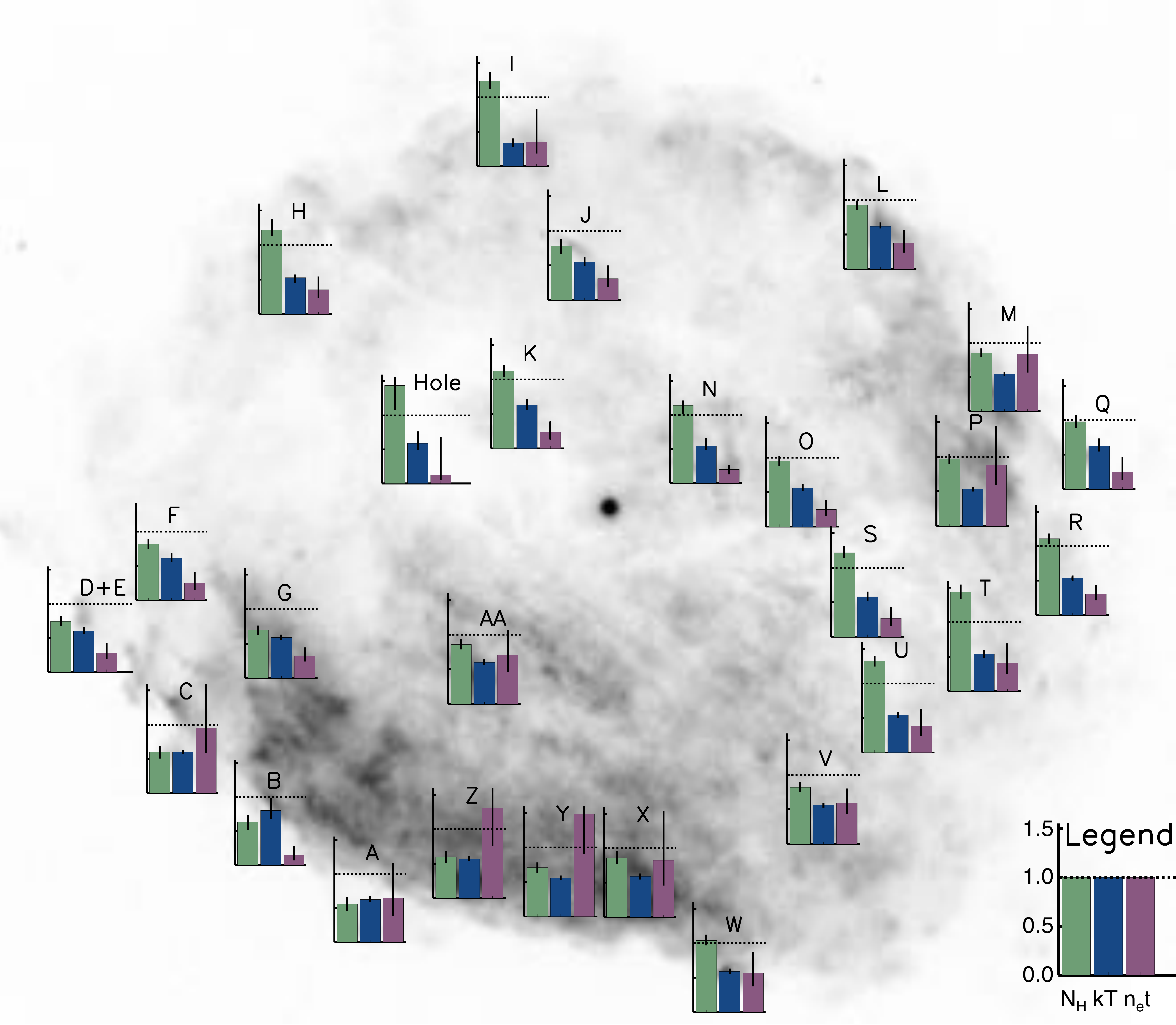}} 
\subfigure{\includegraphics[width=0.45\textwidth]{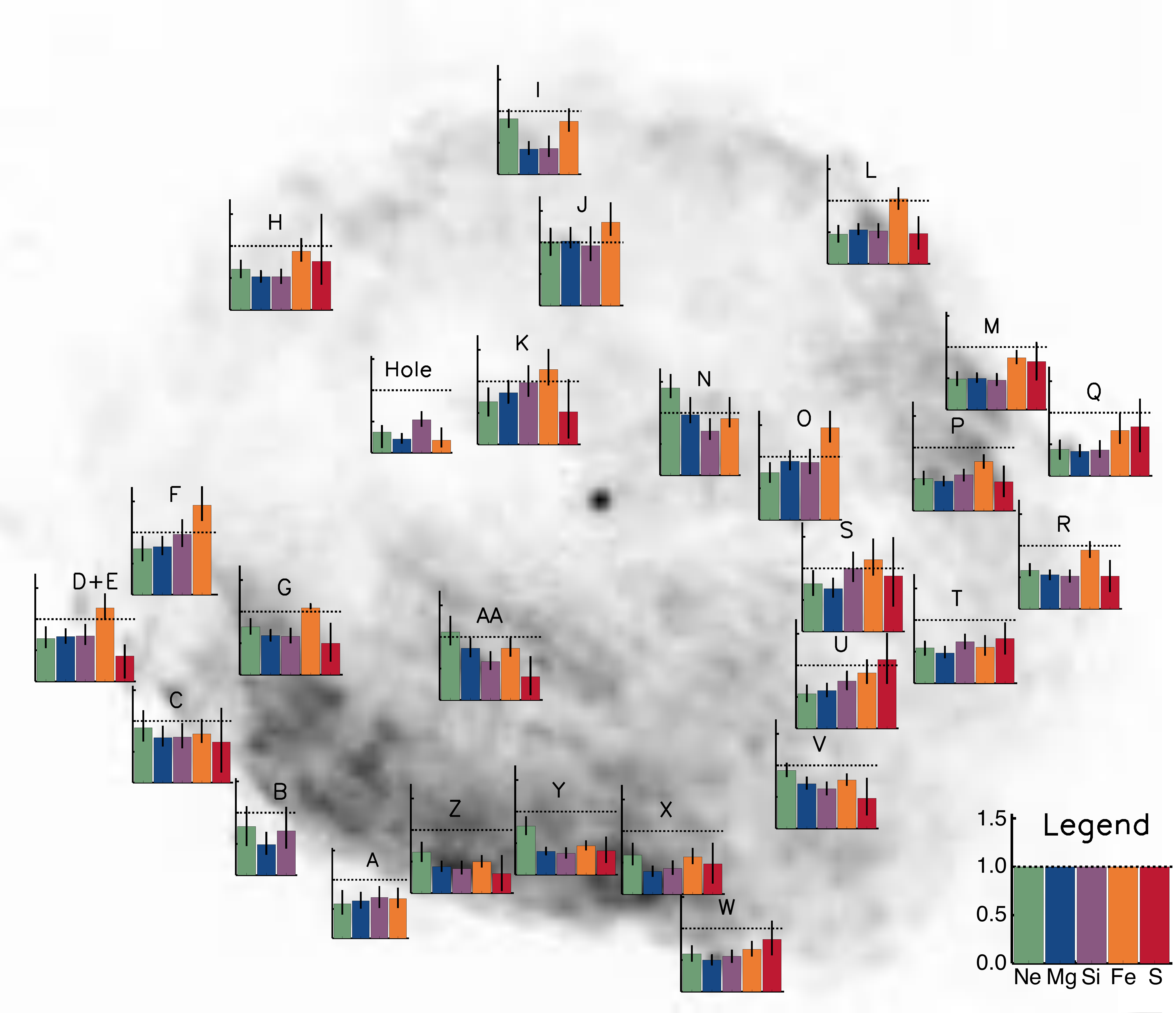}} 
\caption{\footnotesize Best-fit \nh, $kT$ and $n_et$ values (top) and abundances (bottom) overlaid on the X-ray image.  Column density (green) is in units of $10^{22}\mbox{cm}^2$, $kT$ (blue) is in units of keV, and ionization age (purple) is in units of $10^{12}$ s cm$^{-3}$.  Abundances are in solar units.  The vertical tick marks are at 0.5, 1.0, and 1.5 with the dotted line at 1.0.  Error bars represent 90\% confidence intervals.}
\label{fig:bars}
\end{figure}

We fit these spectra using \xspec\ v12.8.1 \citep{Arnaud1996}.  An absorbed, non-equilibrium ionization state (NEI) plane shock model with variable abundances was used \citep[{\tt tbabs*vpshock}, ][]{Borkowski2001}.  An augmented version of ATOMDB \citep{Smith2001,Foster2012} that includes atomic data accounting for inner-shell lines and updated Fe L-shell lines was used \citep{Badenes2006}, along with solar abundances from \citet{Asplund2009}.  We fit these spectra in the range $0.5 - 3.0$ keV.  The spectra from all three observations were fit simultaneously (except where ObsID 970 was excluded for regions outside its field of view), with ObsID 970 multiplied by an extra normalization, $C_{970}$, to allow for slight differences in effective area calibrations between ACIS-I and ACIS-S.  The electron temperature, ionization timescale, normalization, and absorption were allowed to vary.  We also varied the Ne, Mg, Si, S, and Fe abundances (with other elemental abundances fixed at solar).  In eight cases the S abundance could not be constrained; for these regions S was fixed to solar.  The same applies to the Fe abundance in region B, which contains the lowest number of counts. We also fit each spectrum with all elemental abundances fixed to solar except those that significantly improved the fit when they are varied.  The resulting best-fit parameters were consistent with those obtained from allowing all five abundances to vary.  In general, freezing a given abundance only resulted in a better fit if its best-fit value when free was consistent with solar.  For the sake of consistency between the regions and in order to provide uncertainties, we therefore report the results of fitting with all five abundances free (except where unconstrained).  Changing the absorption model had no significant effect on the results.  Switching the solar abundances from \citet{Asplund2009} to \citet{Anders1989} resulted in $\sim$20\% lower values of \nh, marginal increases in the Si abundances, and $\sim$50\% higher Mg abundances.  All other parameters were not significantly affected.  The best-fit model parameters for each region are given in Table \ref{table:specfits} and visualized in Figure \ref{fig:bars}.  


Throughout most of the remnant, the electron temperature is $0.53$ keV $\lesssim kT \lesssim 0.63$ keV, with an average of 0.58 keV.  The two exceptions are regions B and I, with temperatures of $0.80_{-0.11}^{+0.17}$ keV and $0.34\pm0.05$ keV, respectively.  Region B has poor photon statistics, resulting in large uncertainties, consistent with a temperature as low as 0.69 keV.  Region I is more unusual; the temperature is well-constrained and it appears to have an excess of emission at $\lesssim$1 keV, demonstrated by its spectrum (Figure \ref{fig:specfits2}) and solid red color in Figure \ref{fig:rgb}.  These factors suggest the low temperature is real, and not a result of poor statistics.  

The average absorbing column density is \nh $=0.95\times10^{22}\mbox{cm}^{-2}$.  \nh\ is lowest in the southeast and increases to the north and west, spanning a range from $0.56\times10^{22}\mbox{cm}^{-2}$ to $1.44\times10^{22}\mbox{cm}^{-2}$.  Not unexpectedly, the C-shaped `hole' in the northeast has the highest measured column density, along with region T.

The highest ionization timescales, $\sim10^{12}$ s cm$^{-3}$, seem to correspond to the brightest emission, regions X, Y, Z, M, and P, in addition to region C.  The remaining regions have lower ionization timescales, typically a few \tauunit, with an overall average of $\sim5\times$\tauunit.  In general, ionization timescales are higher in the south.

\capstartfalse
\begin{deluxetable*}{cccccccccccccc}[hp]
\tabletypesize{\scriptsize}
\tablecaption{Spectral Fits \label{table:specfits}}
\tablewidth{0pt}
\tablehead{\colhead{Region} & \colhead{Total} & \colhead{\nh} & \colhead{$kT$} & \colhead{$n_et$} & \colhead{norm} & \colhead{Ne} & \colhead{Mg} & \colhead{Si} & \colhead{S} & \colhead{Fe} & \colhead{$C_{970}$} & d.o.f & $\chi^2/$\\
& Counts & ($10^{22}$cm$^{-2}$) & (keV) & (\tauunit) & ($10^{-4}$ cm$^{-5}$) & & & & & & & &d.o.f
}
\startdata
\rule{0pt}{3ex} A & 6295 & 0.56$_{0.09}^{0.09}$ & 0.63$_{0.02}^{0.04}$ & 6.5$_{2.6}^{5.0}$ & 6.83$_{1.03}^{1.25}$ & 0.59$_{0.17}^{0.22}$ & 0.64$_{0.12}^{0.14}$ & 0.70$_{0.17}^{0.18}$ & $\ldots$ & 0.68$_{0.15}^{0.17}$ & 1.18$_{0.05}^{0.06}$ & 162 & 1.23 \\
\rule{0pt}{3ex} B & 2506 & 0.63$_{0.10}^{0.09}$ & 0.80$_{0.11}^{0.17}$ & 1.4$_{0.6}^{1.3}$ & 1.78$_{0.38}^{0.44}$ & 0.78$_{0.30}^{0.31}$ & 0.49$_{0.15}^{0.19}$ & 0.71$_{0.27}^{0.37}$ & $\ldots$ & $\ldots$ & 1.16$_{0.08}^{0.09}$ & 86 & 0.99 \\
\rule{0pt}{3ex} C & 7800 & 0.60$_{0.08}^{0.07}$ & 0.60$_{0.02}^{0.02}$ & 9.6$_{3.6}^{6.2}$ & 9.37$_{1.49}^{1.52}$ & 0.89$_{0.21}^{0.27}$ & 0.73$_{0.13}^{0.18}$ & 0.74$_{0.17}^{0.21}$ & 0.66$_{0.48}^{0.54}$ & 0.79$_{0.13}^{0.23}$ & 0.97$_{0.04}^{0.05}$ & 177 & 1.20 \\
\rule{0pt}{3ex} D+E & 9364 & 0.74$_{0.05}^{0.06}$ & 0.60$_{0.03}^{0.04}$ & 2.8$_{0.8}^{1.2}$ & 16.10$_{2.39}^{2.89}$ & 0.69$_{0.14}^{0.18}$ & 0.72$_{0.10}^{0.12}$ & 0.73$_{0.13}^{0.18}$ & 0.41$_{0.35}^{0.17}$ & 1.18$_{0.18}^{0.22}$ & $\ldots$ & 148 & 1.51 \\
\rule{0pt}{3ex} F & 6889 & 0.82$_{0.06}^{0.06}$ & 0.61$_{0.04}^{0.06}$ & 2.5$_{0.9}^{1.5}$ & 9.27$_{1.66}^{2.28}$ & 0.74$_{0.19}^{0.19}$ & 0.77$_{0.12}^{0.16}$ & 0.97$_{0.19}^{0.23}$ & $\ldots$ & 1.44$_{0.24}^{0.29}$ & 0.86$_{0.04}^{0.04}$ & 170 & 1.12 \\
\rule{0pt}{3ex} G & 16130 & 0.70$_{0.06}^{0.05}$ & 0.59$_{0.02}^{0.03}$ & 3.2$_{0.7}^{1.1}$ & 18.51$_{2.34}^{2.75}$ & 0.76$_{0.11}^{0.12}$ & 0.62$_{0.08}^{0.09}$ & 0.61$_{0.10}^{0.12}$ & 0.50$_{0.28}^{0.31}$ & 1.06$_{0.01}^{0.06}$ & 1.08$_{0.03}^{0.04}$ & 228 & 1.03 \\
\rule{0pt}{3ex} H & 10108 & 1.22$_{0.08}^{0.15}$ & 0.53$_{0.07}^{0.03}$ & 3.5$_{1.1}^{1.8}$ & 36.33$_{5.77}^{15.03}$ & 0.64$_{0.13}^{0.13}$ & 0.52$_{0.08}^{0.09}$ & 0.52$_{0.10}^{0.11}$ & 0.76$_{0.35}^{0.73}$ & 0.92$_{0.15}^{0.19}$ & 1.04$_{0.04}^{0.04}$ & 211 & 1.27 \\
\rule{0pt}{3ex} I & 7840 & 1.24$_{0.11}^{0.11}$ & 0.34$_{0.05}^{0.05}$ & 3.5$_{1.5}^{4.6}$ & 100.84$_{31.70}^{88.17}$ & 0.88$_{0.13}^{0.14}$ & 0.40$_{0.08}^{0.11}$ & 0.41$_{0.12}^{0.19}$ & $\ldots$ & 0.84$_{0.15}^{0.19}$ & $\ldots$ & 143 & 1.12 \\
\rule{0pt}{3ex} J & 7351 & 0.78$_{0.10}^{0.09}$ & 0.55$_{0.05}^{0.05}$ & 3.1$_{1.1}^{1.8}$ & 7.98$_{2.15}^{3.93}$ & 1.01$_{0.21}^{0.21}$ & 1.02$_{0.10}^{0.21}$ & 0.95$_{0.23}^{0.29}$ & $\ldots$ & 1.32$_{0.20}^{0.30}$ & 1.04$_{0.04}^{0.05}$ & 169 & 1.07 \\
\rule{0pt}{3ex} K & 6724 & 1.12$_{0.07}^{0.08}$ & 0.63$_{0.06}^{0.07}$ & 2.4$_{1.0}^{1.5}$ & 13.24$_{3.26}^{0.44}$ & 0.68$_{0.22}^{0.21}$ & 0.82$_{0.16}^{0.19}$ & 0.98$_{0.21}^{0.26}$ & 0.52$_{0.41}^{0.50}$ & 1.19$_{0.24}^{0.31}$ & 1.30$_{0.05}^{0.06}$ & 172 & 1.28 \\
\rule{0pt}{3ex} L & 10680 & 0.93$_{0.06}^{0.05}$ & 0.62$_{0.02}^{0.04}$ & 3.7$_{1.2}^{1.8}$ & 19.34$_{2.32}^{2.58}$ & 0.47$_{0.12}^{0.13}$ & 0.54$_{0.07}^{0.09}$ & 0.52$_{0.10}^{0.11}$ & 0.48$_{0.24}^{0.26}$ & 1.03$_{0.16}^{0.17}$ & 1.02$_{0.04}^{0.05}$ & 206 & 1.27 \\
\rule{0pt}{3ex} M & 13286 & 0.86$_{0.05}^{0.05}$ & 0.55$_{0.02}^{0.01}$ & 8.4$_{2.6}^{4.0}$ & 33.20$_{3.98}^{3.46}$ & 0.49$_{0.10}^{0.11}$ & 0.50$_{0.06}^{0.08}$ & 0.47$_{0.08}^{0.10}$ & 0.77$_{0.29}^{0.30}$ & 0.83$_{0.09}^{0.11}$ & 1.20$_{0.03}^{0.04}$ & 231 & 1.11 \\
\rule{0pt}{3ex} N & 9127 & 1.14$_{0.11}^{0.06}$ & 0.54$_{0.04}^{0.11}$ & 2.0$_{0.6}^{0.5}$ & 20.49$_{8.50}^{5.08}$ & 1.40$_{0.15}^{0.21}$ & 0.97$_{0.12}^{0.27}$ & 0.71$_{0.13}^{0.19}$ & $\ldots$ & 0.91$_{0.13}^{0.33}$ & 1.07$_{0.04}^{0.06}$ & 192 & 1.22 \\
\rule{0pt}{3ex} O & 8841 & 0.95$_{0.06}^{0.06}$ & 0.56$_{0.03}^{0.04}$ & 2.5$_{0.8}^{1.2}$ & 13.00$_{2.66}^{2.95}$ & 0.75$_{0.15}^{0.15}$ & 0.93$_{0.13}^{0.16}$ & 0.91$_{0.17}^{0.20}$ & $\ldots$ & 1.46$_{0.22}^{0.29}$ & 1.14$_{0.04}^{0.05}$ & 190 & 1.31 \\
\rule{0pt}{3ex} P & 12107 & 0.97$_{0.06}^{0.06}$ & 0.53$_{0.02}^{0.02}$ & 8.8$_{2.7}^{5.5}$ & 35.51$_{3.62}^{4.56}$ & 0.51$_{0.09}^{0.11}$ & 0.47$_{0.07}^{0.07}$ & 0.57$_{0.09}^{0.08}$ & 0.46$_{0.23}^{0.24}$ & 0.78$_{0.10}^{0.10}$ & 1.10$_{0.03}^{0.04}$ & 227 & 1.20 \\
\rule{0pt}{3ex} Q & 3991 & 0.98$_{0.08}^{0.08}$ & 0.63$_{0.07}^{0.09}$ & 2.5$_{1.0}^{2.0}$ & 14.53$_{3.15}^{4.88}$ & 0.42$_{0.15}^{0.14}$ & 0.39$_{0.08}^{0.10}$ & 0.41$_{0.11}^{0.14}$ & 0.78$_{0.39}^{0.43}$ & 0.72$_{0.20}^{0.26}$ & $\ldots$ & 100 & 1.45 \\
\rule{0pt}{3ex} R & 12143 & 1.11$_{0.07}^{0.06}$ & 0.54$_{0.03}^{0.02}$ & 3.1$_{0.8}^{1.1}$ & 40.92$_{5.13}^{7.30}$ & 0.61$_{0.08}^{0.10}$ & 0.54$_{0.07}^{0.07}$ & 0.52$_{0.09}^{0.09}$ & 0.52$_{0.24}^{0.24}$ & 0.93$_{0.11}^{0.13}$ & 0.74$_{0.03}^{0.03}$ & 219 & 1.05 \\
\rule{0pt}{3ex} S & 6188 & 1.22$_{0.07}^{0.08}$ & 0.58$_{0.05}^{0.06}$ & 2.7$_{1.0}^{1.5}$ & 14.94$_{3.48}^{5.23}$ & 0.76$_{0.18}^{0.20}$ & 0.68$_{0.13}^{0.16}$ & 1.00$_{0.20}^{0.25}$ & 0.88$_{0.47}^{0.60}$ & 1.14$_{0.24}^{0.32}$ & 1.13$_{0.05}^{0.06}$ & 166 & 1.25 \\
\rule{0pt}{3ex} T & 7989 & 1.44$_{0.09}^{0.09}$ & 0.54$_{0.04}^{0.04}$ & 4.1$_{1.5}^{2.7}$ & 43.32$_{7.37}^{12.30}$ & 0.56$_{0.11}^{0.10}$ & 0.48$_{0.07}^{0.10}$ & 0.66$_{0.11}^{0.11}$ & 0.71$_{0.24}^{0.24}$ & 0.57$_{0.12}^{0.18}$ & 1.09$_{0.04}^{0.05}$ & 198 & 1.16 \\
\rule{0pt}{3ex} U & 8419 & 1.33$_{0.07}^{0.06}$ & 0.54$_{0.03}^{0.03}$ & 3.8$_{1.3}^{2.4}$ & 33.39$_{4.94}^{7.24}$ & 0.55$_{0.12}^{0.13}$ & 0.60$_{0.09}^{0.11}$ & 0.75$_{0.13}^{0.15}$ & 1.09$_{0.37}^{0.41}$ & 0.88$_{0.16}^{0.20}$ & 1.17$_{0.04}^{0.05}$ & 194 & 1.14 \\
\rule{0pt}{3ex} V & 16299 & 0.82$_{0.05}^{0.06}$ & 0.56$_{0.02}^{0.02}$ & 5.9$_{1.5}^{2.0}$ & 29.69$_{0.54}^{6.88}$ & 0.92$_{0.09}^{0.11}$ & 0.71$_{0.08}^{0.10}$ & 0.63$_{0.09}^{0.10}$ & 0.48$_{0.26}^{0.31}$ & 0.77$_{0.08}^{0.09}$ & 1.13$_{0.03}^{0.03}$ & 243 & 1.37 \\
\rule{0pt}{3ex} W & 11462 & 1.04$_{0.06}^{0.07}$ & 0.59$_{0.02}^{0.03}$ & 5.7$_{1.8}^{2.9}$ & 30.15$_{3.66}^{4.74}$ & 0.60$_{0.11}^{0.12}$ & 0.50$_{0.07}^{0.08}$ & 0.56$_{0.09}^{0.09}$ & 0.83$_{0.24}^{0.28}$ & 0.67$_{0.10}^{0.12}$ & 1.12$_{0.04}^{0.04}$ & 231 & 1.20\\
\rule{0pt}{3ex} X & 8455 & 0.86$_{0.08}^{0.08}$ & 0.59$_{0.03}^{0.03}$ & 8.2$_{3.5}^{7.0}$ & 19.93$_{2.54}^{2.79}$ & 0.62$_{0.15}^{0.18}$ & 0.36$_{0.07}^{0.08}$ & 0.41$_{0.09}^{0.11}$ & 0.48$_{0.30}^{0.32}$ & 0.59$_{0.10}^{0.13}$ & 1.15$_{0.04}^{0.05}$ & 196 & 1.47 \\
\rule{0pt}{3ex} Y & 14530 & 0.71$_{0.06}^{0.06}$ & 0.56$_{0.02}^{0.02}$ & 14.8$_{5.7}^{11.9}$ & 31.99$_{3.65}^{4.37}$ & 0.77$_{0.13}^{0.14}$ & 0.37$_{0.05}^{0.06}$ & 0.34$_{0.07}^{0.08}$ & 0.38$_{0.18}^{0.21}$ & 0.46$_{0.06}^{0.07}$ & 1.13$_{0.03}^{0.04}$ & 239 & 1.25 \\
\rule{0pt}{3ex} Z & 10537 & 0.60$_{0.08}^{0.07}$ & 0.57$_{0.02}^{0.03}$ & 13.1$_{5.4}^{11.7}$ & 18.14$_{2.21}^{2.29}$ & 0.65$_{0.14}^{0.15}$ & 0.42$_{0.07}^{0.08}$ & 0.39$_{0.08}^{0.11}$ & 0.31$_{0.26}^{0.28}$ & 0.50$_{0.08}^{0.09}$ & 1.11$_{0.04}^{0.04}$ & 205 & 0.95 \\
\rule{0pt}{3ex} AA & 10600 & 0.86$_{0.06}^{0.06}$ & 0.60$_{0.02}^{0.03}$ & 7.1$_{2.3}^{3.4}$ & 16.35$_{2.22}^{2.31}$ & 1.08$_{0.20}^{0.24}$ & 0.82$_{0.13}^{0.15}$ & 0.61$_{0.12}^{0.15}$ & 0.37$_{0.28}^{0.31}$ & 0.82$_{0.12}^{0.15}$ & 1.11$_{0.04}^{0.04}$ & 202 & 1.12 \\
\rule{0pt}{3ex} Hole & 7514 & 1.44$_{0.35}^{0.11}$ & 0.59$_{0.09}^{0.16}$ & 1.2$_{0.6}^{5.5}$ & 27.67$_{12.96}^{16.56}$ & 0.33$_{0.24}^{0.10}$ & 0.22$_{0.06}^{0.08}$ & 0.53$_{0.10}^{0.12}$ & $\ldots$ & 0.20$_{0.10}^{0.19}$ & 1.54$_{0.08}^{0.09}$ & 182 & 1.20
\enddata
\tablecomments{\footnotesize{Best-fit \vpshock\ parameters to the regional spectra.  The normalization is defined as $10^{-14}(4\pi D^{2})^{-1}\int n_en_HdV$, where D is the distance to \rcw\ and the integral is the volume emission measure.  All abundances are relative to solar.  Sulfur and iron abundances are not given for regions in which they were unconstrained. $C_{970}$ is the extra normalization factor applied to ObsID 970 (not given for the three regions that fall outside the ACIS-S3 field of view).  Errors represent 90\% confidence intervals.}}
\end{deluxetable*}
\capstarttrue

Ne is subsolar in most regions with an average value of 0.71 Ne$_\odot$, typically ranging from 0.4 to 1.0 Ne$_\odot$. Fe has a higher average abundance of 0.88 Fe$_\odot$.  Eight regions have Fe abundances larger than 1 Fe$_\odot$, reaching as high as 1.46 Fe$_\odot$.  The remaining regions have Fe$<1$ Fe$_\odot$ and span a wide range from 0.2 to 0.93 Fe$_\odot$.  Compared to Ne and Fe, Mg and Si have lower overall abundances, typically between 0.4 and 0.8 solar.  The average Mg and Si abundances are 0.60 Mg$_\odot$ and 0.64 Si$_\odot$, respectively, with a few regions going as high as $\sim$1.  Because of the weak S line in many regions, the spectral fits were not always able to constrain the S abundance. It is constrained in about two-thirds of the regions, with values typically between 0.3 S$_\odot$ and 0.8 S$_\odot$ and an average of 0.60 S$_\odot$.  The largest is $\sim$1 S$_\odot$ in region U.  However, uncertainties in the S abundances are much larger than for the other elements, and consequently the measured variations in S across the remnant should be interpreted with some caution.  As expected from the anti-correlation on the EWIs, Mg, Si, and S abundances in the bright southeast regions X, Y, and Z are among the lowest, about 0.4 solar for all three.  Fe is also quite low in these regions, $0.4-0.6$ Fe$_\odot$.

\section{Discussion}
\label{section:discussion}
\subsection{Absorption}
\label{section:absorption}
\begin{figure}[h]
\begin{center}
\includegraphics[width=0.8\columnwidth]{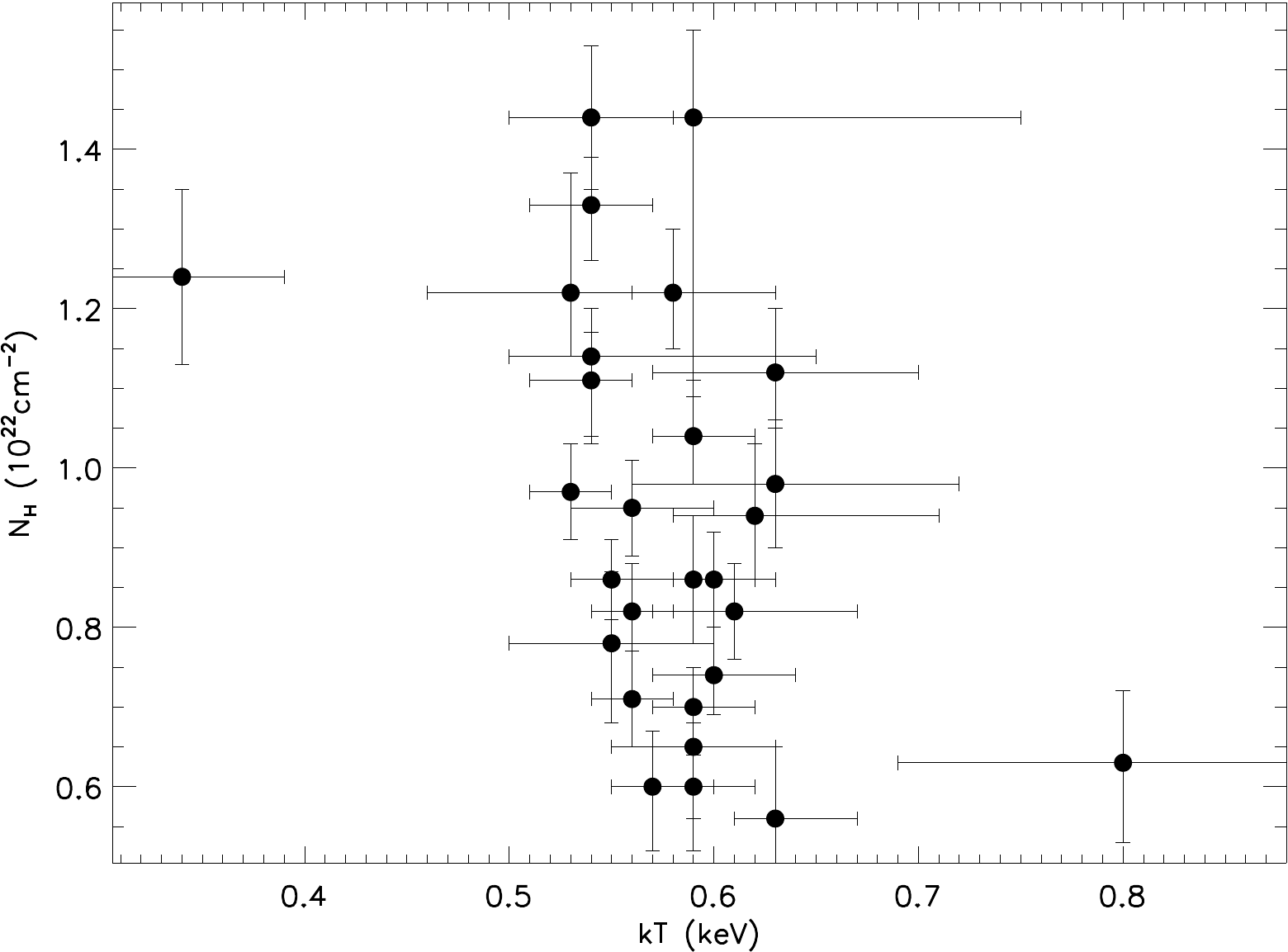} 
\caption{\footnotesize \nh\ vs $kT$ from the single \vpshock\ spectral fits for all regions.  There is no apparent systematic correlation of \nh\ with $kT$.}
\label{fig:nHkTscatter}
\end{center}
\end{figure}

As might be expected from the X-ray image, the spectral fits find generally higher absorbing column densities in the fainter northeast and southwest regions (see Figure \ref{fig:bars}).  It is unlikely that this variation is due to an artifact from a systematic correlation with $kT$ in the model fits, as there is no apparent correlation between the two (Figure \ref{fig:nHkTscatter}); thus these large \nh\ variations appear to be real.  

The C-shaped hole, which is also seen in the infrared at 8 and 24 $\mu$m (Figure \ref{fig:spitzer}), has the highest column density.  Some skepticism of the spectral fit for this region is probably in order, due to the suspiciously low best-fit values of Ne, Mg, and Fe and the high value of $C_{970}$. However, even restricting the parameters during the fit to be similar to the best-fit parameters of neighboring region K or the averages of all regions still results in \nh $ >10^{22}\mbox{cm}^2$. Given the matching feature in the mid-infrared images, it is likely that the `hole' is caused by a foreground dense molecular cloud (see the Appendix for further discussion).

\subsection{CSM}
\label{section:CSM}
In \rcw\, separating shocked CSM emission from that of metal-rich ejecta is not straightforward. 
The average best-fit abundances of Ne, Mg, Si, S, and Fe are all subsolar, which makes it likely that the CSM abundances are also less than solar. We attempt to gain insight into the CSM by examining a few regions that are most likely to be dominated by CSM rather than ejecta.  There are a number of regions that have subsolar abundances for all five elements.  In particular, regions X, Y, and Z show among the lowest abundances and are located where IR and radio observations indicate the blast wave is interacting with local atomic gas \citep{Oliva1999,Reach2006,Pinheiro-Goncalves2011}.  The average abundances for these three regions are Ne $=0.68_{-0.08}^{+0.09}$ Ne$_\odot$, Mg $=0.38\pm0.04$ Mg$_\odot$, Si $=0.38_{-0.05}^{+0.06}$ Si$_\odot$, S $=0.39_{-0.15}^{+0.16}$ S$_\odot$, and Fe $=0.52_{-0.05}^{+0.06}$ Fe$_\odot$.  Whether it is assumed that these, the overall remnant averages, or something in between are representative of the CSM, it is clear that the CSM has distinctly subsolar Ne, Mg, Si, S, and Fe abundances.  These low abundances may also suggest the progenitor had subsolar metallicity. 

We estimate the post-shock electron density in each region based on our measured volume emission measure. We assume $n_e = 1.2 n_H$, where $n_H$ is the number density of hydrogen, and that the path length of each region along the line of sight is similar to its major axis in the plane of the sky.  The distance to \rcw\ is taken to be 3.3 kpc.  For regions X, Y, and Z, which appear to be dominated by CSM emission, we find $n_e\sim30-60f^{-1/2}$ cm$^{-3}$ (where f is the volume filling factor of the X-ray emitting gas).  Similar densities are found in regions M and P, which are located in the bright northwest area and similar in many respects to the above regions.  Densities in the east (regions A-F) are significantly lower, $\sim6-10f^{-1/2}$ cm$^{-3}$.  These eastern emission regions appear as protrusions located beyond the forward shock; combined with the lower densities, this may indicate that these protrusions are clumps of shocked material that are breaking out of the dense CSM shell.

\subsection{Ejecta}
\label{section:ejecta}
Given that the CSM seems to have abundances $\lesssim$0.5 solar, regions with substantially higher abundances likely include contributions from metal-rich ejecta.  Taking the abundances of regions X, Y, and Z to be representative of the CSM, we  roughly classify all of the regions into two categories.  In the first category, the best-fit abundances are at least twice these nominal CSM values for one or more elements, indicating a considerable contribution from metal-rich ejecta.  The second category includes the remaining regions, whose emission is most likely dominated by CSM with very little or no contribution from shocked ejecta. The classification of each region is shown in Table \ref{table:regionclassification}.

\capstartfalse
\begin{deluxetable}{cccc}[h]
\tabletypesize{\footnotesize}
\tablecaption{Classification of Regions \label{table:regionclassification}}
\tablewidth{0pt}
\tablehead{\colhead{Region} & \colhead{Ejecta/CSM} }
\startdata
A & CSM \\
B & CSM \\
C & CSM \\
D+E & Ejecta (Fe) \\
F & Ejecta (Mg, Si, Fe) \\
G & Ejecta (Fe) \\
H & CSM \\
I & CSM \\
J & Ejecta (Mg, Si, Fe) \\
K & Ejecta (Mg, Si, Fe) \\
L & CSM \\
M & CSM \\
N & Ejecta (Ne, Mg) \\
O & Ejecta (Mg, Si, Fe) \\
P & CSM \\
Q & Ejecta (S) \\
R & CSM \\
S & Ejecta (Si, S, Fe) \\
T & CSM \\
U & Ejecta (S) \\
V & CSM \\
W & Ejecta (S)\\
X & CSM \\
Y & CSM \\
Z  & CSM \\
AA & Ejecta (Mg) \\
Hole & CSM
\enddata
\end{deluxetable}
\capstarttrue

We classify twelve regions as candidate ejecta regions. We fit the spectra of these regions with a two-component CSM+ejecta model.  The model is the same as in the previous fits, but with an extra thermal component to represent the CSM.  For each region, the absorbing column density (applied to both the CSM and ejecta components) is fixed to the value from the single-component fit.  
We assume the CSM is uniform and fix the abundances and temperature to the averages of regions X, Y, and Z from the single-component fits: Ne $=0.68_{-0.08}^{+0.09}$ Ne$_\odot$, Mg $=0.38\pm0.04$ Mg$_\odot$, Si $=0.38_{-0.05}^{+0.06}$ Si$_\odot$, S $=0.39_{-0.15}^{+0.16}$ S$_\odot$, Fe $=0.52_{-0.05}^{+0.06}$ Fe$_\odot$, and $kT=0.57$ keV.  It is possible that the CSM does not have a uniform temperature distribution within the remnant; however our single-component fits indicate little variation in temperature across the remnant. Allowing the CSM temperature to vary in our two-component model fits results in poorly constrained ejecta temperatures, abundances, and ionization ages (which tend to be higher, larger, and lower, respectively), along with a systematically lower CSM temperature ($\sim$0.35 keV) that is inconsistent with the temperature of regions X, Y, and Z. For the CSM component, we first tried a \vpshock\ model, but the best-fit ionization timescales tended to be $\gtrsim10^{12}$ s cm$^{-3}$, so we instead used an equilibrium model ({\tt vequil}) to represent the CSM.  A \vpshock\ component with free temperature, ionization age, and Ne, Mg, Si, S,  and Fe abundances represents the ejecta.  For five of the regions (D+E, Q, U, K, and W), the single-component Ne abundance was similar to or smaller than the nominal CSM value, and the two-component fit was thus improved significantly by fixing Ne $= 0$ in the ejecta component.  This was also the case for S in regions D+E and K.  As with the single-component fits, S was fixed to solar in regions where it was unconstrained.  Results from these fits are shown in Table \ref{table:specfits2}.  While the CSM appears to be in equilibrium, the ionization timescales of the ejecta are only a few \tauunit. The average ejecta temperature, at 0.66 keV, is higher than that of the CSM, and it has a wider range, from 0.53 keV to 0.85 keV.  The ejecta abundances are typically within a factor of two of solar, except in region O, which has abundances of $\sim$2-7 solar.  

Taking into account the EWIs and the spectral analyses, we can provide a picture of the overall distribution of ejecta in \rcw.  Only a few regions seem to contain significant Ne ejecta, all of which are located (in projection) in the interior and also contain emission from Mg ejecta. Mg ejecta is present in a small area on the eastern edge of the remnant (regions D+E and F, where the remnant appears to be slightly elongated), with the remainder spread out in the north and west interior and the very compact region AA.  The distribution of Si ejecta is nearly identical to that of Mg.  Other than a small clump near region H, it appears that S ejecta is concentrated on the far western side of the remnant.  The most widespread element is Fe, which seems to be present nearly everywhere except the far south, typically in larger quantities than the other elements.  The highest concentrations of Fe ejecta essentially mirror the distribution of Mg ejecta.  While metal-rich ejecta appears to contribute to the emission in many regions, we do not find any features that are strongly dominated by this ejecta emission as has been found in other young SNRs (with the possible exception of region O).

\capstartfalse
\begin{deluxetable*}{ccccccccccccc}[hp]
\tabletypesize{\scriptsize}
\tablecaption{CSM+Ejecta Spectral Fits \label{table:specfits2}}
\tablewidth{0pt}
\tablehead{\colhead{Region} & \colhead{norm$_{CSM}$} & \colhead{$kT_{ejecta}$} & \colhead{$n_et$} & \colhead{norm$_{ejecta}$} & \colhead{Ne} & \colhead{Mg} & \colhead{Si} & \colhead{S} & \colhead{Fe} & \colhead{$C_{970}$} & d.o.f & $\chi^2/$\\
& ($10^{-4}$ cm$^{-5}$) & (keV) & (\tauunit) & ($10^{-4}$ cm$^{-5}$) & & & & & & & & d.o.f
}
\startdata
\rule{0pt}{3ex} D+E &  6.40$_{3.16}^{2.21}$ & 0.75$_{0.06}^{0.09}$ & 1.4$_{0.4}^{0.8}$ & 7.26$_{1.74}^{3.35}$ & 0 & 0.88$_{0.20}^{0.21}$ & 0.82$_{0.22}^{0.22}$ & 0 & 2.34$_{0.56}^{0.51}$ & $\ldots$ & 150 & 1.84 \\
\rule{0pt}{3ex} F &  4.05$_{3.23}^{2.63}$ & 0.64$_{0.08}^{0.10}$ & 1.4$_{0.7}^{1.1}$ & 5.81$_{2.05}^{3.39}$ & 0.92$_{0.31}^{0.43}$ & 0.95$_{0.22}^{0.34}$ & 1.19$_{0.30}^{0.49}$ & $\ldots$ & 1.82$_{0.42}^{0.63}$ & 0.86$_{0.04}^{0.04}$ & 170 & 1.10 \\
\rule{0pt}{3ex} G &  5.01($<$10.31) & 0.59$_{0.02}^{0.05}$ & 2.3$_{0.9}^{1.2}$ & 13.94$_{4.10}^{6.50}$ & 0.92$_{0.24}^{0.32}$ & 0.73$_{0.17}^{0.23}$ & 0.70$_{0.18}^{0.24}$ & 0.55$_{0.43}^{0.54}$ & 1.21$_{0.23}^{0.28}$ & 1.08$_{0.03}^{0.04}$ & 228 & 1.03 \\
\rule{0pt}{3ex} J &  3.25($<$5.79) & 0.53$_{0.09}^{0.06}$ & 2.1$_{1.0}^{1.5}$ & 5.57$_{1.82}^{2.63}$ & 1.33$_{0.41}^{0.58}$ & 1.36$_{0.43}^{0.68}$ & 1.20$_{0.39}^{0.70}$ & $\ldots$ & 1.65$_{0.43}^{0.73}$ & 1.04$_{0.04}^{0.05}$ & 169 & 1.05 \\
\rule{0pt}{3ex} K &  4.36$_{3.48}^{3.20}$ & 0.83$_{0.11}^{0.12}$ & 1.1$_{0.3}^{0.8}$ & 6.12$_{2.04}^{2.97}$ & 0 & 0.97$_{0.20}^{0.34}$ & 1.10$_{0.27}^{0.44}$ & 0 & 2.32$_{0.52}^{0.83}$ & 1.31$_{0.06}^{0.05}$ & 174 & 1.39 \\
\rule{0pt}{3ex} N &  0.30($<$4.25) & 0.55$_{0.06}^{0.04}$ & 1.9$_{0.5}^{0.6}$ & 19.90$_{4.34}^{4.24}$ & 1.41$_{0.16}^{0.14}$ & 0.98$_{0.11}^{0.22}$ & 0.71$_{0.12}^{0.12}$ & $\ldots$ & 0.93$_{0.14}^{0.22}$ & 1.07$_{0.04}^{0.04}$ & 192 & 1.22 \\
\rule{0pt}{3ex} O &  9.31$_{1.47}^{1.34}$ & 0.54$_{0.07}^{0.09}$ & 0.2$_{0.1}^{0.1}$ & 5.42$_{1.20}^{1.48}$ & 1.79$_{0.36}^{0.56}$ & 3.46$_{1.05}^{1.64}$ & 4.55$_{1.71}^{2.77}$ & $\ldots$ & 6.82$_{2.53}^{5.11}$ & 1.15$_{0.04}^{0.04}$ & 190 & 1.27 \\
\rule{0pt}{3ex} Q &  5.37$_{4.21}^{3.31}$ & 0.83$_{0.14}^{0.14}$ & 1.1$_{0.4}^{1.6}$ & 5.59$_{2.24}^{3.58}$ & 0 & 0.53$_{0.15}^{0.23}$ & 0.49$_{0.17}^{0.26}$ & 0.87$_{0.63}^{0.98}$ & 1.60$_{0.52}^{0.67}$ & $\ldots$ & 101 & 1.57 \\
\rule{0pt}{3ex} S &  0.0($<$2.79) & 0.58$_{0.04}^{0.06}$ & 2.7$_{1.0}^{1.2}$ & 14.84$_{3.81}^{3.83}$ & 0.76$_{0.17}^{0.19}$ & 0.68$_{0.11}^{0.15}$ & 1.00$_{0.19}^{0.27}$ & 0.88$_{0.45}^{0.59}$ & 1.14$_{0.21}^{0.33}$ & 1.13$_{0.05}^{0.05}$ & 166 & 1.25 \\
\rule{0pt}{3ex} U &  14.49$_{5.76}^{4.71}$ & 0.66$_{0.06}^{0.08}$ & 1.4$_{0.5}^{0.0}$ & 9.73$_{3.85}^{5.43}$ & 0 & 1.10$_{0.34}^{0.77}$ & 1.23$_{0.40}^{0.89}$ & 1.51$_{0.77}^{1.43}$ & 2.50$_{0.67}^{1.44}$ & 1.17$_{0.04}^{0.05}$ & 195 & 1.30 \\
\rule{0pt}{3ex} W &  17.91$_{0.67}^{4.01}$ & 0.85$_{0.10}^{0.06}$ & 1.3$_{0.3}^{0.8}$ & 6.87$_{1.99}^{3.21}$ & 0 & 0.74$_{0.16}^{0.23}$ & 0.75$_{0.18}^{0.27}$ & 0.98$_{0.50}^{0.75}$ & 1.83$_{0.40}^{0.52}$ & 1.13$_{0.04}^{0.04}$ & 232 & 1.34 \\
\rule{0pt}{3ex} AA &  0.22($<$11.86) & 0.60$_{0.02}^{0.05}$ & 6.9$_{4.2}^{2.4}$ & 16.12$_{9.86}^{2.60}$ & 1.08$_{0.10}^{0.62}$ & 0.82$_{0.11}^{0.12}$ & 0.61$_{0.11}^{0.13}$ & 0.37$_{0.37}^{0.35}$ & 0.83$_{0.12}^{0.35}$ & 1.11$_{0.04}^{0.04}$ & 202 & 1.12
\enddata
\tablecomments{\footnotesize{Best-fit parameters from the {\tt vequil+vpshock} fits to the ejecta regional spectra.  The normalizations and $C_{970}$ have the same definitions as given in Table \ref{table:specfits}.  Abundances are relative to solar.  For several regions that had Ne and/or S abundances less than the nominal CSM abundances in the single-component fits, it was necessary to fix the ejecta Ne and/or S to zero.  As in the single-component fits, S was fixed to 1 where unconstrained.  Errors represent 90\% confidence intervals.}}
\end{deluxetable*}
\capstarttrue

\begin{figure}[htbp!]
\subfigure {\includegraphics[width=0.9\columnwidth]{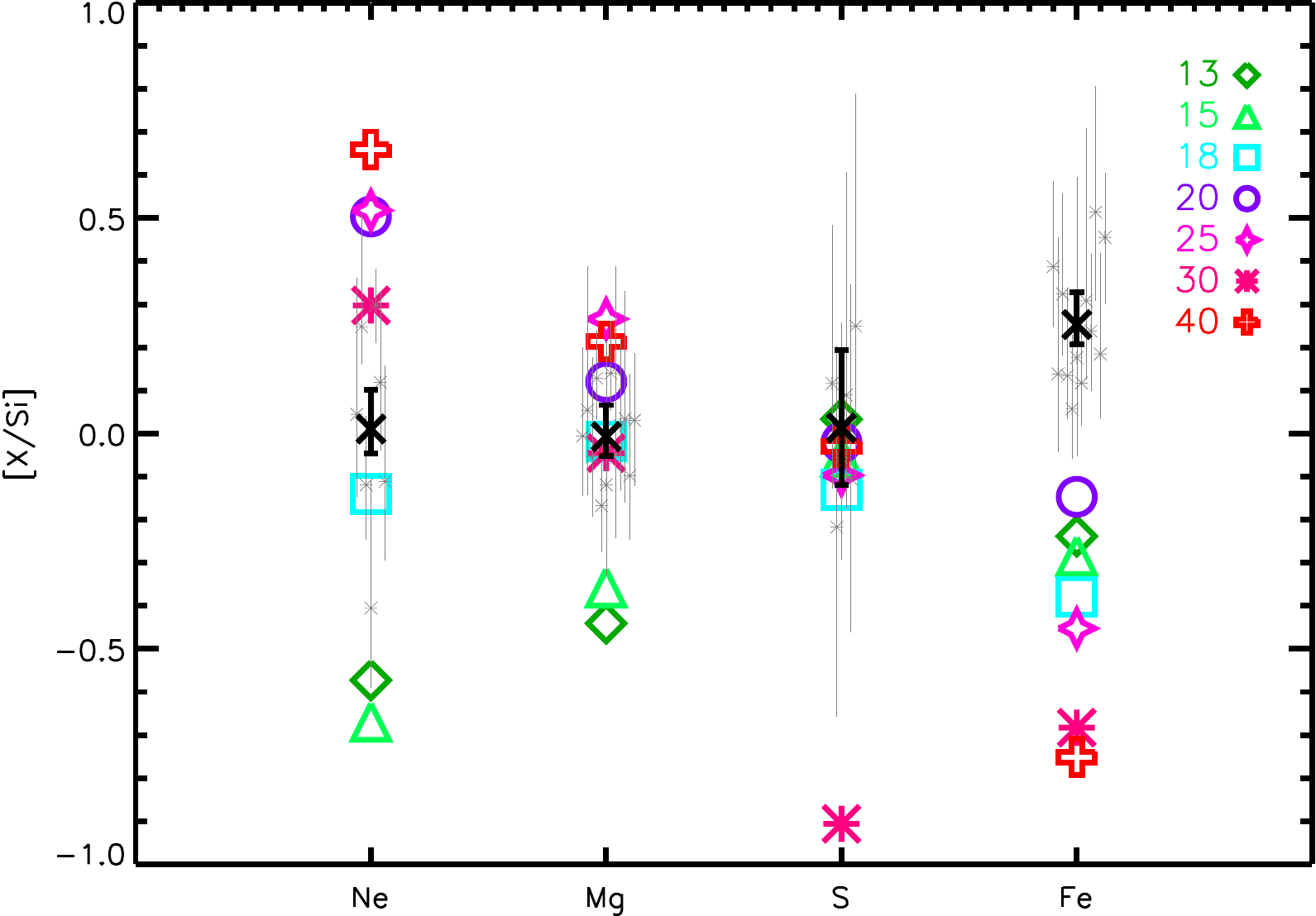}} 
\subfigure {\includegraphics[width=0.9\columnwidth]{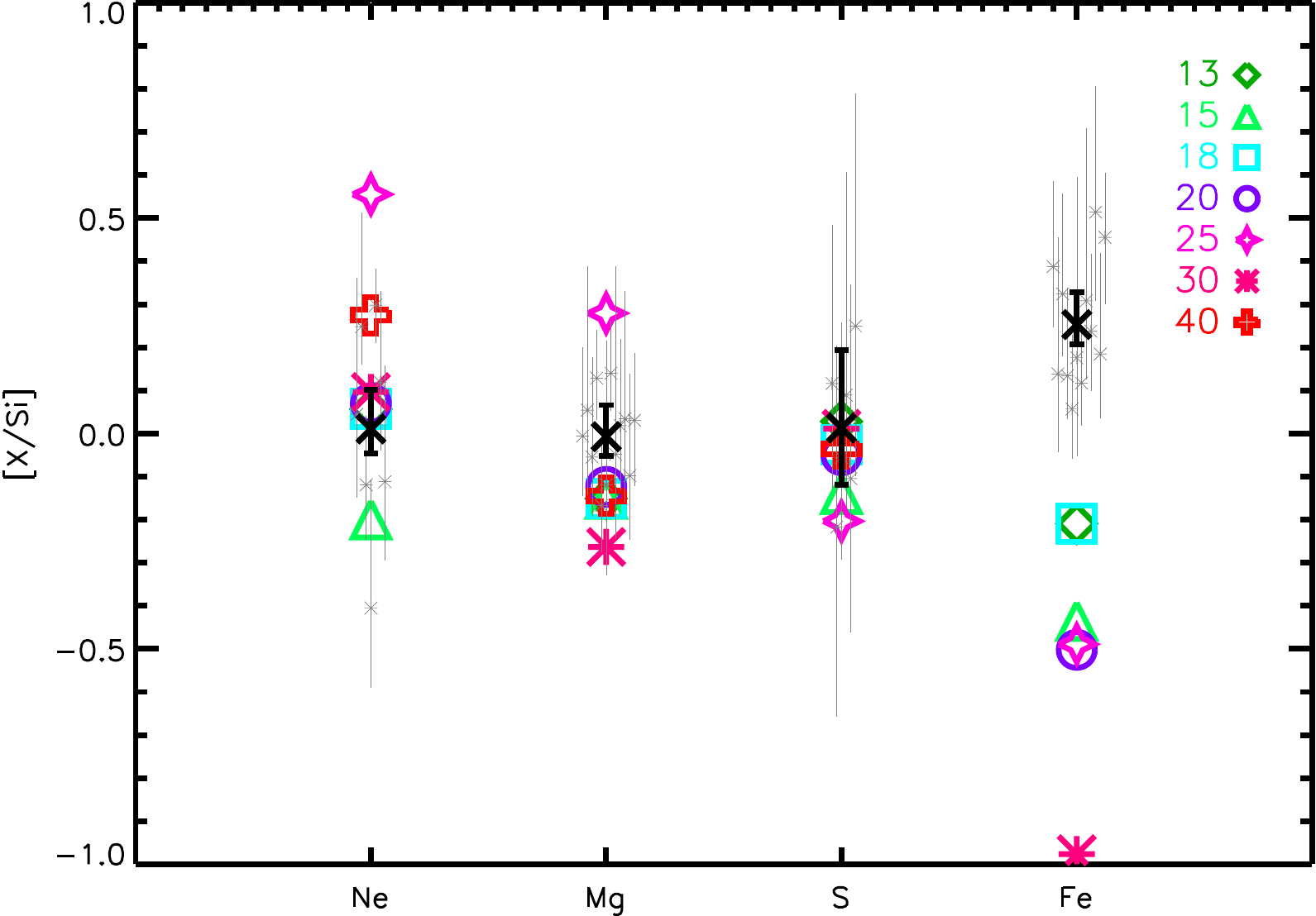}} 
\caption{\footnotesize Comparison of the abundance ratios predicted by the \citet{Nomoto2006} models for different progenitor masses (in units of solar mass) and metallicities (top = solar metallicity, bottom = 0.5 solar metallicity) to the best-fit abundance ratios of the 12 ejecta regions, as measured from the two-component spectral fits. The small gray points and bars represent the individual abundance ratios of each region and their associated errors (each slightly offset horizontally for visibility).  The black $\times$ are the average of these ratios for each element.}
\label{fig:nomoto}
\end{figure}

We can compare the measured abundance ratios with those from supernova nucleosynthesis models.  O/Fe, along with Si/Fe, typically are the most useful ratios for distinguishing between nucleosynthesis models.  However, the high absorption prevents any oxygen measurements in \rcw, and abundances measured from the Fe L shell lines suffer from incomplete atomic data and as a result have larger uncertainties.   We instead use abundances relative to Si, though the results are similar when comparing to Fe.  As it is only relevant to compare ejecta abundances, not CSM, we compare only the ejecta abundances from the CSM+ejecta model spectral fits.  
Figure \ref{fig:nomoto} compares the ejecta abundances with respect to Si to the predicted ratios from the core-collapse SN nucleosynthesis models of \citet{Nomoto2006} for a range of progenitor masses.  All models are of spherical explosions with energies of $10^{51}$ ergs.  We show models with both solar and half solar metallicity progenitor stars.  For a solar metallicity progenitor, the Ne, Mg, and S abundance ratios together prefer the 18 M$_\odot$ model, driven primarily by the Ne and Mg abundances.  If the progenitor had a metallicity of about half solar (as suggested by the CSM abundances), the results are similar, but also roughly match the 20 M$_\odot$ model.  The measured Fe/Si ratio is quite high and neither the average value or the ratios of any individual regions are consistent with any of the models; however, the Fe abundance is measured by weak L-lines, and thus may be less reliable or involve larger uncertainties than the other abundances. 
Overall, the ejecta abundances suggest that \rcw\ had a progenitor with approximately $18-20$ M$_\odot$.  This should be taken with some caution, as 1) it is supported primarily by Ne and Mg, whereas the S abundances are consistent with both higher and lower mass models, and the Fe/Si ratios are substantially higher than predicted for any model, and 2) it assumes that regions O, P, and U are representative of the CSM and that the measured abundances are therefore ``pure" ejecta.

The preferred progenitor mass is near the upper boundary for Type IIP supernovae \citep{Heger2003}, suggesting \rcw\ might be the remnant of a Type IIP supernova from a red supergiant.
The existence of a neutron star also argues for a progenitor mass of $\lesssim25$ M$_\odot$ \citep{Fryer1999,Heger2003}.  This is in contrast to the larger progenitor mass suggested by the dense CSM shell, which would suggest a higher mass-loss rate and is thus more compatible with a Type IIL/b origin, as also proposed by \citet{Chevalier2005}.  A scenario in which the progenitor was part of a binary system may somewhat alleviate this discrepancy, as it would alter the mass-loss rate.  Such a binary scenario is also one of the two leading hypotheses for explaining the strange nature of the CCO \citep{Pizzolato2008,Bhadkamkar2009}. Unfortunately, the superposition of CSM and metal-rich ejecta emission throughout the remnant, along with the variations in absorbing column density, make a reliable estimate of the shocked CSM mass very difficult in the current data. 
Further constraints on the nature of the progenitor via ejecta abundance measurements or estimates of the CSM mass will require deeper X-ray observations, as the two-component spectral fits necessary to separate the ejecta emission from CSM emission are hampered by low statistics and high absorption. 

\section{Conclusions}
\label{section:conclusions}

We analyzed three \chandra\ observations of the Galactic supernova remnant \rcw.  Investigation of the X-ray images and the spectral analysis of 27 regions find that large areas of \rcw\ are subject to significant absorption, which varies across the remnant and is stronger in the north and west. The electron temperature, however, is fairly uniform at $\sim$0.58 keV.  The highest ionization timescales are in the southeast and northwest regions, which are associated with the brightest emission at all wavelengths and are known to be sites of shock interaction with ambient atomic and molecular gas.  

Both the equivalent width images and spectral analyses of the regions indicate that the Ne, Mg, Si, S, and Fe abundance distributions in \rcw\ are inhomogeneous and can vary from region to region, from about 0.3 to 1.4 solar.  The EWIs show that Mg, Si, and S emission appears to be anti-correlated with the bright southeast region. 
The lowest measured abundances are generally found in this part of the remnant, $\lesssim0.5$ solar  for Mg, Si, S, and Fe, and $\sim$0.7 for Ne.  In addition, the average best-fit abundances of Ne, Mg, Si, S, and Fe for the entire SNR are all subsolar; this implies that the CSM abundances are also less than solar.

From CSM+ejecta model fits to a number of the regional spectra, we find ejecta abundances of a few times solar, $\sim1-5$, with Fe being the most abundant.  All five elements (Ne, Mg, Si, S, and Fe) appear to be scattered over the face of the remnant, including near the edges and in (or projected on) the interior.  This may indicate that at least in some areas, core products (Si, S, Fe) have propagated into the remnant outskirts, i.e. some amount of ejecta mixing or overturning has possibly occurred.  Comparisons of ejecta abundance ratios from the CSM+ejecta spectral fits with nucleosynthesis model predictions, despite finding larger Fe/Si ratios than predicted for any progenitor mass, overall suggest a progenitor mass of $\sim$18-20 M$_\odot$, corresponding with a relatively high-mass Type IIP origin \citep{Heger2003}.  

The main challenge in characterizing the abundances in \rcw\ is disentangling ejecta emission from shocked CSM.   Despite some contribution from metal-rich ejecta in a number of regions, the X-ray emission is typically dominated by shocked CSM.  This may indicate an unusually dense or massive CSM, whose emission drowns out much of the ejecta emission, and/or a relatively small quantity of metal-rich ejecta that is only present in significant quantities along some lines-of-sight.  CSM densities estimated from spectral fits to the brightest regions (in the southeast and northwest) are $n_e\sim30-60f^{-1/2}$ cm$^{-3}$ (where $f$ is the volume filling factor of the X-ray emitting gas).  These densities are several times lower in the east, where the emission extends beyond the forward shock (Figure \ref{fig:rgb}), possibly indicating that the remnant is expanding into a more tenuous ambient medium in this direction. 

The apparently large amount of CSM may indicate a high mass-loss rate for the progenitor, and therefore a Type IIL/b origin, as proposed by \citet{Chevalier2005}.  This is in tension with both the low progenitor metallicity implied by the CSM abundances, which would lead to lower mass-loss rates, and the somewhat lower progenitor mass implied by the ejecta abundance ratios.  The fact that a neutron star was created also implies a progenitor mass of $\lesssim$25 M$_\odot$, as a more massive star would have resulted in a black hole \citep[c.f.][]{Fryer1999,Heger2003}.  Solving this discrepancy may have implications on the mysterious nature of the CCO.  Note that many stellar evolution and SN models, including \citet{Heger2003}, assume an isolated, non-rotating star.  A binary system is a major ingredient in many of the proposed explanations for the CCO's strange variability \citep[e.g.][]{Pizzolato2008,Bhadkamkar2009}, and binary interactions could have played a role in the mass loss of the progenitor, affecting both the CSM and the final progenitor mass.

\rcw\ is an unusual system. It displays an apparent dearth of metal-rich ejecta compared to similarly-aged SNRs, and it contains a unique and unexplained CCO. 
The current analysis demonstrates the existence of some metal-rich ejecta, but the strong CSM emission prevents precision measurements of the ejecta abundances.  Conversely, the presence of metal-rich ejecta, though minor, along with strong variations in absorbing column density and other physical properties, make it difficult to constrain overall properties of the CSM, such as the total swept-up mass.  Deeper observations are needed to more precisely separate the CSM and ejecta emission and thereby obtain better constraints on the properties of both components, and through them the origin of the remnant and its CCO.

\acknowledgments
The authors would like to thank Robert A. Benjamin for helpful remarks on molecular cloud absorption.
This work was supported by the Penn State ACIS Instrument Team Contract SV4-74018, issued by the Chandra X-ray Center, which is operated by the Smithsonian Astrophysical Observatory for and on behalf of NASA under contract NAS8-03060. The Guaranteed Time Observations (GTO) included here were selected by the ACIS Instrument Principal Investigator, Gordon P. Garmire, of the Huntingdon Institute for X-ray Astronomy, LLC, which is under contract to the Smithsonian Astrophysical Observatory; Contract SV2-82024. 

\bibliography{/Users/kafrank/Dropbox/master_refs}

\include{rcw103_appendix}

\end{document}

%% file: rcw103_appendix.tex
\appendix
\section{Region Descriptions} 
\label{section:regions}

Many of the individual regions can be grouped together based on abundances and ionization timescales from the single \vpshock\ spectral fits, which indicate similar physical conditions within each group.  We briefly discuss each of these groups, as well as regions which do not fit into any groups.  

\subsection{Group 1}
\label{section:group1}
The three southern regions X, Y, and Z represent the brightest X-ray emission.  They are also colocated with the brightest 24 $\mu$m, optical, and radio emission.  IR imaging and spectroscopy has revealed shocked atomic gas in the same location \citep[cf.\ Figure 17 in][]{Reach2006}. The abundances are all among the lowest anywhere in the remnant, and the ionization timescales among the highest ($n_et=8-15\times$\tauunit), suggesting high densities, as do the estimated densities from the model normalizations.  This bright area appears to be where the main blast wave is impacting dense ambient material and shock-heating local dust.  The nominal CSM abundances are taken to be the average of these three regions, Ne $=0.68_{-0.08}^{+0.09}$ Ne$_\odot$, Mg $=0.38\pm0.04$ Mg$_\odot$, Si $=0.38_{-0.05}^{+0.06}$ Si$_\odot$, S $=0.39_{-0.15}^{+0.16}$ S$_\odot$, and Fe $=0.52_{-0.05}^{+0.06}$ Fe$_\odot$.

\subsection{Group 2}
\label{section:group2}
The brightest emission in the north comes from regions M and P, which are more similar to the bright regions in the south (X, Y, and Z) than to neighboring regions in the north or west.  They have high ionization timescales and are coincident with shocked atomic gas seen in the IR.  Both have Ne abundances less than the CSM values and only slightly enhanced Mg ($\sim$0.5 Mg$_\odot$), Si ($\sim$0.5 Si$_\odot$ and $\sim$0.6 Si$_\odot$), and Fe ($\sim$0.8 Fe$_\odot$).  Region M, easily the most significant peak in the S EWI, also has enhanced S ($0.77_{-0.29}^{+0.30}$S$_\odot$).  These two regions may contain small amounts of Mg, Si, and Fe ejecta (and S in region M), but are overall quite similar to CSM-dominated regions in the south where the blast wave is shocking ambient atomic gas.

\subsection{Group 3}
\label{section:group3}
Regions C, V, and AA are located in the southern half of the remnant.  Region AA is a bright knot in the eastern interior, region V is a fainter, more absorbed, region to the west, and region C is one of the larger protrusions on the southeast edge.  
Ionization timescales for this group are larger than average, $n_et\sim6-9.5\times$ \tauunit.  All regions have Ne abundance values of $\sim$1 and enhanced Fe, Si, and Mg abundances ($\sim$0.8 Fe$_\odot$, $0.6-0.75$ Si$_\odot$, and $0.7-0.8$ Mg$_\odot$ respectively).  The S abundance is near or slightly above the CSM value, $\sim0.4-0.6$ S$_\odot$.   
The CSM+ejecta spectral fit to region AA measures $\sim$solar abundances of Ne, along with enhanced Mg ($0.82_{-0.11}^{+0.12}$ Mg$_\odot$), Si ($0.61_{-0.11}^{+0.13}$ Si$_\odot$), and Fe ($0.83_{-0.12}^{+0.35}$ Fe$_\odot$).
These regions likely contain minor amounts of Ne, Si, Mg, and Fe ejecta, along with the usual large amount of CSM.

\subsection{Group 4}
\label{section:group4}
The two southern-most protrusions, regions A and W, along with region T in the west, have similar spectral properties.  A and W are colocated with molecular emission (traced by the 4.5 $\mu$m emission, Figure \ref{fig:spitzer}).  Ionization timescales are $\sim4-6.5\times$ \tauunit.  In all three regions, Ne is less than the CSM value, Mg and Si are slightly enhanced ($0.5-0.7$ solar), and Fe is slightly elevated, though consistent with the CSM value. As expected from the S EWI, in regions T and W, S is moderately enhanced ($0.71\pm0.24$ S$_\odot$ and $0.83_{-0.24}^{+0.28}$ S$_\odot$, respectively), while region A has no detectable S line.  Region T is the most highly absorbed region, tied with the hole, with $N_H=1.44\pm0.09\times$\nhunit.  Region W is distinct from the other protrusions in that the emission is markedly harder (blue in Figure \ref{fig:rgb}), similar to the metal-rich regions found in N49 \citep{Park2012}.  The CSM+ejecta model fit to the region W spectra confirms subsolar quantities of Mg and Si ejecta, and higher S and Fe ejecta abundances of $0.98_{-0.50}^{+0.75}$ S$_\odot$ and $1.83_{-0.40}^{+0.52}$ Fe$_\odot$, respectively.  There appears to be some Si, Mg, and Fe in all three regions and S in regions T and W, but no Ne.
The combination of higher absorption and a prominent S line in region W is likely what set it apart from the other protrusions.

\subsection{Group 5}
\label{section:group5}
Despite being widely separated, regions G, H, L, and R all have similar spectral characteristics.  Regions G and L are both bright knots in the southeast interior and northwest rim, respectively, while R and H are fainter, more diffuse features on the northeast and western edges.  Their ionization timescales are $3-4\times$\tauunit. 
Ne abundances are all $\lesssim$ the CSM value.  Mg and Si are only slightly enhanced in each region, measuring $0.5-0.6$ solar.  Fe is $\sim$solar.  S is constrained in all four regions, and is somewhat higher than the CSM value, especially  in region H, where S $=0.76_{-0.35}^{+0.73}$ S$_\odot$.  The two-component spectral fit to region G finds ejecta abundances of Ne $=0.92_{-0.24}^{+0.32}$ Ne$_\odot$, Mg $=0.73_{-0.17}^{+0.23}$ Mg$_\odot$, Si $=1.23_{-0.40}^{+0.89}$ Si$_\odot$, S $=0.55_{-0.43}^{+0.54}$ S$_\odot$, and Fe $=1.21_{-0.23}^{+0.28}$ Fe$_\odot$.  Overall, regions G, H, L, and R seem to contain Fe ejecta, along with minor amounts of Mg, Si, and possibly S (particularly in region H).

\subsection{Group 6}
\label{section:group6}
The largest group includes regions D+E, F, J, K, O, and S.  Region D+E, two prominent protrusions which were combined for the spectral analysis, and nearby region F are located at the very soft eastern edge in the vicinity of ambient unshocked material (illustrated by the 8 $\mu$m emission).  Regions J, K, O, and S are all on the opposite side of the remnant, but in the interior rather than the edge.  All have ionization ages of $\sim$$2.5-3.0\times$\tauunit\ and correspond to bright knots or filaments of X-ray emission with supersolar Fe abundances.  Si abundances in all regions are very close to solar, except region D+E with Si $=0.73_{-0.13}^{+0.18}$ Si$_\odot$.  Mg abundances are lower, $0.7-1.0$ Mg$_\odot$, but still enhanced.  The Ne abundances are all consistent with the CSM value, except region J (Ne $=1.01\pm0.21$ Ne$_\odot$).  S is only constrained in regions D+E and K, where it is consistent with the CSM value (though with large uncertainty), and region S, where it is slightly higher at $0.88_{-0.45}^{+0.59}$ S$_\odot$.  CSM+ejecta model fits find $\sim$solar Ne abundances (where not zero), except in regions J and O with Ne $=1.33_{-0.41}^{+0.58}$ Ne$_\odot$ and $1.79_{-0.36}^{+0.56}$ Ne$_\odot$, respectively.  Mg ranges from $0.7-1.4$ Mg$_\odot$, Si from $0.8-1.2$ Si$_\odot$, and Fe from $1.6-2.3$ Fe$_\odot$, except in region O.  Region O is an outlier in this group, with the highest abundances measured anywhere in the remnant, Mg $=3.46_{-1.05}^{+1.64}$ Mg$_\odot$, Si $=4.55_{-1.71}^{+2.77}$ Si$_\odot$, and Fe $=6.82_{-2.53}^{+5.11}$ Fe$_\odot$, along with a very low ionization age of $0.2\pm0.1\times$\tauunit. S is unconstrained or zero in every region except S, where it is the same as in the single-component fit.
The high Fe abundances in all of these regions, in both the one- and two-component spectral fits, coupled with the enhanced Si and Mg, suggest they contain some Mg and significant Si and Fe ejecta, Ne in only some regions and very little S.  
These regions are the most similar to the metal-rich ejecta features found in other SNRs, especially region O.

\vspace{-0.071in}
\subsection{Unique Regions}
\label{section:unique}
The first unique region is located in the highly absorbed southwest of the remnant with $n_et=3.8_{-1.3}^{+2.4}\times$\tauunit.  As indicated in the EWIs, region U  has little Ne (less than the CSM value), slightly enhanced Mg ($0.60_{-0.09}^{+0.11}$ Mg$_\odot$), and moderately enhanced Si ($0.75_{-0.13}^{+0.15}$ Si$_\odot$) and Fe ($0.88_{-0.16}^{+0.20}$ Fe$_\odot$).  Region U is the only region with a supersolar S abundance in the single-component fits, at $1.09_{-0.29}^{+0.41}$ S$_\odot$, though the large uncertainties make it consistent with the lower S abundances seen in other regions.  From the  two-component spectral fit we find ejecta abundances of Mg $=1.10_{-0.34}^{+0.77}$ Mg$_\odot$, Si $=1.23_{-0.40}^{+0.89}$ Si$_\odot$, Fe $=2.50_{-0.67}^{+1.44}$ Fe$_\odot$, and S $=1.51_{-0.77}^{+1.43}$ S$_\odot$, the highest S measured anywhere in the remnant.  Region U likely contains significant S and Fe ejecta, as well as Mg and Si, but lacks Ne.

Region I is located in the area of very soft emission at the northern rim, near an area of unshocked local material and shocked dust (green and red in Figure \ref{fig:spitzer}).  It has a similar ionization timescale to region U.  The Mg, Si, and S EWIs suggest no enhanced abundances from these elements, and this is confirmed in the spectral analysis.  Ne and Fe abundances, however, are both somewhat enhanced, $0.88_{-0.13}^{+0.14}$  Ne$_\odot$and $0.84_{-0.15}^{+0.19}$ Fe$_\odot$, respectively.  The temperature of region I is unusually low, $kT = 0.34\pm0.05$ keV rather than the more typical $kT\sim0.55$ keV seen in other regions.  

Region Q is a small bright knot protruding from the western edge.  It is notably soft and located in the same position as molecular emission seen at other wavelengths. 
The ionization timescale is $n_et=2.5_{-1.0}^{+2.0}\times$\tauunit. The Ne, Mg, and Si abundances are all similar to the CSM values, but Fe and S are both elevated ($\sim$0.75 solar).  The CSM+ejecta model spectral fit confirms this picture, with Mg and Si ejecta abundances of $\sim$0.5 solar and S and Fe ejecta abundances of $0.87_{-0.63}^{+0.98}$ S$_\odot$ and $1.60_{-0.52}^{+0.67}$ Fe$_\odot$, respectively.  Region Q may be a small knot of Fe and S ejecta that is proceeding slightly ahead of the main shock.

A bright knot near the center of the remnant, region N is aligned with bright or moderately bright features in the \nehefe, \nely, and Mg EWIs.  It has a low ionization timescale ($n_et=2.0_{-0.6}^{+0.5}\times$\tauunit) but high abundances, $1.8-2.5$ times the CSM values for Ne, Mg, Si, and Fe (S is unconstrained).  Ne, in particular, at $1.4_{-0.15}^{+0.21}$ Ne$_\odot$, is the highest measured anywhere in the remnant from the single-component spectral fits.   As expected from the EWIs, both Mg and Fe are also relatively high, $\sim$solar.  Si is slightly lower, but still enhanced.  A CSM+ejecta fit to the spectra finds $\sim$ solar abundances of Mg and Fe, a Si abundance of $0.71\pm0.12$ Si$_\odot$, and about the same Ne as in the single-component fits, suggesting the presence of ejecta of all four elements, especially Ne.

Region B is one of the small protrusions on the southeast edge of the remnant.  It contains the fewest counts of any region; as a result, the spectral properties are poorly constrained.  With this caveat in mind, the measured ionization timescale is very low ($n_et=1.4_{-0.6}^{+1.3}\times$\tauunit), and the temperature is abnormally high compared to other regions.  The Si abundance, and possibly Ne and Mg, appear to be marginally enhanced compared to the nominal CSM abundances.  If this is the case, it would indicate a small amount of ejecta may contribute to the X-ray emission.  Fe and S abundances are both unconstrained in this region.

The last region is the hole, which has spectral properties unlike any other region.  In addition to being highly absorbed (see \S\ref{section:absorption}), the abundances, except Si, are all extremely low; Ne, Mg, and Fe are lower than any other region. The ionization timescale is also very low, though with considerable uncertainty.  The extremely low abundances combined with an abnormally high Obs ID 970 normalization factor suggest that these measurements should be interpreted with caution.  The high column density, however, is not dependent on the other spectral properties.  Given the corresponding feature in the IR images, it is likely that a foreground cloud is responsible for the low X-ray and IR emission in this region, consistent with the presence of dark clouds mentioned by \citet{Reach2006}.  However, the X-ray emission in the hole is higher than expected given the strong absorption at 8 and 24 $\mu$m.  Out-of-time events and scattering due to the wings of the PSF can be ruled out as sources of the observed X-ray emission, but it is plausible that the hole emission is contaminated by photons from other areas of the remnant being scattered by intervening dust in the ISM. 